\documentclass[a4paper,fleqn,usenatbib]{mnras}

\usepackage{newtxtext,newtxmath}

\usepackage[T1]{fontenc}
\usepackage{ae,aecompl}

\usepackage{graphicx} 
\usepackage{float}
\usepackage{amsmath}	
\usepackage{amssymb}	
\usepackage{dcolumn}
\usepackage{epsfig}
\usepackage{color}
\usepackage{pdflscape}
\usepackage{upgreek}
\usepackage{mathptmx}
\usepackage{grffile}
\usepackage{array,float}
\usepackage{multirow}
\usepackage{lscape}
\usepackage{hyperref}
\hypersetup{draft}
\usepackage{xargs}
\usepackage{xcolor}
\usepackage[normalem]{ulem}
\usepackage{tikz}
\usetikzlibrary{shapes,arrows}
\usepackage{threeparttable}
\usepackage{verbatim}
\usepackage{longtable}

\usepackage{soul,xcolor}

\newcolumntype{d}[1]{D{.}{\cdot}{#1}}
\newcolumntype{.}{D{.}{.}{-1}}

\newcommand{\lsun}{L$_\odot$}
\newcommand{\msun}{M$_\odot$}

\newcommand{\lmratio}{\emph{L}$_{\rm{bol}}$/\emph{M}$_{\rm{fwhm}}$}

\newcommand{\mum}{$\umu$m}

\newcommand{\kms}{km\,s$^{-1}$}
\newcommand{\cmthree}{cm$^{-3}$}

\newcommand{\hii}{H{\sc ii}}
\newcommand{\uchii}{UC\,H{\sc ii}}

\newcommand{\nhthree}{NH$_3$~}

\tikzstyle{decision} = [diamond, draw, fill=blue!20, text width=4.5em, text badly centered, node distance=3cm, inner sep=0pt, text=black, draw=black]
\tikzstyle{block} = [rectangle, draw, fill=blue!20, text width=5em, text centered, minimum height=4em, text=black, draw=black]
\tikzstyle{line} = [draw, -latex', text=black, draw=black, fill=black]

\title[Methanol maser associated dust clumps]{ATLASGAL -- physical parameters of dust clumps associated with 6.7\,GHz methanol masers}

\author[S.\,J.\,Billington et al.]{
	S.\,J.\,Billington,$^{1}$\thanks{E-mail: sjbb2@kent.ac.uk} J.\,S.\,Urquhart,$^{1}$ C.\,K\"{o}nig,$^{2}$  T.\,J.\,T.\,Moore,$^{3}$ D.\,J.\,Eden,$^{3}$ S.\,L.\,Breen,$^{4}$
	\newauthor W.\,-J.\,Kim,$^{5}$   M.\,A.\,Thompson,$^{6}$ S.\,P.\,Ellingsen,$^{7}$ K.\,M.\,Menten,$^{2}$ F.\,Wyrowski,$^{2}$ 
	\newauthor and S.\,Leurini$^{8,2}$ \\
	\\
	$^{1}$ Centre for Astrophysics and Planetary Science, University of Kent, Canterbury, CT2\,7NH, UK \\
	$^{2}$ Max-Planck-Institut f\"{u}r Radioastronomie, Auf dem Higel 69, D-53121 Bonn, Germany \\
	$^{3}$ Astrophysics Research Institute, Liverpool John Moores University, Liverpool Science Park, 146 Brownlow Hill, Liverpool, L3\,5RF, UK \\
	$^{4}$ Sydney Institute fo Astronomy (SIfA), School of Physics, University of Sydney, NSW 2006, Australia \\
	$^{5}$ Instituto de Radioastronom\'{i}a Milim\'{e}trica (IRAM), Granda, Spain \\
	$^{6}$ Science and Technology Research Institute, University of Hertfordshire, College Lane, Hatfield, AL10\,9AB, UK \\
	$^{7}$ School of Mathematics and Physics, University of Tasmania, Private Bag 37, Hobart, Tasmania 7001, Australia \\
	$^{8}$ INAF-Osservatorio Astronomico di Cagliari, Via della Scienza 5, I-09047, Selargius (CA)}

\date{Accepted XXX. Received YYY; in original form ZZZ}

\pubyear{2019}

\begin{document}
\label{firstpage}
\pagerange{\pageref{firstpage}--\pageref{lastpage}}
\maketitle

\begin{abstract}

We have constructed the largest sample of dust-associated class II 6.7\,GHz methanol masers yet obtained. New measurements from the the Methanol MultiBeam (MMB) Survey were combined with the 870\,\micron\ APEX Telescope Large Area Survey of the Galaxy (ATLASGAL) and the 850\,\micron\ JCMT Plane Survey (JPS). Together with two previous studies we have now identified the host clumps for 958 methanol masers across the Galactic Plane, covering approximately 99\,per\,cent of the MMB catalogue and increasing the known sample of dust-associated masers by over 30\,per\,cent. We investigate correlations between the physical properties of the clumps and masers using distances and luminosities drawn from the literature.  Clumps hosting methanol masers are significantly more compact and have higher volume densities than the general population of clumps. We determine a minimum volume density threshold of $n$(H$_2$)$ \geq 10^4$\,cm$^{-3}$ for the efficient formation of intermediate- and high-mass stars. We find 6.7\,GHz methanol masers are associated with a  distinct part of the evolutionary process (\lmratio\ ratios of between 10$^{0.6}$ and 10$^{2.2}$) and have well defined turning on and termination points. We estimate the lower limit for the mass of embedded objects to be $\geq6$\,\msun\ and the statistical lifetime of the methanol maser stage to be $\sim3.3\times$10$^{4}$\,yrs. This suggests that methanol masers are indeed reliable tracers of high mass star formation, and indicates that the evolutionary period traced by this marker is relatively rapid.

\end{abstract}



\section{Introduction}

Massive stars ($\geq8$\,\msun) play a fundamental role in the evolution of the universe and the interstellar medium (ISM), through large amounts of feedback, ultraviolet (UV) radiation fields and dispersal of heavy elements. They have a substantial effect on their host galaxies and potentially fuel future generations of star formation \citep{Elmegreen1992, kennicutt2005}. Given the extensive effect that high-mass stars have, not only on their local environments but also on the  whole of their host galaxy, it is important to understand the underlying processes of the earliest stages of their formation. 

Our knowledge of massive star formation is still fairly limited as relatively few of these objects are located closer than a few kpc, and they almost exclusively form in tight clusters or loosely bound associations, making it difficult to distinguish between individual protostellar cores and stars observationally. More importantly, high-mass stars are relatively rare and evolve far more quickly than lower-mass stars, and so will reach the main sequence while still deeply embedded within their initial environment. These initial environments, where the earliest stages of high-mass star formation takes place, are massive clumps of dust and gas which can be traced through their thermal dust emission at (sub)millimetre wavelengths. At these wavelengths, thermal dust emission is optically thin and is therefore an excellent tracer of column density and global clump mass \citep{Schuller2009}. Thermal dust emission also has the advantage of being sensitive towards the colder pre-stellar stages of star formation, and so allow the study of the whole evolutionary sequence of massive star formation. 

A few other tracers exist that can be attributed to the formation of massive stars. One in particular is the class\,II (radiatively pumped) 6.7\,GHz methanol (CH$_3$OH) masers (hereafter referred to as `methanol masers' or just `masers') first reported by \cite{Menten1991}. Multiple studies have found that these masers are exclusively associated with high-mass YSO candidates (e.g. \citealt{Minier2003, Xu2008, Breen2013}), subsequent submillimetre surveys (e.g. \citealt{Hill2005, Breen2010, Urquhart2013, Urquhart2015}) have found that the majority of masers are found to be almost ubiquitously associated with dense clumps across the Galactic plane ($\sim$99\,per\,cent; \citealt{Urquhart2015}). This provides a simple and convenient method of identifying potential regions of embedded massive star formation in a range of different environments across the Galaxy. The Methanol Multibeam (MMB; \citealt{green2009_full}) survey has produced an unbiased catalogue of 972 class\,II\,6.7\,GHz methanol masers across the Galactic plane ($186\degr < \ell < 60\degr$; Table\,\ref{table:mmb_coverages}), and has been utilised in a number of follow up studies of Galactic structure and coincidence with other maser types (e.g. \citealt{Green2017, Breen2018}). It has been found that methanol masers are one of the most populous star formation species and are associated with a number of other maser species, such as water (H$_2$O) masers \citep{Titmarsh2014,Titmarsh2016} and 12.2\,GHz methanol maser emission \citep{Blaszkiewicz2004,Breen2011a}. Maser sources are seen to increase in luminosity as they, and their parent star forming environments evolve, and that 6.7\,GHz methanol masers trace a well defined stage of massive star formation \citep{Breen2010,Urquhart2015}.

The APEX Telescope Large Area Survey of the GALaxy (ATLASGAL; \citealt{Schuller2009, csengeri2014}) is an unbiased dust continuum survey of the inner Galactic plane and has identified $\sim$10\,000 dense clumps across the inner Galaxy (\citealt{contreras2013, Urquhart2014}). A comparison between the MMB survey catalogue and the ATLASGAL compact source catalogue (CSC) was presented in \citet{Urquhart2013}. This was accompanied by a set of dedicated follow-up dust continuum observations, which targeted MMB sources located within the ATLASGAL region not associated with an ATLASGAL counterpart and MMB sources not covered by ATLASGAL (i.e. $|b| > 1.5\degr$ or $186\degr \leq \ell  \leq 280\degr$; \citealt{Urquhart2015}). Combined, these studies associated 99\,per\,cent of all MMB methanol masers with dense star forming clumps. These two studies focused on the 707 methanol masers identified at the time ($186\degr \leq \ell  \leq 20\degr$). The environment, physical properties and Galactic distribution of maser associated dust clumps were investigated with the methanol masers being preferentially associated with more massive and luminous dust clumps.

Furthermore, other studies have also found that there is a relationship among the different maser transitions and species in these regions. \cite{Ellingsen2007b} presented a ``straw man'' model for the evolutionary sequence of masers in star formation regions. This model has been refined by more recent studies (e.g. \citealt{Breen2010}). However, how this model relates to the surrounding material and environment has not yet been fully developed. We hope to develop this model by looking at the statistical lifetimes of methanol masers, similar to previous works \citep{VanDerWalt2005}, albeit with a larger and more complete sample of masers.

\begin{figure}
    \includegraphics[width=0.47\textwidth]{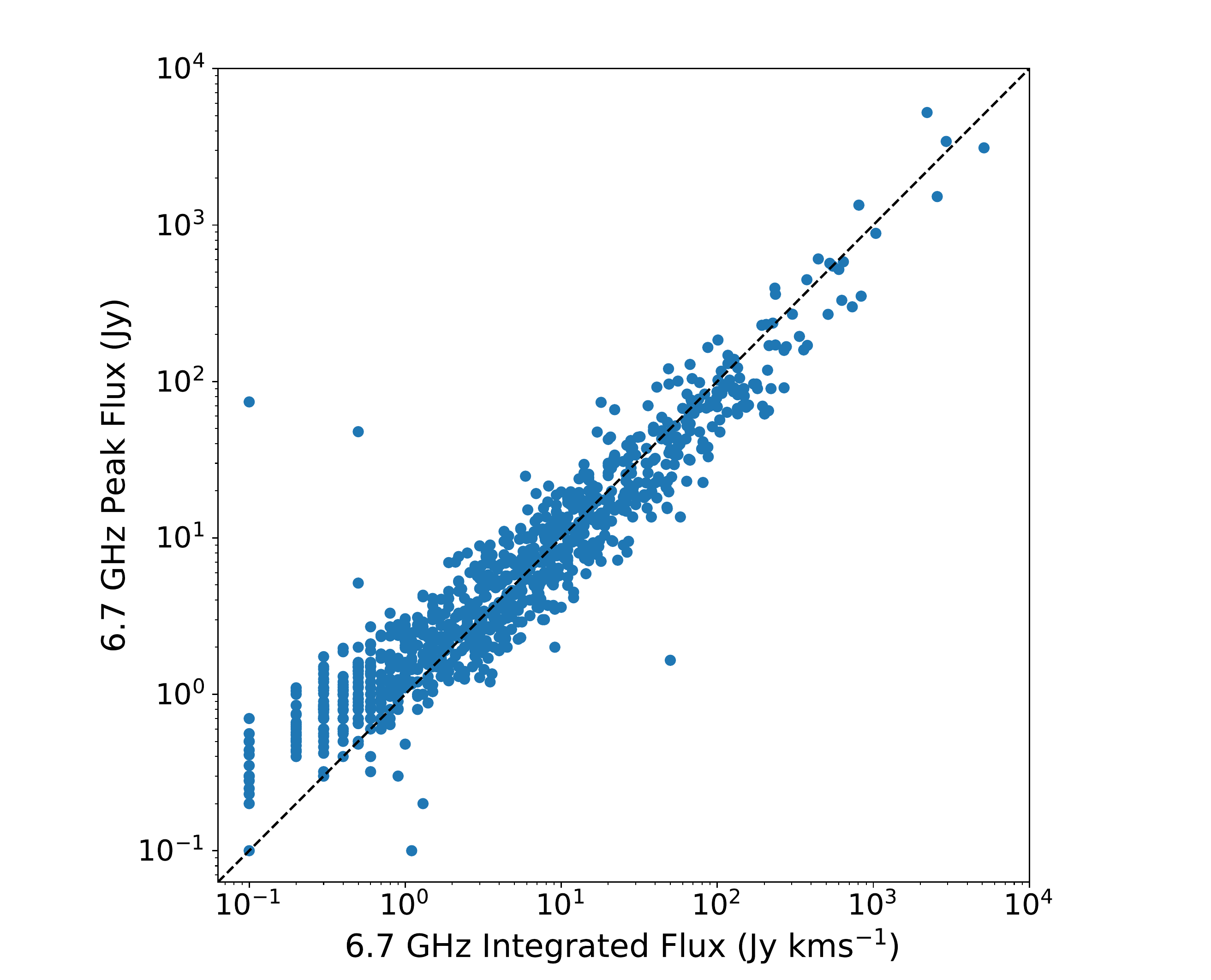}
    \caption{Distribution of methanol maser peak flux densities versus the maser integrated flux densities for the entire MMB catalogue \protect\citep{Breen2015}. The 1:1 line is also shown.}
    \label{fig:mmb_peak_int_fluxes}
\end{figure}

In this paper we match the methanol masers reported in the most recent MMB catalogue (265 sources; \citealt{Breen2015}) with dense clumps identified by the ATLASGAL and JPS (JCMT Plane Survey; \citealt{Moore2015,Eden2017}) surveys, to produce the largest and most complete census of the physical environments where high-mass stars are forming across the Milky Way. This particular study will focus on expanding these previous works by taking advantage of the availability of integrated maser fluxes \citep{Breen2015}, distances and bolometric luminosities that were not available in the two previous studies conducted (\citealt{Urquhart2013, Urquhart2015}). We compare the physical parameters of this methanol maser associated sample of clumps with the properties of the general population of clumps to investigate when this stage begins and ends, and how the properties differ as the clumps evolve. We also use this large sample to critically test and refine correlations between the physical properties of the host clumps and the methanol masers that have been previously reported in the literature \citep{Hill2005,Breen2010}.

The structure of this paper is as follows: in Section\,\ref{sect:surveys}, we describe the survey data which will be used throughout the study, while Section\,\ref{sect:associations} describes the matching process between methanol masers and dust continuum sources. In Section\,\ref{sect:physical_parameters} we estimate the physical properties of maser associated dust clumps, along with uncertainties for these physical parameters. Section\,\ref{sect:discussion} presents a discussion of various aspects of our sample, including how masers are related to the evolutionary stage of their host clumps. We also derive lower limits for mass and density for maser emission production, and give estimate statistical lifetimes for methanol maser emission. A summary of our findings is presented in Section\,\ref{sect:conclusions}.

\begin{table}
	\caption{\label{table:mmb_coverages} Summary of the different longitude ranges covered in the various MMB survey papers and the number of 6.7\,GHz methanol maser sites detected in each range.}
	\begin{tabular}{ccl}
		\hline
		\hline
		Longitude range & \# of maser sites & Reference\\
		\hline
		$345\degr \leq \ell \leq 6\degr$    & 183          & \cite{Caswell2010} \\
		$6\degr \leq \ell \leq 20\degr$     & 119          & \cite{Green2010}   \\
		$330\degr \leq \ell \leq 345\degr$  & 198          & \cite{Caswell2011a}\\
		$186\degr \leq \ell \leq 330\degr$  & 207          & \cite{Green2012}  \\
		$20\degr \leq \ell \leq 60\degr$    & 265          & \cite{Breen2015}   \\
		\hline
		$186\degr \leq \ell \leq 60\degr$   & 972	       &\\
		\hline
	\end{tabular}
\end{table}

\begin{figure*}
	\includegraphics[width=0.47\textwidth]{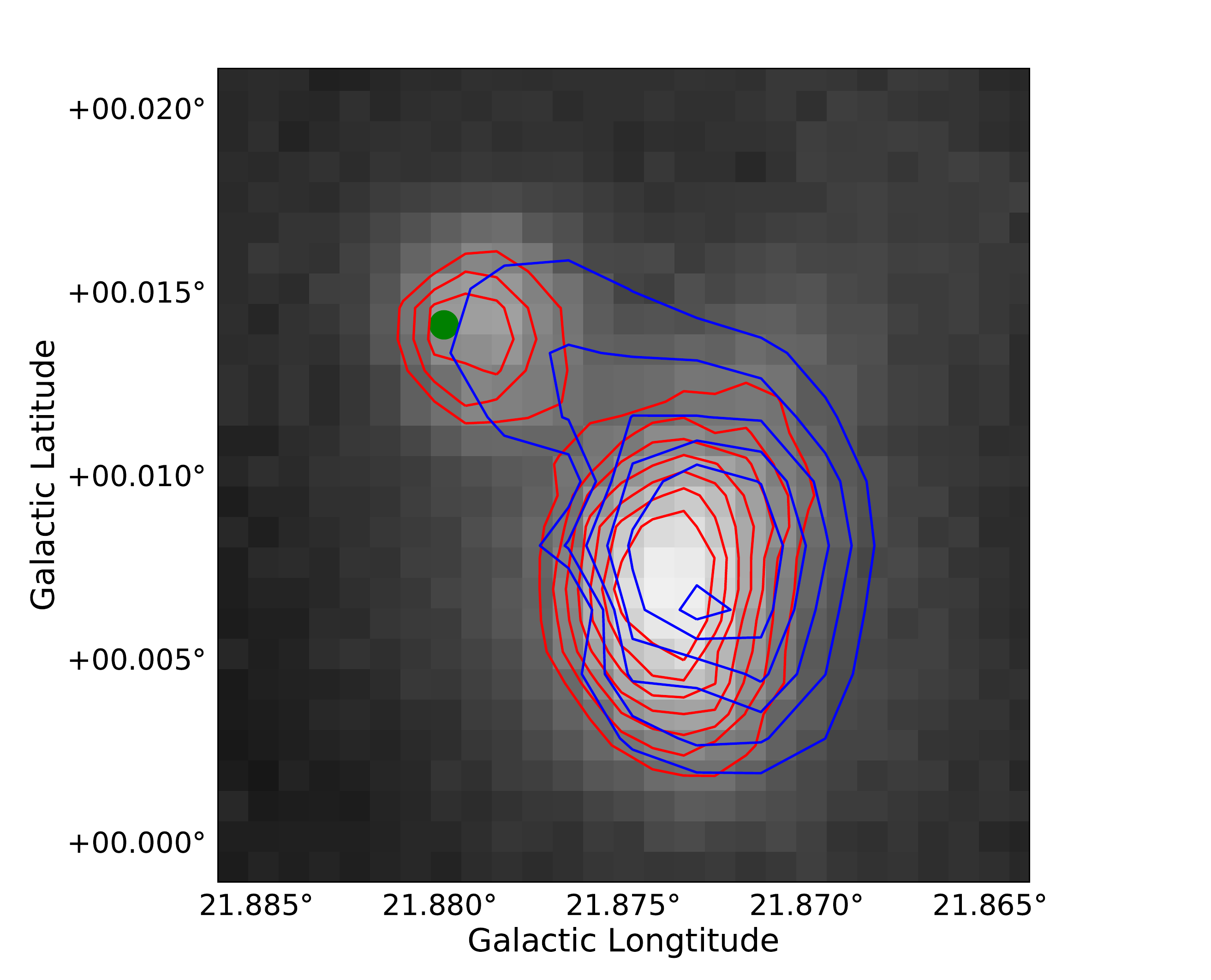}
	\includegraphics[width=0.47\textwidth]{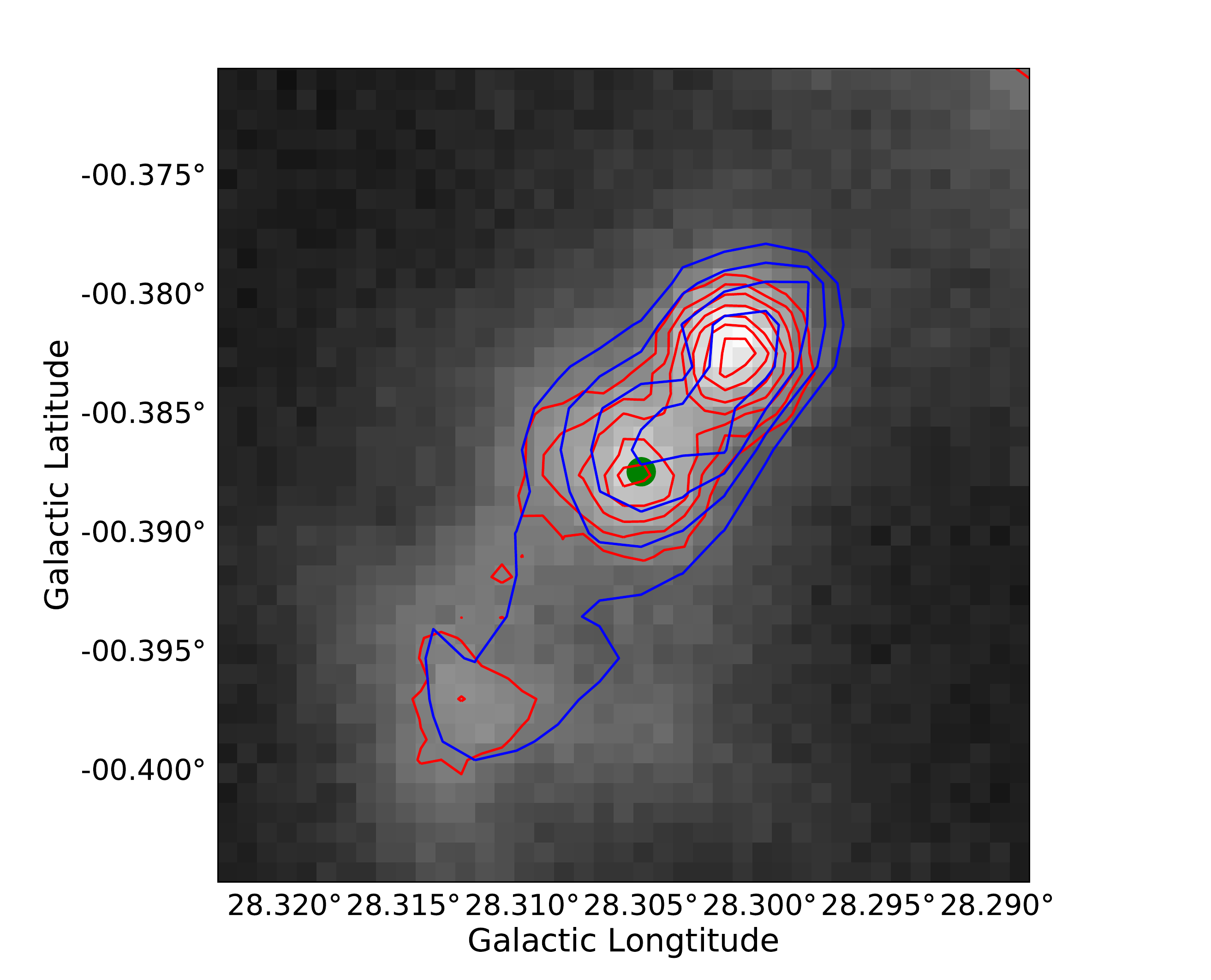}
	
	\caption{Example maps of JPS 850\,\micron\ emission from two regions, AGAL021.873+00.007 \& AGAL028.301$-$00.382. The red and blue contours trace the dust emission detected by the JPS and ATLASGAL emission respectively, the contours start at 3$\sigma$ and increase in steps of 0.5$\sigma$. The position of the methanol masers are indicated by filled green circles. These maps illustrate how the higher sensitivity and resolution of the JPS data is better able to trace the structure of the clumps compared to ATLASGAL.}
	\label{fig:jps_substructure_example}
\end{figure*}

\section{Survey Descriptions}
\label{sect:surveys}

In order to determine the dust properties of associated dense clumps we have opted to use both the JPS and ATLASGAL surveys. ATLASGAL has produced a complete sample of clump properties across the inner Galactic plane, however, we use JPS to help refine these properties for a select number of clumps. An overview of these two surveys, together with the MMB survey is given in the following subsections.

\subsection{Methanol Multibeam (MMB) survey}

The Methanol Multibeam (MMB) survey (\citealt{green2009_full}) has surveyed $186\degr \leq \ell  \leq 60\degr$ and $|b| < 2\degr$ of the Galactic plane in search of the 6.7\,GHz methanol masers using the Parkes 64\,m radio telescope. Detected maser emission sites were followed-up using the Australia Telescope Compact Array (ATCA) or the Multi-Element Radio Linked Interferometer Network (MERLIN; \citealt{Thomasson1986}) to determine accurate positions ($<1$\,arcsec) for those sources that did not have previously determined positions. The total number of masers detected is 972 and these have been reported in a series of papers, each of which has focused on a specific range of Galactic longitude. In Table\,\ref{table:mmb_coverages} we present a summary of the longitude ranges, number of masers detected and the publication where these are reported. It is likely that the entire MMB survey accounts for the majority of 6.7\,GHz methanol maser sources across the Galaxy, as it has been shown that away from the Galactic plane, methanol maser emission is rare \citep{Yang2017,Yang2019}.

Until recently, the integrated flux density for methanol maser sources has been unavailable in the literature, and previous studies have used the peak maser fluxes in order to estimate the corresponding maser luminosities. The integrated flux densities are now available \citep{Breen2015} and shall be used in this work to derived the maser luminosities. Figure\,\ref{fig:mmb_peak_int_fluxes} presents the distribution of the maser peak fluxes against the maser integrated fluxes for the entire MMB catalogue \citep{Breen2015}. The two flux samples are similar above 1\,Jy, however,  below this value we find that the peak flux densities are systematically larger. A consequence of this is that studies that have used the peak flux density to estimate the luminosities will have systematically overestimated the luminosities for the weaker masers \citep{Urquhart2015}. This effect was also noted by \cite{Breen2016a} for the 12.2\,GHz methanol maser transition (see Fig.\,2 of \citealt{Breen2016a}). They determined that this deviation is due to weaker masers exhibiting fewer spectral features, and therefore, having lower integrated flux densities than their stronger counterparts.

\subsection{ATLASGAL survey}

The APEX Telescope Large Area Survey of the GALaxy (ATLASGAL; \citealt{Schuller2009}) is the largest, most sensitive ground-based submillimetre wavelength survey to date. It traces dust emission at 870\,\micron\ across the Galactic plane and covers 420 sq. degrees  (|$\ell$| < 60\degr, |$b$| < 1.5\degr\ and $280\degr \leq \ell  \leq 300\degr$ with $-2\degr \leq b \leq 1\degr$). The ATLASGAL survey was carried out using the Large APEX Bolometer Camera (LABOCA; \citealt{siringo2009}), an array of 295 composite bolometers centred at a wavelength of 870\,\micron\ with a passband of 150\,\micron; this instrument was specifically designed for fast mapping of large areas of the sky at moderate resolution and with high sensitivity. The 12\,m diameter APEX telescope has a full width at half-maximum (FWHM) resolution of 19.2\,arcsec. 

Approximately 10\,000 dense clumps have been identified from the ATLASGAL maps (\citealt{contreras2013, Urquhart2014}). Many of these have since been followed-up with molecular line observations to obtain radial velocities {\citep{Wienen2012, Csengeri2016, wienen2018, Urquhart2018,Urquhart2019}}, which when combined with a Galactic rotation curve can be used to calculate kinematic distances and derive physical properties (\citealt{Wienen2015}). Images from the HiGAL survey (\citealt{molinari2010a,Molinari2016}) have been used to fit the spectral energy distributions (SED), to determine the dust temperatures and bolometric luminosities of the embedded protostellar objects (\citealt{Konig2017}).  Many of these techniques have been applied to determine the physical properties of a large fraction ($\sim$8\,000; \citealt{Urquhart2018}) of the clumps located outside the Galactic centre region (i.e., $|\ell| > 5\degr$); the Galactic centre region has been excluded due to issues with source confusion and the difficulty obtaining reliable kinematic distances. However, reliable distances and physical properties are now available for a large fraction of the ATLASGAL catalogue.

\subsection{JCMT Plane Survey}
\label{sect:jps}

The JCMT Plane Survey (JPS; \citealt{Moore2015}) is an 850\,\micron\ continuum survey of six fields in the northern inner Galactic plane with a longitude range of $7\degr \leq \ell \leq 63\degr$. Each field covers approximately 12 sq. degrees (with $|b| < 1\degr$ and $\ell$ centred at 10\degr, 20\degr, 30\degr, 40\degr, 50\degr\ and 60\degr). The survey was conducted using the Submillimetre Common-User Bolometer Array\,2 (SCUBA2; \citealt{Holland2013}) on the James Clerk Maxwell Telescope. The JPS survey has produced a compact source catalogue of $\sim$8\,000 sources \citep{Eden2017} that has a completeness of 95\,per\,cent at 0.04\,Jy\,beam$^{-1}$ and 0.3\,Jy for the peak and integrated fluxes respectively. The JCMT has a dish size of 15\,m with an angular resolution (FWHM) of 14.4\,arcsec.

The JCMT survey provides a useful complementary data set to the ATLASGAL survey due to the similar wavelength range, modestly better resolution than ATLASGAL (14.4\,arcsec compared to 19.2\,arcsec) and significantly better sensitivity ($\sim$8 times improved). These factors mean that the JPS data are better able to identify substructure within larger ATLASGAL clumps. We show two examples in Fig.\,\ref{fig:jps_substructure_example} to illustrate this point. In both of these examples the ATLASGAL contours (blue) hint at the presence of multiple components, however, the difference in peak flux was not sufficient (i.e., $\Delta {\rm SNR} <  3$) for the extraction algorithm to be able to separate them into distinct objects. The improved sensitivity of JPS means that although the fluxes for the different components are similar, the difference in the signal to noise ratio is sufficient (i.e., $\Delta {\rm SNR} > 3$) for source extraction to identify them as individual structures. 

\cite{Eden2017} compared the peak and integrated fluxes between the detected ATLASGAL and JPS sources, and found them to be linearly correlated. They reported a small systematic offset with lower JPS peak flux densities, which was shown by \cite{Moore2015} to be due to the smaller beam size of the JCMT. 




\setlength{\tabcolsep}{3pt}
\begin{table*}
	\caption{\label{table:mmb_matches}Matched MMB sources with the ATLASGAL and JPS catalogues; maser offsets and continuum fluxes are shown for each survey. Maser names appended with a * are where we have opted to use the measured JPS flux.}
	\begin{threeparttable}
	\begin{tabular}{llccclccc}
		\hline
		\hline
		MMB name & ATLASGAL CSC  & ATLASGAL-MMB & ATLASGAL            & ATLASGAL   & JPS catalogue &   JPS-MMB   &   JPS              & JPS \\
		& name          & offset    & peak flux           & integrated flux   & name  &   offset    &   peak flux        & integrated flux \\
		&               & (\,arcsec) & (Jy\,beam$^{-1}$)   & (Jy)       &          &   (\,arcsec) & (Jy\,beam$^{-1}$) & (Jy) \\
		\hline
		G020.081-00.135 &  AGAL020.081-00.136 &            1.76 &        7.68 &      18.43 &  JPSG020.080-00.135 &       3.88 &           6.2 &        14.64 \\
		G020.237+00.065 &  AGAL020.236+00.064 &            6.31 &        1.45 &        5.1 &  JPSG020.237+00.066 &       2.16 &           1.1 &         4.24 \\
		G020.239+00.065 &  AGAL020.236+00.064 &           12.83 &         $-$ &        $-$ &  JPSG020.237+00.066 &       2.16 &           1.1 &         4.24 \\
		G020.364-00.013 &  AGAL020.362-00.012 &             5.3 &        1.95 &       4.84 &  JPSG020.362-00.013 &       6.99 &          1.65 &         3.92 \\
		G020.733-00.059 &  AGAL020.731-00.059 &            9.87 &        2.81 &      19.19 &  JPSG020.732-00.060 &      12.32 &          2.41 &         8.57 \\
		G020.926-00.050 &                 $-$ &             $-$ &         $-$ &        $-$ &  JPSG020.926-00.050 &       5.63 &          0.18 &         0.34 \\
		G020.963-00.075 &  AGAL020.962-00.074 &            4.42 &        0.42 &        1.4 &  JPSG020.963-00.074 &       8.56 &          0.41 &         1.33 \\
		G021.023-00.063* &  AGAL021.000-00.057 &           83.52 &        0.48 &       2.32 &  JPSG021.023-00.063 &        4.9 &          0.33 &         0.71 \\
		G021.407-00.254 &  AGAL021.406-00.252 &            7.17 &        0.55 &        1.2 &  JPSG021.407-00.255 &       5.35 &          0.43 &         0.88 \\
		G021.562-00.033 &  AGAL021.561-00.032 &            6.38 &        1.03 &       4.04 &  JPSG021.562-00.032 &       3.83 &          0.78 &         2.15 \\
			\hline
		\end{tabular}
		\begin{tablenotes}
			\centering
			\item Notes: Only a small portion of the data is provided here, the full table is only available in electronic format.
		\end{tablenotes}
	\end{threeparttable}	
\end{table*}

\section{MMB Associations}
\label{sect:associations}

We have attempted to associate each of the 265 masers listed in the latest methanol maser catalogue published by the MMB survey (\citealt{Breen2015}), with either an ATLASGAL or JPS detection from their respective catalogues. This was done by identifying any methanol maser that is within the 3$\sigma$ boundary of a dense clump found in the ATLASGAL and JPS catalogues (\citealt{Urquhart2014, Eden2017}, respectively). Images have been created of every region so that each match could be confirmed visually. This has resulted in 232 methanol masers being matched to ATLASGAL sources and 146 to JPS sources, with 131 masers having associated emission in both ATLASGAL and JPS. We find that there are 11 masers which are associated with emission seen in JPS with no counterpart in ATLASGAL. Matches between the MMB survey and ATLASGAL/JPS can be found in Table\,\ref{table:mmb_matches}. 

\subsection{Angular correlation of the $20\degr \leq \ell \leq 60\degr$ masers}
\label{sect:matching_statistics}
\begin{table}
	\caption{\label{table:mmb_non_matches}Detected sources from the MMB survey which have no corresponding dust continuum emission at 850 or 870\,\micron. Note the last two entries in this table are not covered by either the JPS or ATLASGAL surveys and so are only included in this table for completeness. Maser names appended with * are where the lower submillimetre limits are derived from the JPS 850\,\micron\ emission.}
	\begin{threeparttable}
		\begin{tabular}{lcccc}
			\hline
			\hline
			MMB Name               & $S_{{\rm peak},6.7\,{\rm GHz}}$ & $S_{{\rm int},6.7\,{\rm GHz}}$ & $S_{{\rm peak},{\rm submm}}$ & \\
			& (Jy)          & (Jy\,\kms)   & (Jy\,beam$^{-1}$)            \\
			\hline
			MMB005.677$-$00.027    & 0.79          & 0.4                 & <0.12 \\
			MMB014.521$+$00.155    & 1.40          & 2.4                 & <0.12 \\
			MMB023.126+00.395      & 1.10          & 1.0                 & <0.04 \\
			
			MMB029.993-00.282*      & 1.70          & 0.9                 & <0.03 \\
			MMB032.516+00.323*      & 1.60          & 0.5                 & <0.03 \\
			MMB033.486+00.040      & 3.30          & 3.7                 & <0.06 \\
			MMB044.644-00.516      & 0.50          & 0.1                 & <0.05 \\
			MMB059.833+00.672*      & 0.00          & 17.7                & <0.03 \\
			
			MMB303.869$+$00.194    & 0.90          & 0.6                 & <0.12 \\
			MMB337.517$-$00.348    & 1.50          & 0.3                 & <0.10 \\
			MMB350.470$+$00.029    & 1.44          & 0.2                 & <0.08 \\
			MMB356.054$-$00.095    & 0.52          & 0.2                 & <0.14 \\
			\hline
			MMB026.422+01.685      & 3.80          & 2.6                 & $-$  \\  
			MMB035.200-01.736      & 519.90        & 601.3               & $-$  \\
			\hline
		\end{tabular}
	\end{threeparttable}
\end{table}

\begin{figure}
	\includegraphics[width=0.47\textwidth]{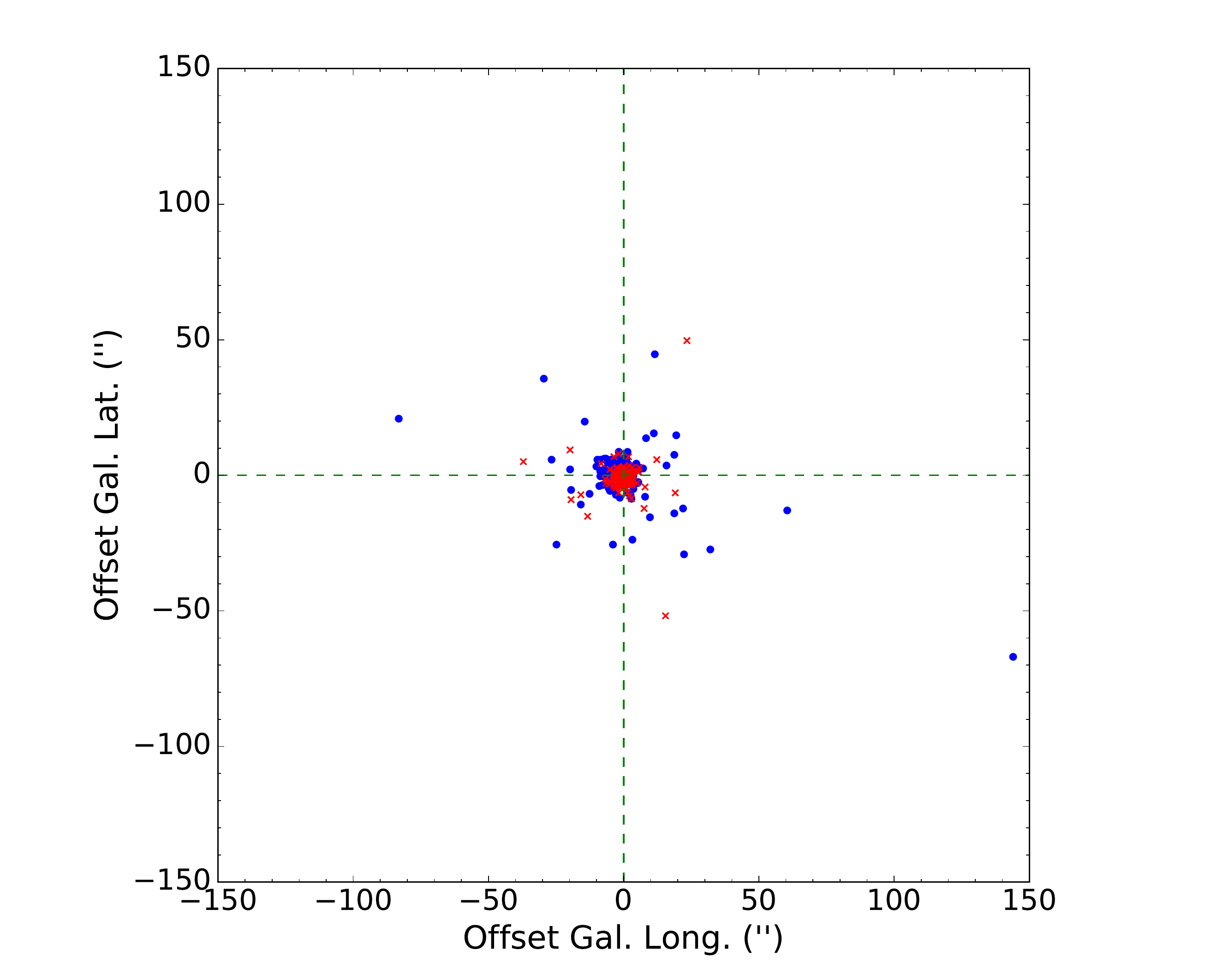}
	\includegraphics[width=0.47\textwidth]{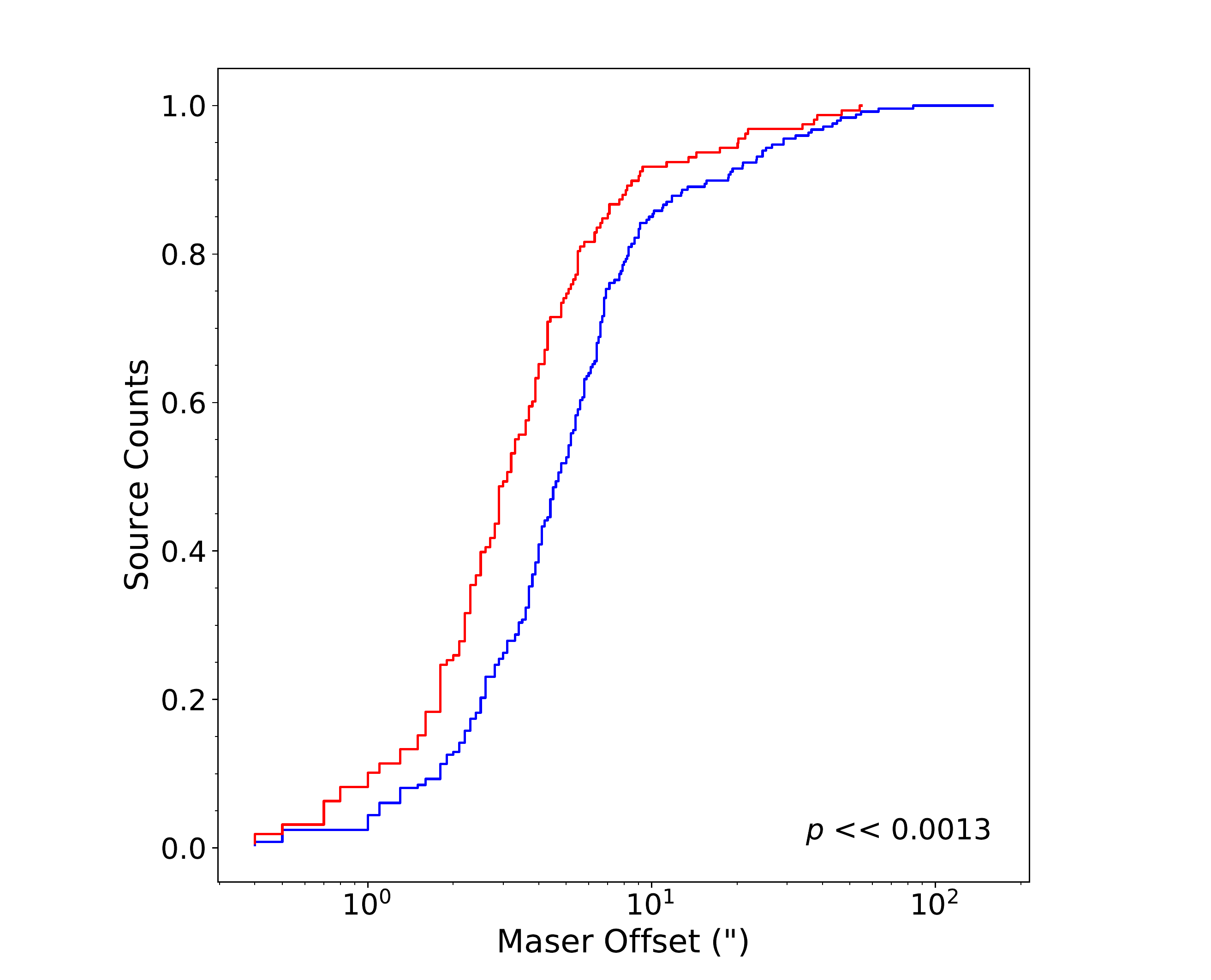} 
	\caption{Offset distributions of the matching statistic for the latest addition of the MMB survey ($20\degr < \ell < 60\degr$). The upper panel presents the 2D offsets between methanol maser detections and the ATLASGAL and JPS source catalogues. The lower panel shows the cumulative distribution functions of the MMB-JPS and MMB-ATLASGAL offsets. The $p$-value from a KS test is shown in the lower right of the plot. In both plots the distribution of the MMB-JPS and MMB-ATLASGAL offsets are shown in red and blue respectively.}
	\label{fig:atlas_jps_maser_offsets}
\end{figure}

\begin{figure}
	\centering
	\begin{tikzpicture}
	\draw[black, thick] (-1,0) circle (2cm);
	\node at (-1,2.3) {ATLASGAL};
	\node at (-1.7,0) {111};
	\draw[black, thick] (1,0) circle (2cm);
	\node at (1,2.3) {JPS};
	\node at (1.7,0) {15};
	\node at (0,0) {131};
	\end{tikzpicture}
	\caption{Venn diagram presenting the matching statistics for detected masers in the region $l = 20\degr - 60\degr$. The total sample of methanol masers in this region is 265, for which we have found a match to either ATLASGAL of JPS for 257 of these. Manual extraction has been used for 10 of the ATLASGAL matches and 4 of the JPS matches. For the 6 sources with no corresponding detection in ATLASGAL and for a further 8 sources where JPS better describes the clump distribution, we have opted to use the JPS flux values.} 
	\label{fig:venn_matches}
\end{figure}
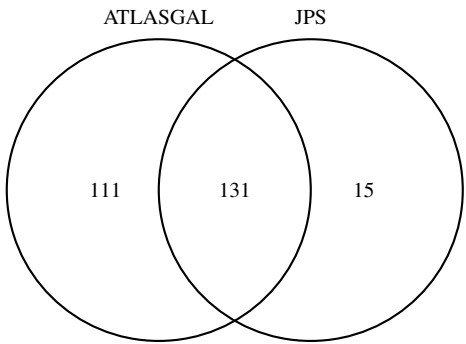

The initial matches were identified using a search radius of 50\,arcsec and this was sufficient to match $\sim$90\,per\,cent of the methanol masers. Visual inspection of the submm emission maps identified another four masers that are located within the boundary of extended ATLASGAL clumps; these matched masers lie further away from the peak of the thermal dust emission. In the upper panel of Fig.\,\ref{fig:atlas_jps_maser_offsets} we show the 2-dimensional angular offsets between the methanol masers and the peak submillimetre (submm) positions given in the JPS and ATLASGAL catalogues. It is clear from this plot that the positions of the vast majority of the masers are tightly correlated with the peak of the submm emission. In the lower panel of Fig.\,\ref{fig:atlas_jps_maser_offsets} we show the cumulative distribution functions (CDFs) of the MMB-JPS and MMB-ATLASGAL offsets; this clearly shows that the offsets between the MMB and JPS sources are smaller than the offsets between MMB and ATLASGAL. Comparing these two distributions using a KS test gives a $p$-value $\ll$ 0.0013 confirming that the differences in maser offsets between the two submillimetre surveys is statistically significant.

The difference in the angular offset between the MMB-JPS and MMB-ATLASGAL associations is likely to be due to the increased resolution and sensitivity of JPS (as discussed in Sect.\,\ref{sect:jps}). To verify this we visually inspected all sources where there is a significant offset between the ATLASGAL and JPS peak emission (see Fig.\,\ref{fig:jps_substructure_example} for examples). In cases where the JPS data has resolved an ATLASGAL clump into multiple components, and where one component provides a significantly smaller offset ($< 16.7$\,arcsec; this corresponds to three times the median of the offset distribution) we have opted to associate the maser to the JPS source as these values are likely to be more reliable in determining the clump properties associated with nearby maser emission. This is the case for 10 methanol maser sources. Inspecting these matches it appears that for 2 of the methanol maser sources (G048.902$-$00.273 \& G049.482$-$00.402), the respective source extraction algorithms employed to create the ATLASGAL and JPS catalogues, have chosen different centroid positions for the same dust emission. For these two cases we have opted to use the ATLASGAL parameters for consistency. We identify the 8 methanol masers that are matched to an ATLASGAL and JPS source where we have decided to use the JPS parameters by appending a * to the MMB name in Table\,\ref{table:mmb_matches}.

We visually inspected the ATLASGAL and JPS maps for every maser that was not matched in either catalogue. This was to identify emission coincident with methanol maser positions that were below the detection threshold used for the automatic source extraction algorithms (SNR $> 5$). This process has identified a further 14 masers found to be associated with ATLASGAL (10) and JPS (4) emission. For these regions we have used the SAOImage DS9 software to place an aperture over the emission to extract the peak and integrated fluxes. Therefore, in total we have matched 257 methanol masers to thermal dust emission with 242 of these maser being associated with ATLASGAL emission and 146 to JPS emission, Fig.\,\ref{fig:venn_matches} presents a Venn diagram of the matching statistics.

No dust continuum emission was detected for 8 of the 265 masers in this region. However, 2 of these lie beyond the Galactic latitude coverage of JPS and ATLASGAL and so no information is available; these are MMB026.422+01.685 and MMB035.200$-$01.736. Therefore we have only failed to match 6 methanol masers with a submillimetre counterpart. We provide the MMB names, maser fluxes and an upper limit for the submillimetre emission in Table\,\ref{table:mmb_non_matches} for these 8 masers and the 7 masers we were unable to associate with submillimetre emission in our previous studies (\citealt{Urquhart2013,Urquhart2015}).

\begin{table*}
	\caption{\label{table:mmb_atlasgal_parameters}Matches between MMB and ATLASGAL for the newest addition to the MMB survey, with corresponding ATLASGAL catalogue parameters. Columns: (1) MMB defined maser name; (2) ATLASGAL CSC name; (3)-(4) Maser peak and integrated flux densities; (5) Molecular line velocities; (6)-(7) Maser peak and median velocities; (8) Maser velocity range; (9) ATLASGAL-MMB Offset.}
	\begin{threeparttable}
	\begin{tabular}{llccccccc}
		\hline
		\hline
		MMB name & Clump name  & MMB peak flux & MMB integrated flux & $v_{\rm{lsr}}$ & $v_{\rm{peak}}$ &  $v_{\rm{median}}$ &  $v_{\rm{range}}$   & MMB offset \\
		&           & (Jy)          & (Jy\,\kms)  & (\kms)               & (\kms)                    & (\kms)                       &  (\kms)        & (\,arcsec)  \\
		\hline
		G020.081-00.135 &  AGAL020.081-00.136 &         1.7 &         0.7 &     41.1 &       43.5 &   43.05 &     1.70 &            1.76 \\
		G020.237+00.065 &  AGAL020.236+00.064 &        84.9 &        -1.0 &     70.8 &       71.9 &   73.00 &     9.80 &            6.31 \\
		G020.239+00.065 &  AGAL020.236+00.064 &         6.9 &        -1.0 &      $-$ &       70.4 &   65.30 &    11.40 &           12.83 \\
		G020.364-00.013 &  AGAL020.362-00.012 &         2.9 &         5.3 &     52.2 &       55.9 &   54.55 &     9.50 &             5.3 \\
		G020.733-00.059 &  AGAL020.731-00.059 &         1.3 &         1.4 &       56 &       60.7 &   59.40 &     8.80 &            9.87 \\
		G020.926-00.050 &                 $-$ &         4.1 &         6.8 &      $-$ &       27.4 &   26.60 &     6.20 &             $-$ \\
		G020.963-00.075 &  AGAL020.962-00.074 &         2.2 &         1.0 &     30.9 &       34.6 &   34.20 &     1.20 &            4.42 \\
		G021.023-00.063 &  AGAL021.000-00.057 &         2.2 &         1.1 &     31.1 &       31.1 &   31.85 &     5.70 &           83.52 \\
		G021.407-00.254 &  AGAL021.406-00.252 &        14.5 &        14.2 &     91.1 &       89.0 &   89.00 &    12.60 &            7.17 \\
		G021.562-00.033 &  AGAL021.561-00.032 &          13 &        18.3 &    113.6 &      117.2 &  114.75 &    11.90 &            6.38 \\		
		\hline
		\end{tabular}
		\begin{tablenotes}
			\centering
			\item Notes: Only a small portion of the data is provided here, the full table is only available in electronic format.
		\end{tablenotes}
	\end{threeparttable}
\end{table*}

\begin{table*}
	\caption{\label{table:mmb_atlasgal_parameters_all} Matches between MMB and ATLASGAL for the entire MMB catalogue, with corresponding ATLASGAL source parameters.}
	\begin{threeparttable}
	\begin{tabular}{llccccccccc}
			\hline
			\hline
			MMB name &          Clump name & Dust        & Distance & Radius        & Maser           & Surface     & Volume              &    Clump     & Clump     &  \lmratio \\
			&                     & temperature &          &                & luminosity     & density     & density             &  luminosity  & mass      & ratio     \\
			&                     & (K)         & (kpc)    & (pc)           & (\lsun)        & (log[\msun\,pc$^{-2}$]) & (log[\cmthree]) & (log[\lsun])  & (log[\msun]) & (log[\lsun/\msun]) \\
			\hline
   G005.618-00.082 &   AGAL005.617-00.082 &       15.05 &     16.2 &           0.79 &          -5.07 &        2.76 &        4.65 &        4.38 &          3.81 &    0.58 \\
   G005.630-00.294 &   AGAL005.629-00.294 &          28 &        3 &           0.12 &          -6.73 &        2.77 &        5.06 &        3.16 &          1.75 &    1.41 \\
   G005.657+00.416 &      G005.656+00.417 &         $-$ &      $-$ &            $-$ &            $-$ &         $-$ &         $-$ &         $-$ &           $-$ &     $-$ \\
   G005.677-00.027 &                  $-$ &         $-$ &      $-$ &            $-$ &            $-$ &         $-$ &         $-$ &         $-$ &           $-$ &     $-$ \\
   G005.885-00.393 &   AGAL005.884-00.392 &       34.54 &        3 &           0.11 &           -7.2 &        2.49 &        6.15 &        5.33 &          2.72 &    2.61 \\
   G005.900-00.430 &   AGAL005.899-00.429 &       25.48 &        3 &           0.15 &          -6.42 &         2.9 &        5.69 &        4.75 &          2.71 &    2.04 \\
   G006.189-00.358 &   AGAL006.188-00.357 &       22.86 &      $-$ &            $-$ &            $-$ &        2.73 &         $-$ &         $-$ &           $-$ &     $-$ \\
   G006.368-00.052 &   AGAL006.368-00.051 &       24.33 &     14.3 &           0.77 &          -5.41 &        2.08 &         4.2 &         4.7 &          3.32 &    1.38 \\
   G006.539-00.108 &   AGAL006.551-00.097 &       25.45 &        3 &           0.25 &          -7.36 &        2.33 &        4.77 &        4.41 &          2.44 &    1.96 \\
   G006.588-00.192 &  AGAL006.588-00.192* &         $-$ &      $-$ &            $-$ &            $-$ &         $-$ &         $-$ &         $-$ &           $-$ &     $-$ \\
			\hline
		\end{tabular}
		\begin{tablenotes}
			\centering
			\item Notes: Only a small portion of the data is provided here, the full table is only available in electronic format.
		\end{tablenotes}
	\end{threeparttable}
\end{table*}

\begin{figure}
	\includegraphics[width=0.47\textwidth]{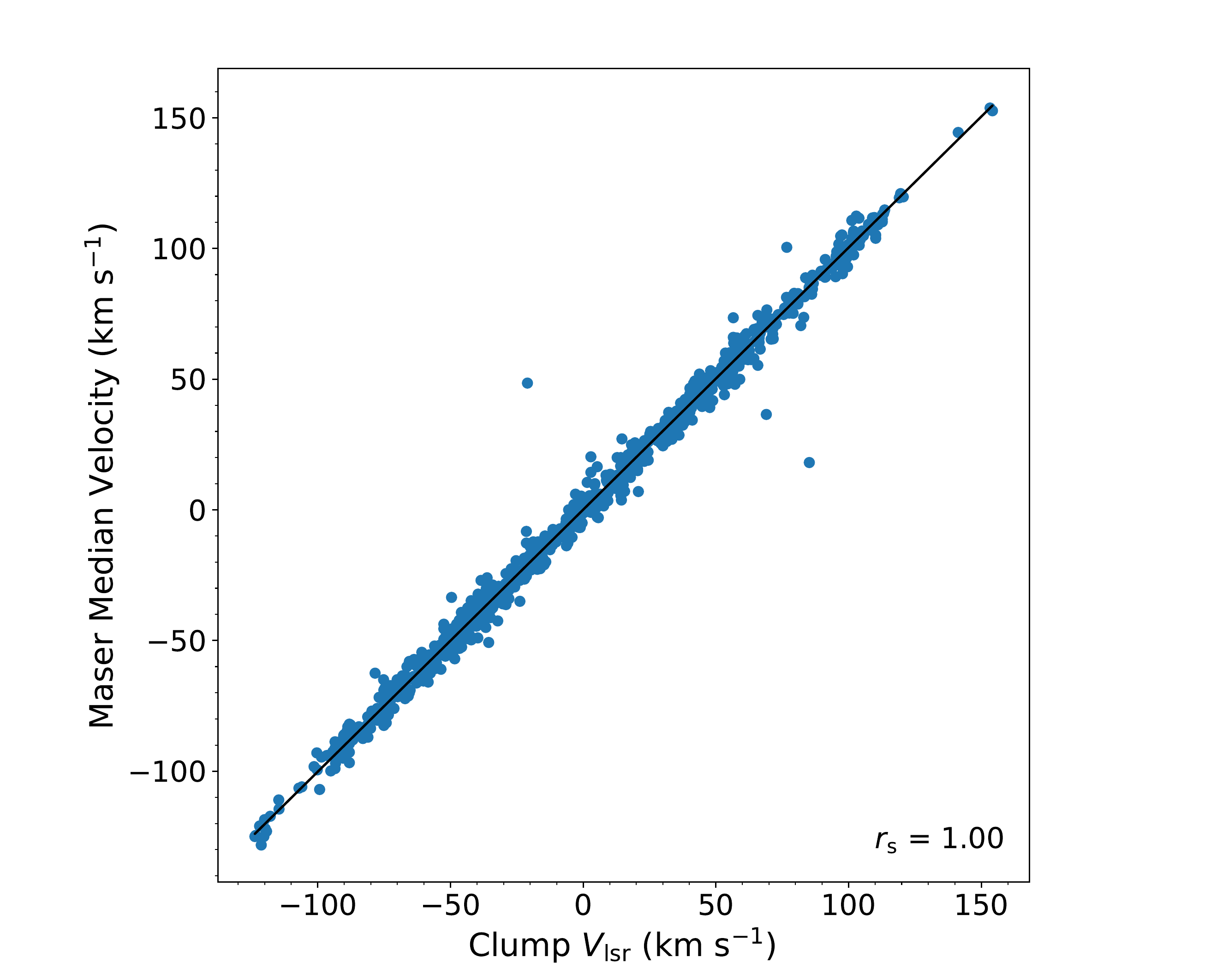} \\
	\includegraphics[width=0.47\textwidth]{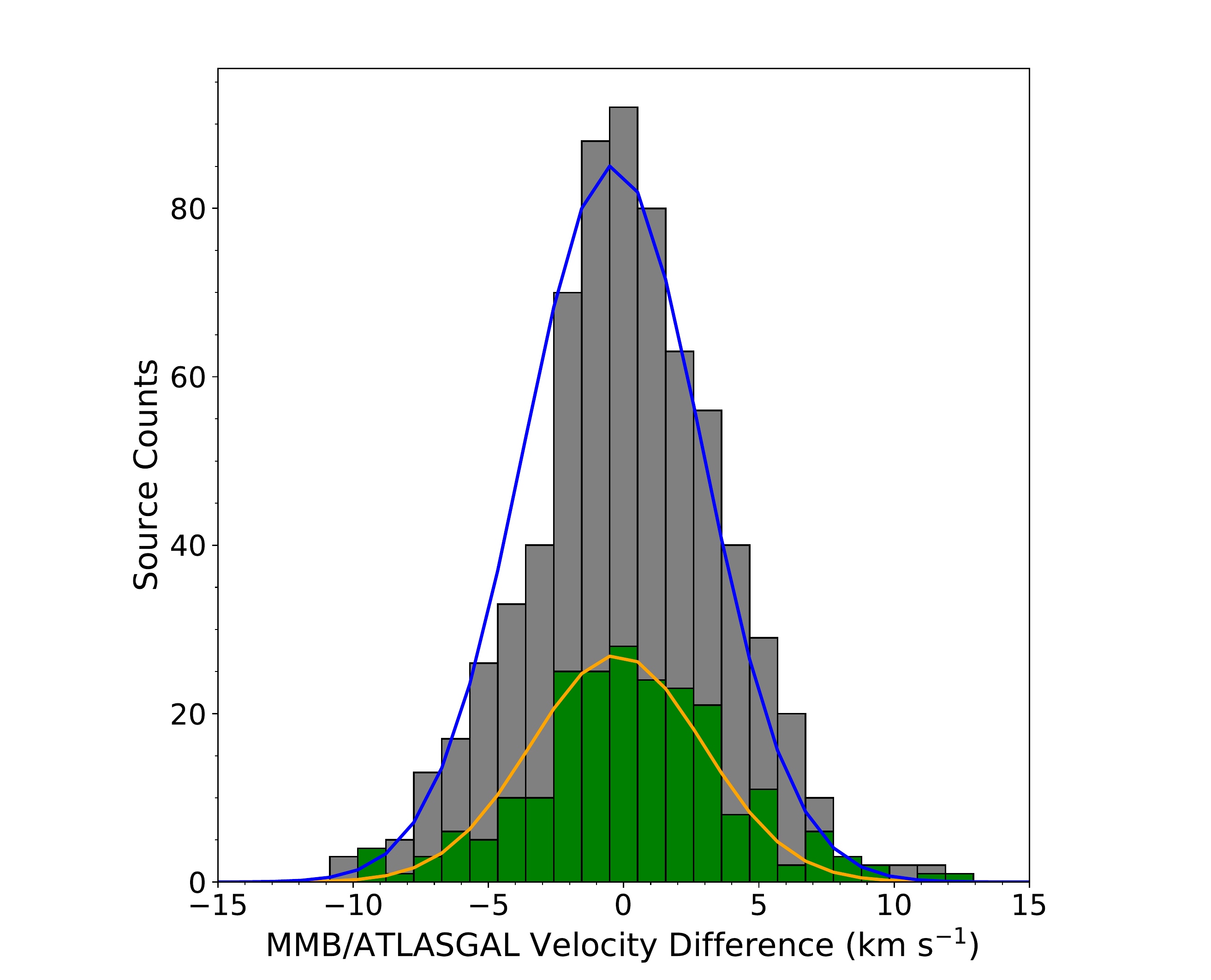}
	\caption{The upper panel presents the distribution of the median maser velocities against the clump velocities for the entire matched sample ($60\degr \geq \ell \geq 186\degr$), with a best fit line shown. The results from a Spearman's rank correlation test is shown in the lower right of the plot. The lower panel presents histograms of the offsets between the median maser velocities and the molecular line velocities for every match in the MMB survey is shown in grey with the matches for the latest addition to the MMB survey shown in green. Gaussian curves have been fitted to the data to determine the mean and standard deviations of the distributions.}
	\label{fig:velocity_difference}
\end{figure}

\subsection{Velocity correlation of the $20\degr \leq \ell \leq 60\degr$ masers}
\label{sect:velo_matching}

Maser velocity information is available for all 265 sources in the latest addition of the MMB survey catalogue. Velocity values are also available for 219 dense clumps in this region, taken from the ATLASGAL catalogue (\citealt{Urquhart2018}). We have compared the velocities of the new methanol masers with the corresponding molecular line velocities to confirm maser associations in velocity space as well as positional space. The upper panel of Fig.\,\ref{fig:velocity_difference} presents these clump velocities against the corresponding spatially matched maser velocities. The majority of masers have small offsets in velocity space and so it is likely that these sources are in fact associated with the dust continuum emission. We have compared the velocity differences between the latest addition of the MMB survey with previously matched maser sources \citep{Urquhart2015} using a KS test. This produces a $p$-value much higher than our significance threshold and so there is likely no difference between these two samples.

The histograms presented in the lower panel of Fig.\,\ref{fig:velocity_difference} show the velocity difference distributions of both the previous and latest matches of the MMB survey. Both histograms have been fitted using a Gaussian model. As there appears to be no difference between these two samples we have opted to use the statistics of the full sample in order to identify any outliers in velocity space. The mean and median velocity difference in this sample is 0.17\,\kms\ and 0.15\,\kms\ respectively, with a standard error of 0.15\,\kms. We find that the standard deviation of this model to be 4.5\,\kms\ and set a confidence threshold of 3$\sigma$ (13.5\,\kms) to identify outliers in this distribution. All of the maser sources within our new sample ($20\degr < \ell < 60\degr$) have a velocity difference of less than this value and we identify no outliers in our distribution.

\subsection{Matching statistics for the whole MMB catalogue}

In the previous section we have matched 257 methanol masers with dust emission. Combining the these new matches with those identified in our previous studies has resulted in 958 of the 972 masers within the MMB catalogue being associated with dust emission; this is an association statistic of 98.5\,per\,cent. There are 14 of the 972 maser sources that have not been matched to any corresponding dust continuum emission. All of these non-matches can be found in Table\,\ref{table:mmb_non_matches}. The majority of these simply have no associated dust emission, likely due to the sensitivity of the JPS and ATLASGAL surveys.

\begin{figure}
	\includegraphics[width=0.47\textwidth]{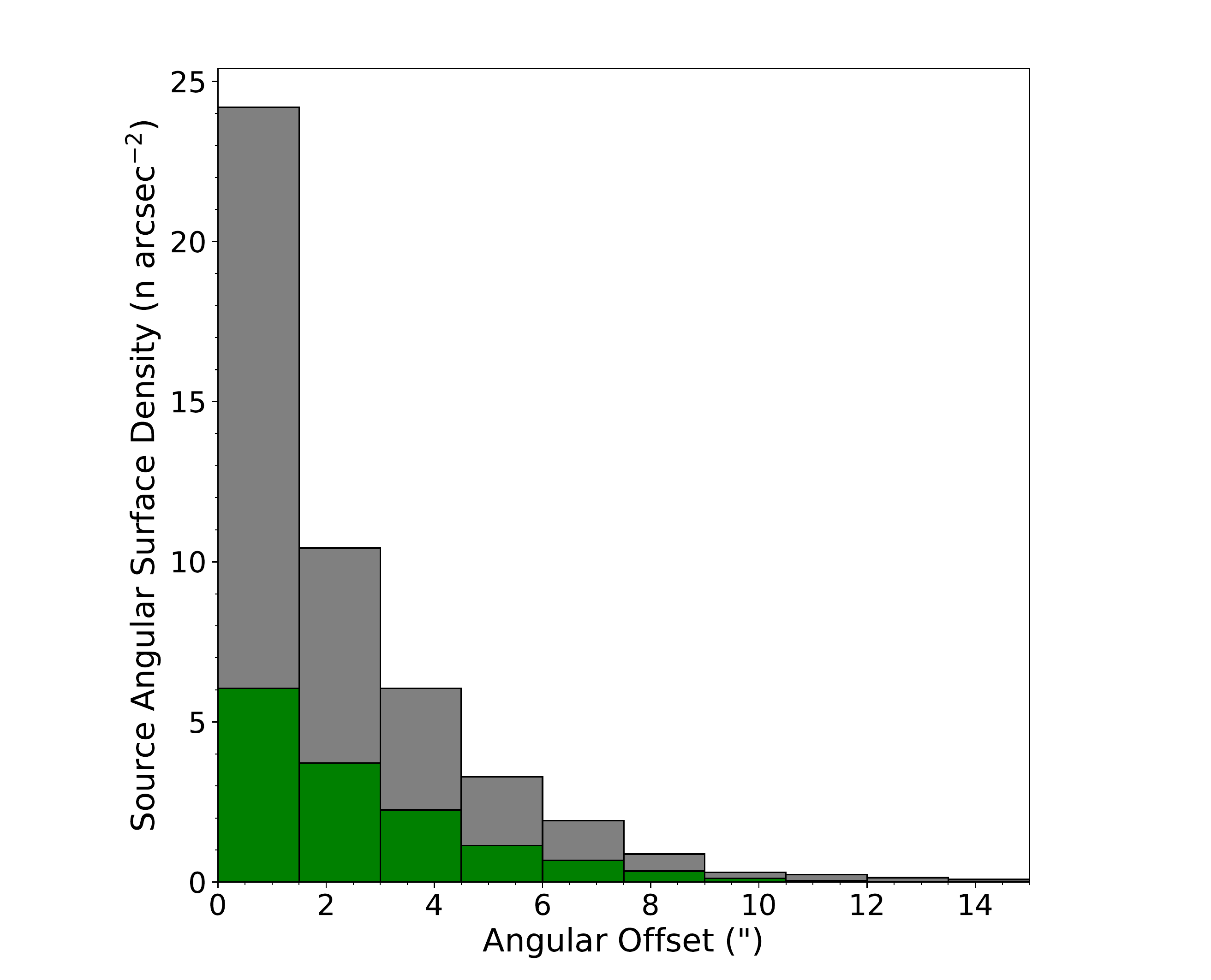}
	\caption{Source angular surface density as a function of angular offset between the methanol masers and the peak of the submm clumps for all the whole matched sample (\citealt{Urquhart2013, Urquhart2015} and this work). Surface densities for the newest addition to the MMB survey ($20\degr < \ell < 60\degr$) are shown in green with the rest of the Galactic sample presented in grey. There are 15 maser associations beyond the limits of this histogram.}
	\label{fig:all_offsets}
\end{figure}

In Fig.\,\ref{fig:all_offsets} we show the source angular surface density distribution of all matched MMB sources (958) as a function of the angular offset. For comparison this plot also shows the distribution for the masers matched in this work. The distributions look indistinguishable from each other and this is confirmed by a KS test ($p$-value = 0.11). The mean of this distribution is 7.6\,arcsec, with a standard deviation of 10.9\,arcsec. We find that there are 33 associations with a higher offset than three times this standard deviation. Although it is likely that all detected associations are real, objects that are at a considerable angular distance from the peak continuum emission are very likely to be embedded in regions of enhanced density, however, are unlikely to be resolved as discrete clumps within JPS or the ATLASGAL survey. Therefore, this study shall only consider sources which have an angular offset of less than this 3$\sigma$ (32.7\,arcsec), ensuring that the derived physical parameters are representative of the material which is actually associated with methanol maser emission.

Figure\,\ref{fig:velocity_difference} presents the velocity offset distribution of the maser associations, for both the newest addition to the MMB survey and for every match in the entire MMB survey. There are 922 maser sources in the entire sample that have a radial velocity measurement and for which we can determine a velocity difference using the median maser velocity. As with the angular offsets, we have chosen to only include sources in our sample that have a velocity offset of less than 3$\sigma$ (13.5\,\kms); 11 sources within the entire matched sample (958 masers) have velocity offsets larger than this value and so are excluded from our statistical analysis of physical properties. 

The maser velocity ranges from this sample have a mean and standard deviation of 7.10 and 5.50\,\kms\ respectively. The maximum velocity range is found to be 28.5\,\kms. \cite{Sridharan2002} found that velocity ranges were confined to 15\,\kms, although this is likely be due to their smaller sample size. The velocity ranges we find are much more consistent with \cite{Caswell2009}. In general, we find that the more luminous masers also have larger velocity ranges, which has been noted in previous studies (e.g. \citealt{Breen2011a}), this relationship is described further in Sect.\,\ref{sect:luminosities}.

Therefore, taking into account the conditions above, along with the number of non associations, our sample consists of 839 matched masers, all of which have an angular and velocity offset of less than 3$\sigma$. Table\,\ref{table:mmb_atlasgal_parameters} presents the matches between the MMB survey and corresponding dust emission for the newest addition to the MMB survey and Table\,\ref{table:mmb_atlasgal_parameters_all} presents these matches for the entire MMB survey.

\begin{figure}
	\includegraphics[width=0.47\textwidth]{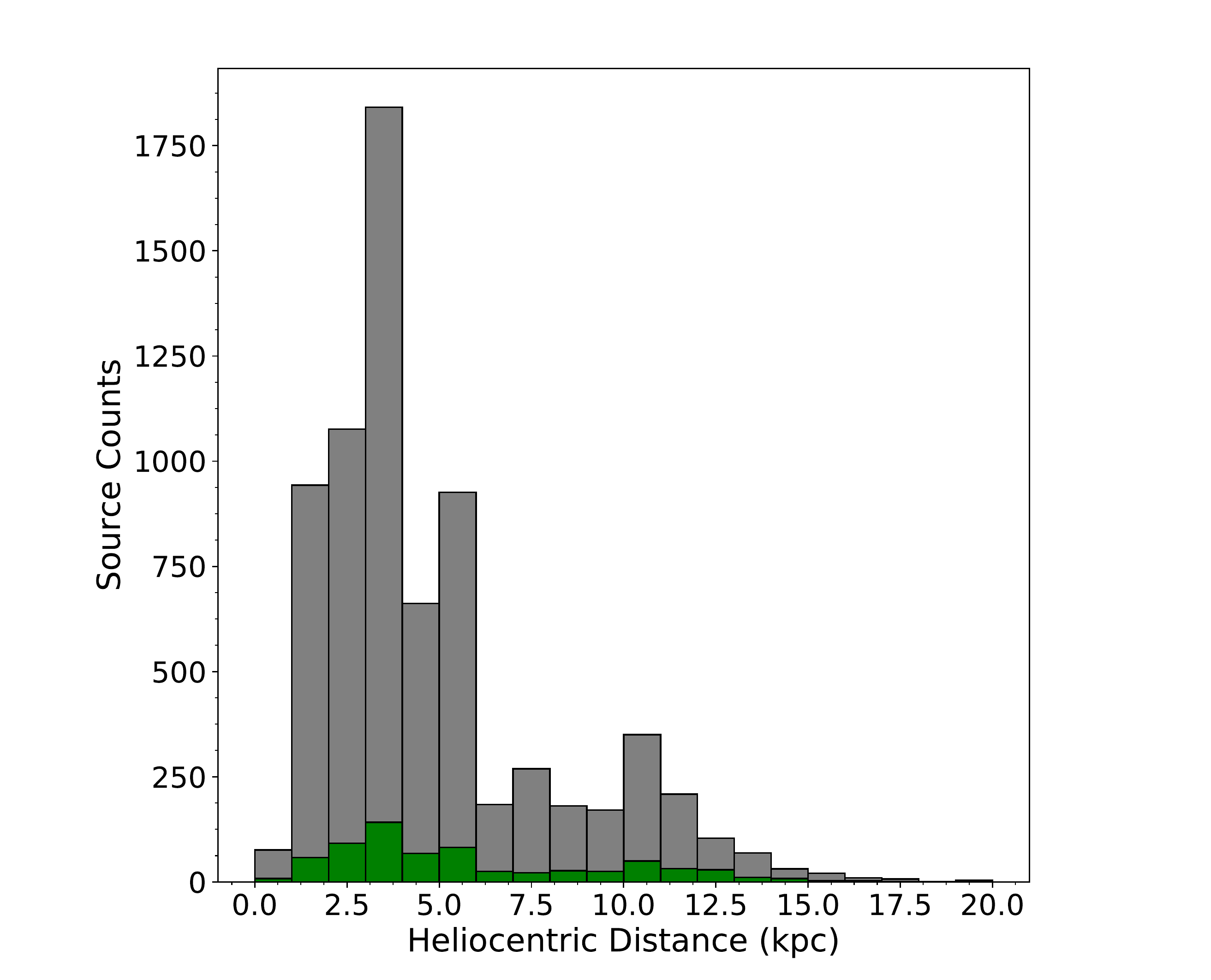}
	\caption{Histogram presenting the distribution of clump heliocentric distances. The entire ATLASGAL sample is shown in grey, with the MMB associated clumps shown in green. The bin size used is 1\,kpc. There are two sources that have a heliocentric distance beyond 20\,kpc. }
	\label{fig:helio_distances}
\end{figure}

\begin{figure}
    \includegraphics[width=0.47\textwidth]{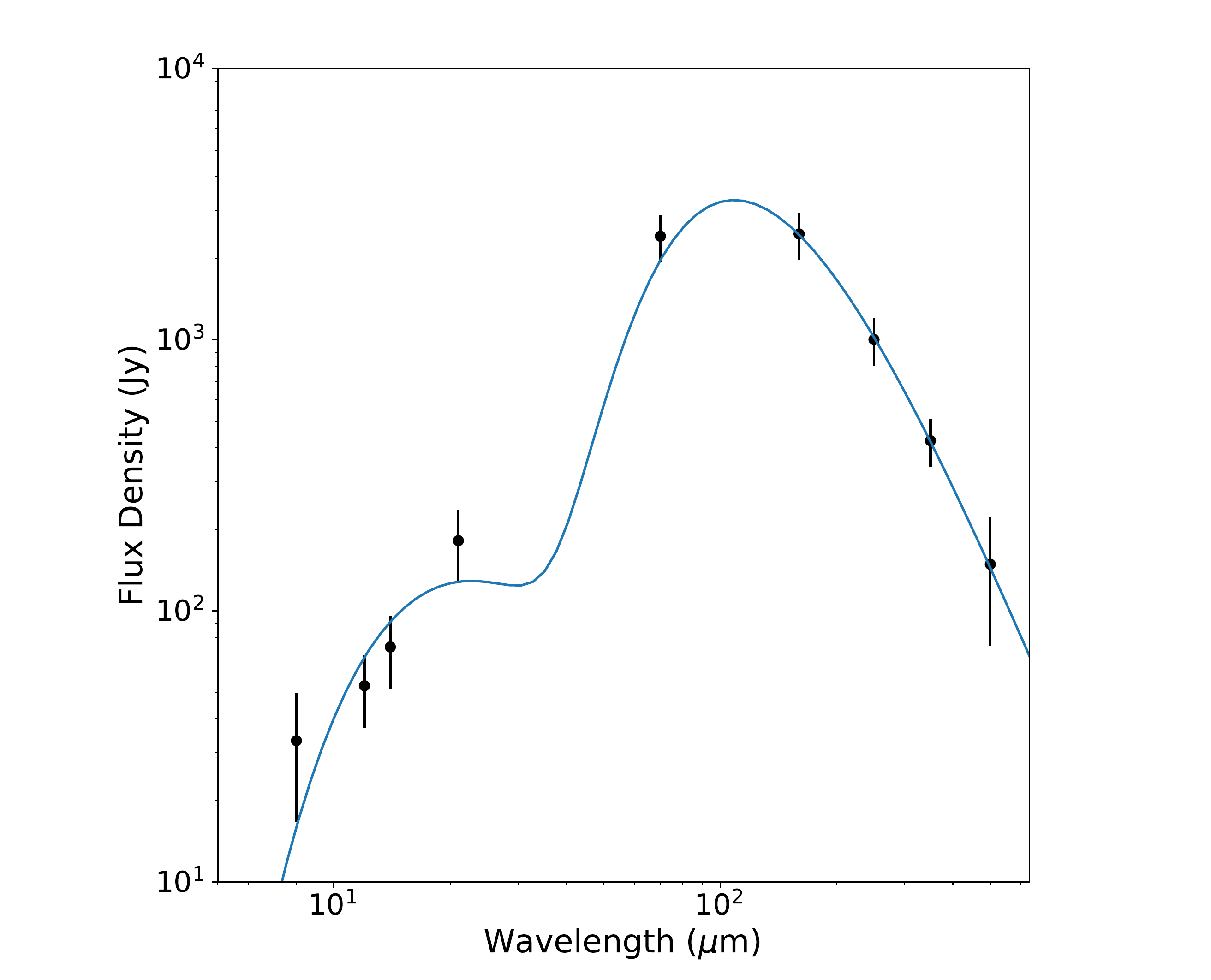}
    \caption{A spectral energy distribution and corresponding fit for an example region (G189.030+00.783). Flux density error bars are shown.}
    \label{fig:sed_fit}
\end{figure}

\begin{figure}
	\includegraphics[width=0.47\textwidth]{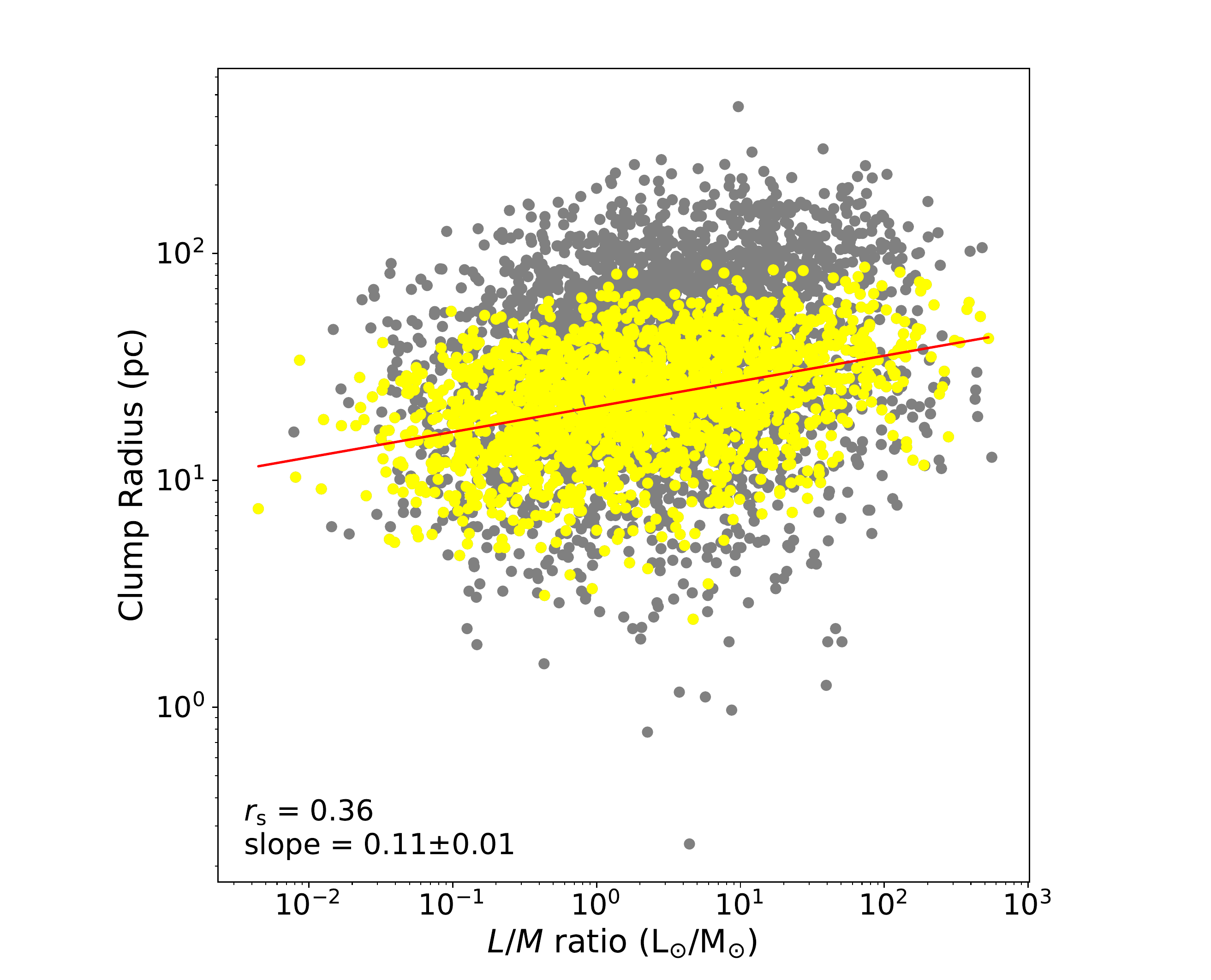}
	\includegraphics[width=0.47\textwidth]{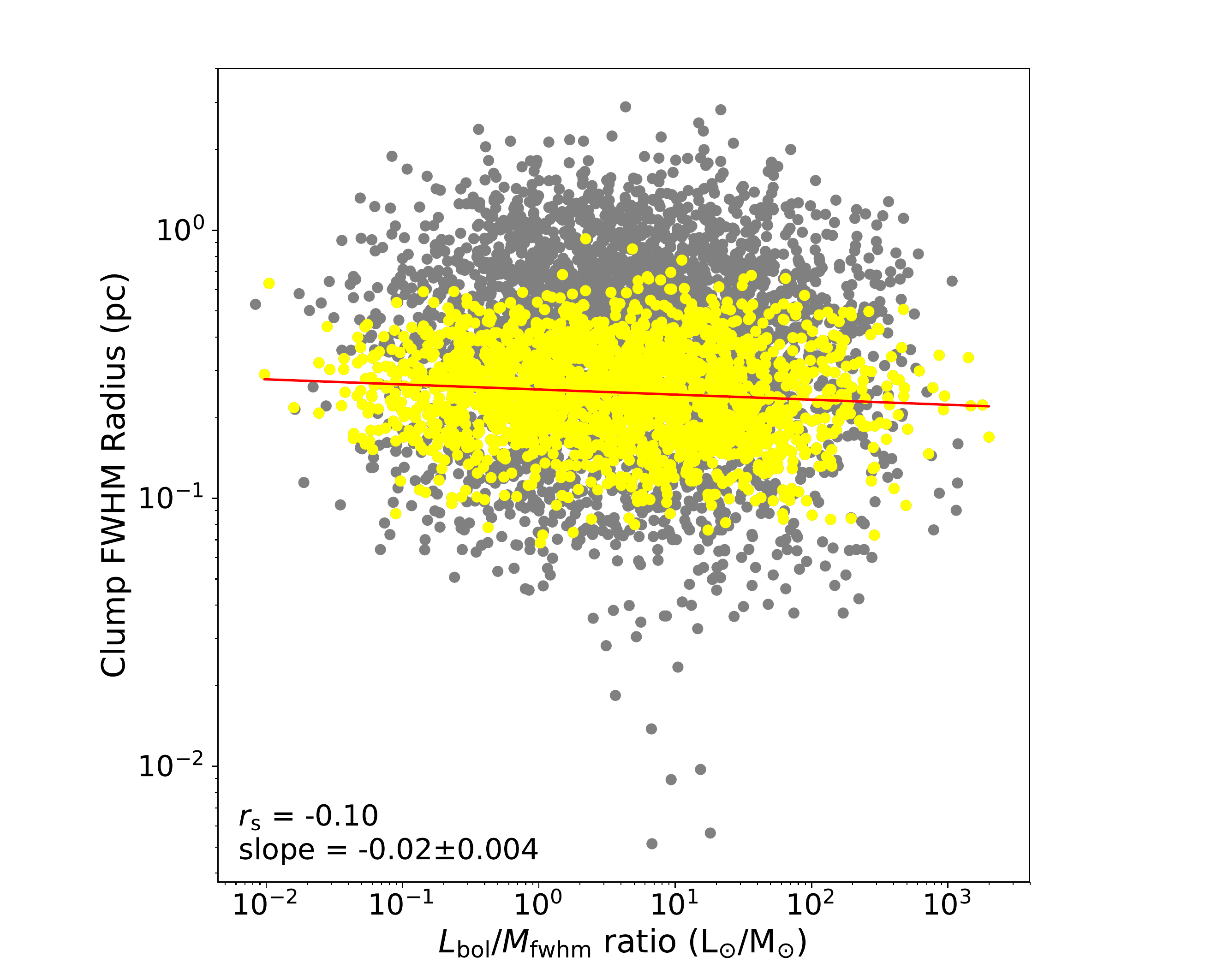}
	\caption{The upper panel presents the clump radius vs clump $L/M$ ratio for previously calculated values of radius, luminosity and mass within the ATLASGAL survey. The lower panel presents clump FWHM radius against clump \lmratio\ ratio, for the newly derived clump sizes (see Sect.\,\ref{sect:clump_sizes}). Both plots present the entire ATLASGAL sample and a distance limited sample (2 to 4\,kpc) in grey and yellow respectively. The results from a Spearman's rank correlation coefficient test for the distance limited samples (2 to 4\,kpc) is shown in the lower left of each panel, the $p$-value for both distributions is below our significance threshold.}
	\label{fig:radius_vs_lms}
\end{figure}

\begin{figure*}
	\includegraphics[width=0.47\textwidth]{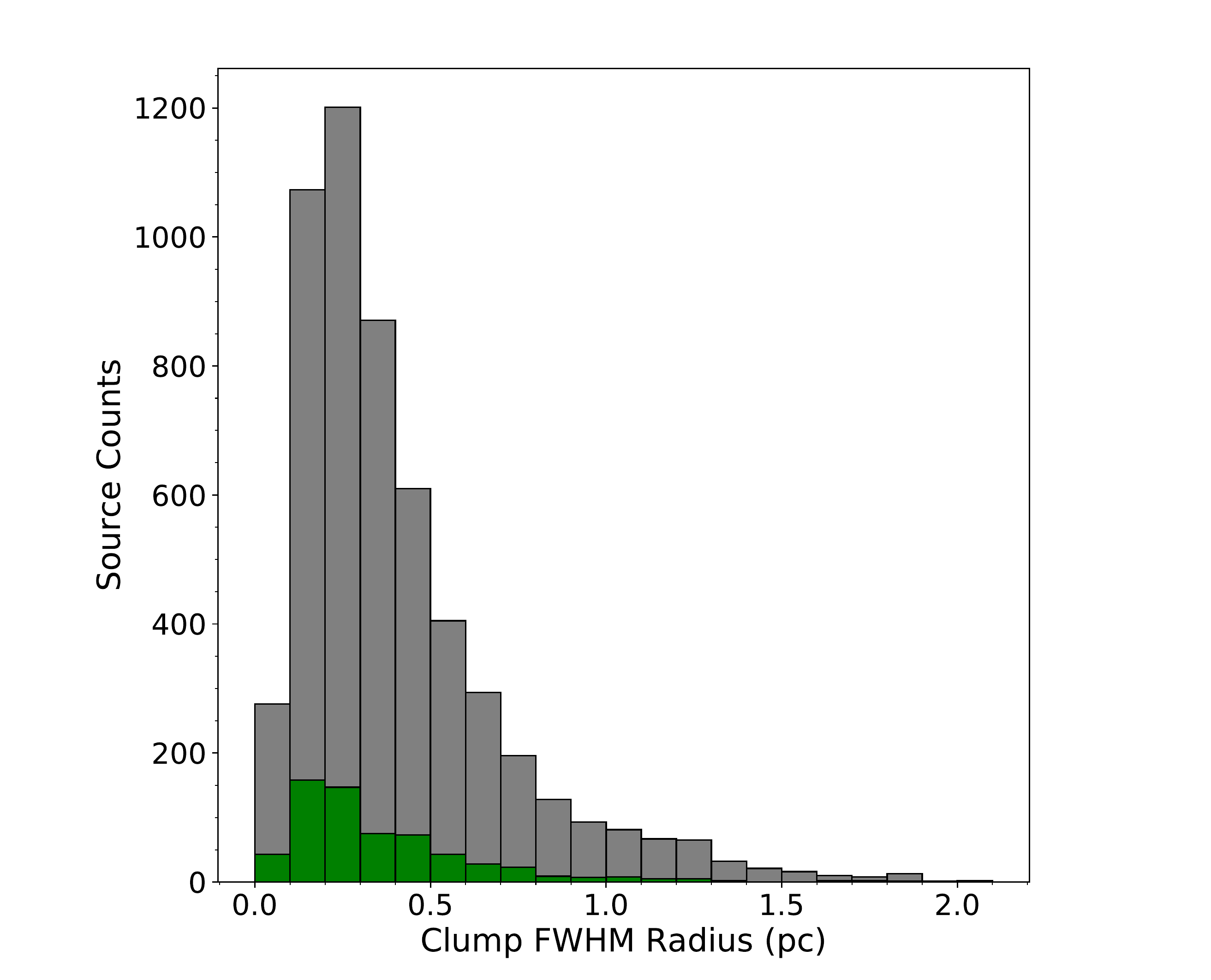}
	\includegraphics[width=0.47\textwidth]{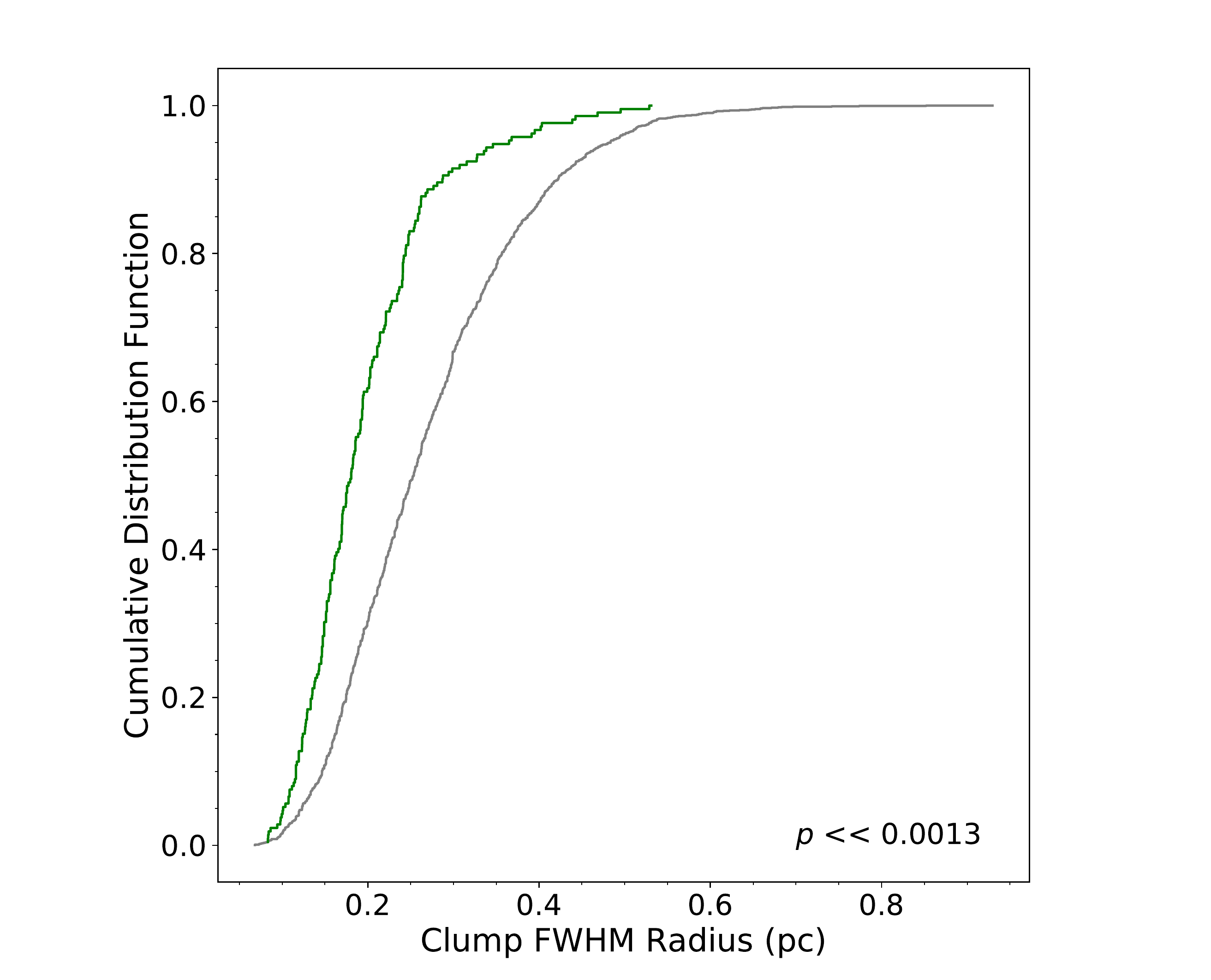} \\
	\includegraphics[width=0.47\textwidth]{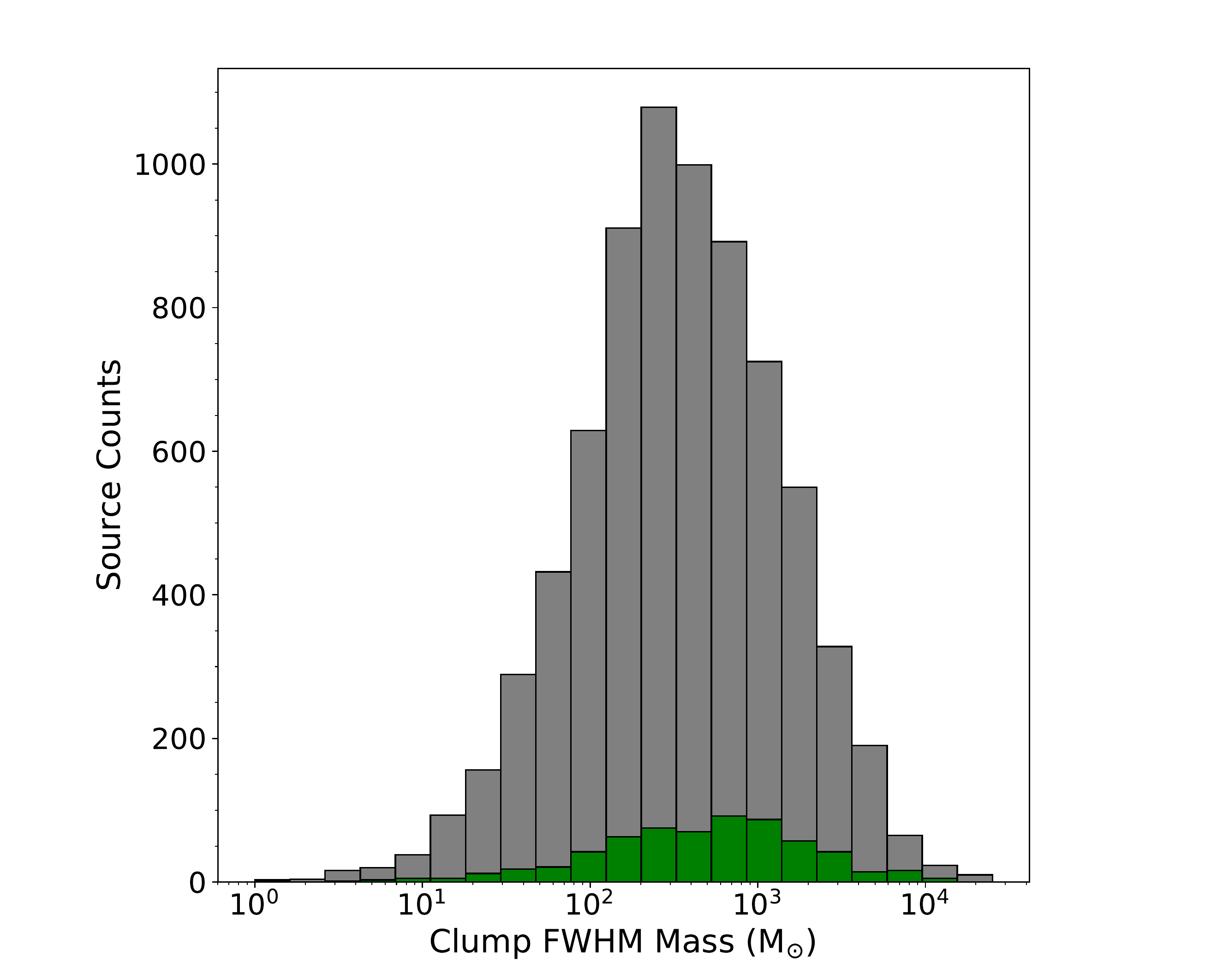}
	\includegraphics[width=0.47\textwidth]{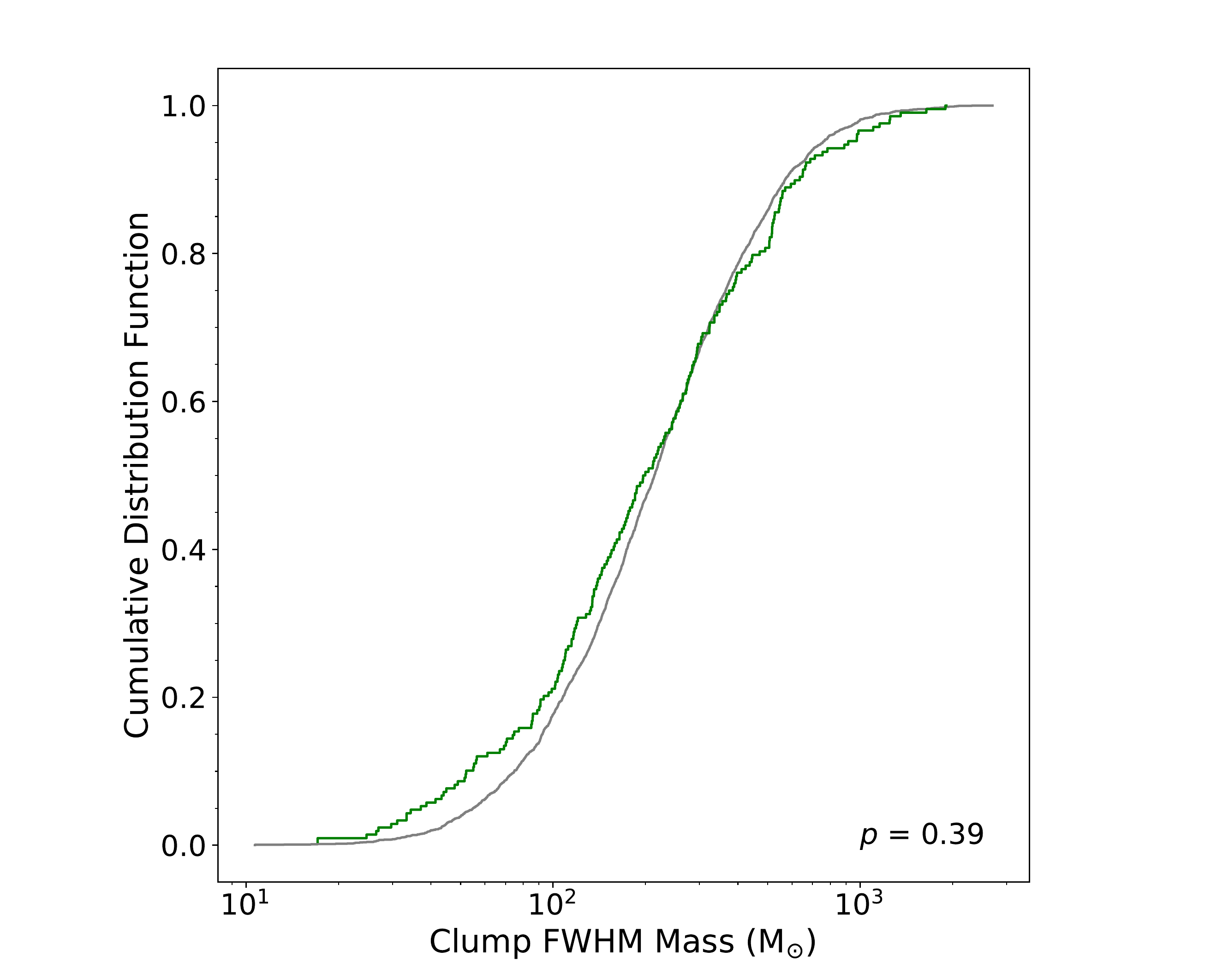} \\
	\includegraphics[width=0.47\textwidth]{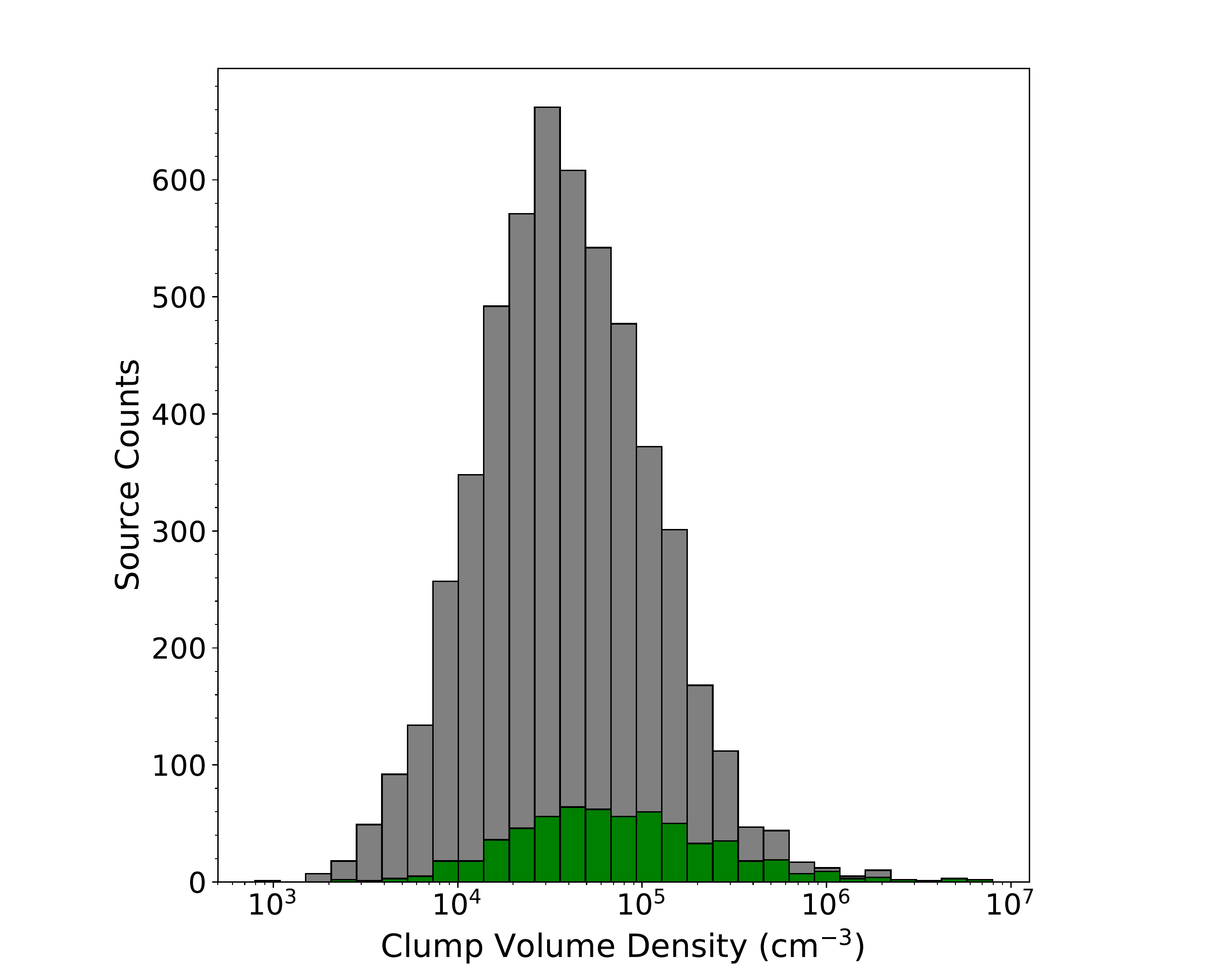}
	\includegraphics[width=0.47\textwidth]{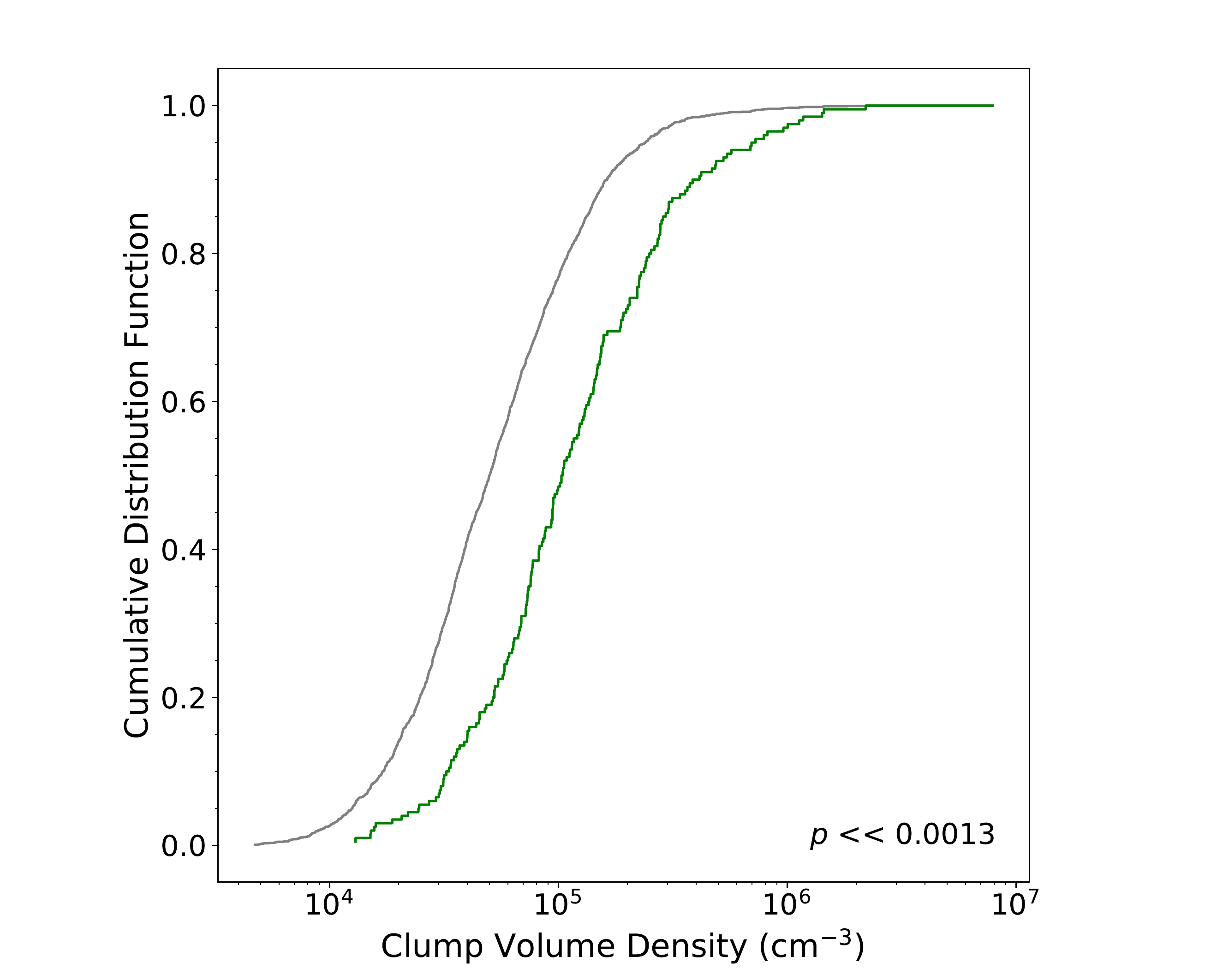} \\
	\caption{Radius, mass and volume density parameter distributions are presented in the upper, middle and lower panels respectively. The histograms in the left panels present the entire distributions of the maser associated clumps and ATLASGAL sample in green and grey respectively, whereas the cumulative distribution functions in the right panels present a distance limited sample. The $p$-value results from a KS test are shown in the lower right of the CDF panels.}
	\label{fig:full_sample_radius_mass_volden}	
\end{figure*}

\begin{figure}
    \includegraphics[width=0.47\textwidth]{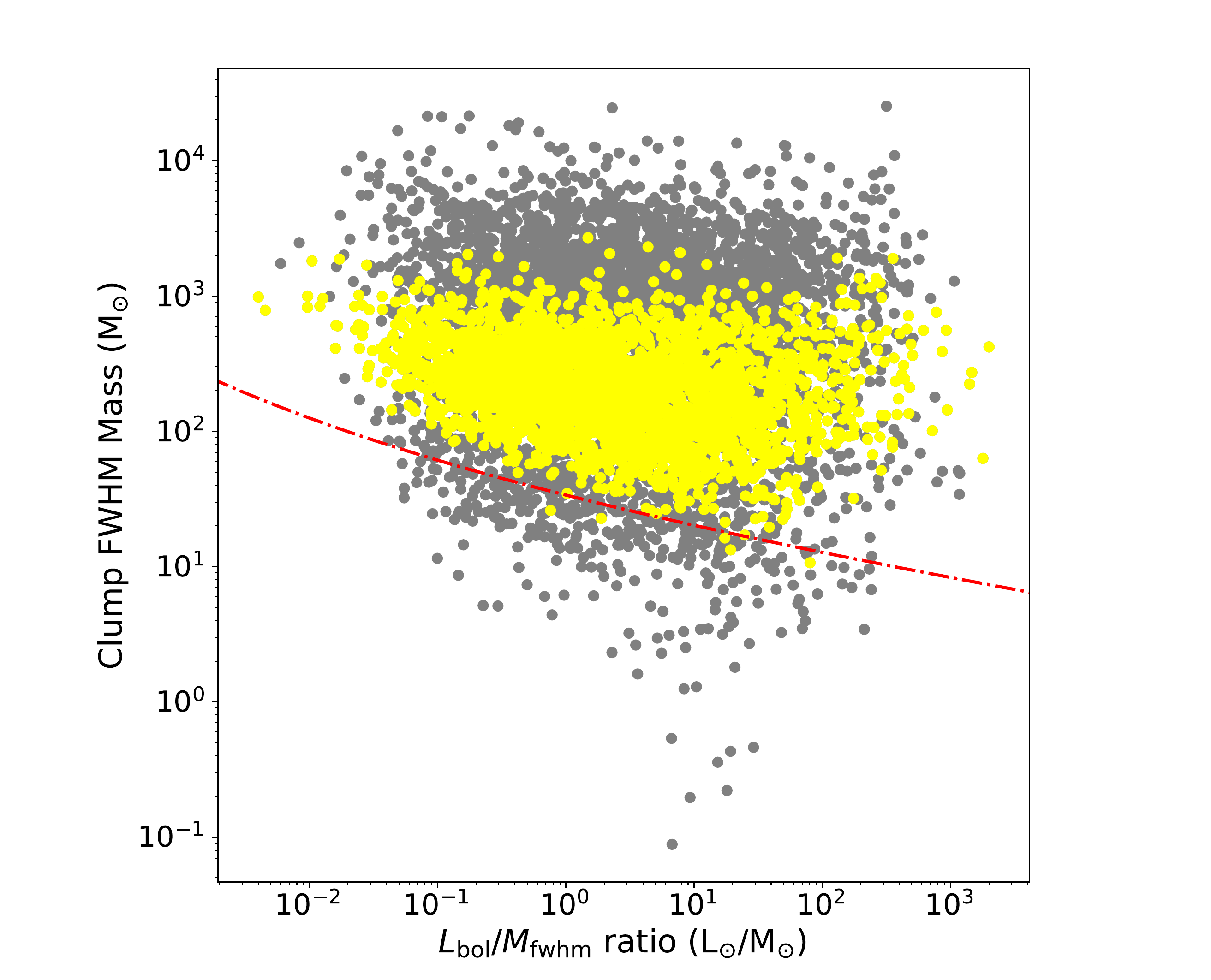}
    \caption{Clump FWHM mass vs. \lmratio\ ratio. A distance limited sample (2 to 4\,kpc) is shown in yellow with the full ATLASGAL sample shown in grey. The red dash-dotted line presents the 5$\sigma$ observational sensitivity limit of the ATLASGAL survey.}
    \label{fig:fwhm_mass_vs_lms}
\end{figure}

\section{Physical Parameters of Maser Clumps}
\label{sect:physical_parameters}

In this section we will compare the physical parameters of the masers with those of their host clumps and also compare the properties of these host clumps themselves with the properties of the full population of dense clumps that encompass the whole range of evolutionary stages, as identified by the ATLASGAL survey. The physical parameters for the vast majority of clumps located have been determined in a previous study \cite{Urquhart2018}. This study is concentrated on $\sim$8\,000 clumps located away from the Galactic centre (GC; i.e. $ 5\degr<  |\ell| < 60\degr$).


Maser emission is detected in 866 sources of the entire ATLASGAL catalogue (8.5\,per\,cent). There are 25 maser emission sites which have been associated with either JPS emission only or where the dust continuum fluxes have been manually extracted (as described in Sect.\,\ref{sect:matching_statistics}), we have opted to not include these in our sample due to the lack of distance measurements and therefore poorly constrained physical parameters. Distances are available to 660 methanol maser associated clumps, which have been obtained from \cite{Urquhart2018}. Figure\,\ref{fig:helio_distances} presents a histogram of the distribution of clump heliocentric distances, the entire ATLASGAL sample is shown in grey with the MMB associated clumps shown in green. 

Therefore, the maser associated sample presented in this paper includes 660 dust clumps across the Galactic plane not including the central 10$\degr$ (i.e. $355\degr < \ell < 5\degr$). Also, throughout this study we have opted to use a distance limited sample (2 to 4\,kpc; a total of 222 clumps), along side the full sample of clumps, to test for any distance bias affecting our statistical results.

In addition to the physical properties that we will describe in the following subsections, we also define the \lmratio\ ratio, which will be described in Sect.\,\ref{sect:masses_vol_dens}. It has been shown in previous studies that the luminosity-to-mass ratio is a good statistical indicator of evolutionary stage \citep{Molinari2008} and we will use this ratio throughout the remainder of this study to investigate the properties of the clumps with respect to their evolutionary stage. This parameter also has the advantage of being distance independent, and so eliminates an important source of uncertainty, but can also be determined for clumps where a distance is not yet available. The \lmratio\ ratio is calculated from the luminosities given in \cite{Urquhart2018} and the $M_{\rm fwhm}$, with units of \lsun\,\msun$^{-1}$. 
In the following subsections we describe the methods used to derive the physical parameters and provide a summary of the statistical physical properties of the dense clumps in Table\,\ref{table:complete_stats}.

\begin{table*}
	\begin{center}
		\caption{\label{table:complete_stats}Total statistical parameters for all maser associated clumps, determined within our matching criteria, within the Galactic longitude range $-60\degr < \ell < 60\degr$, excluding the central 10$\degr$ towards the Galactic centre. The values in brackets show the statistics from a distance limited sample (2 to 4\,kpc).}
		
		\begin{tabular}{lccccccc}
			\hline
			\hline
			Parameter                          & No. & Mean & Standard error & $\sigma$ & Median & Min  & Max    \\
			\hline 
			Distance (kpc)                     & 660 (222) & 5.85 (3.07) & 0.15 (0.03) & 3.82 (0.48) & 4.70 (3.10) & 0.10 (2.10) & 24.20 (3.90)  \\
			FWHM Radius (pc)                   & 631 (212) & 0.36 (0.20) & 0.01 (0.01) & 0.27 (0.08) & 0.26 (0.18) & 0.01 (0.08) & 1.85 (0.53)   \\	
			Temperature (K)                    & 642 (210) & 24.16 (24.06) & 0.19 (0.36) & 4.88 (5.23) & 23.66 (23.35) & 13.41 (13.41) & 46.41 (46.41) \\
			Log[Maser Luminosity (\lsun)]      & 634 (220) & -6.01 (-6.34) & 0.04 (0.06) & 0.99 (0.87) & -5.97 (-6.29) & -9.25 (-8.09) & -3.40 (-3.49) \\ 
			Log[Luminosity (\lsun)]            & 642 (210) & 4.03 (3.69) & 0.04 (0.06) & 0.96 (0.81) & 4.02 (3.57) & -0.22 (1.74) & 6.91 (5.83) \\
			Log[Clump Mass (\msun)]            & 632 (208) & 2.65 (2.30) & 0.03 (0.03) & 0.66 (0.41) & 2.72 (2.31) & -1.05 (1.23) & 4.40 (3.28) \\		
			Log[\lmratio (\lsun / \msun)] & 632 (208) & 1.39 (1.40) & 0.02 (0.04) & 0.59 (0.61) & 1.37 (1.35) & -0.07 (0.12) & 3.17 (3.17) \\			
			Log[$n$(H$_2$)](\cmthree)          & 613 (200) & 4.86 (5.07) & 0.02 (0.03) & 0.56 (0.44) & 4.82 (5.02) & 3.40 (4.11) & 6.90 (6.90) \\
			Free Fall Time (Myr)               & 613 (200) & 0.22 (0.16) & 0.01 (0.01) & 0.14 (0.08) & 0.19 (0.15) & 0.02 (0.02) & 1.00 (0.44) \\
			\hline		
		\end{tabular}
	\end{center}
\end{table*}

\subsection{Spectral Energy Distributions}

The physical parameters used in this study are taken from \cite{Urquhart2018} as mentioned in Sect.\,\ref{sect:physical_parameters}, which covers the longitude range $60\degr \geq \ell \geq 300\degr$. To determine the environmental properties for the maser sources within the MMB survey which lie beyond this longitude range (i.e. $300\degr \geq \ell \geq 186\degr$), we have performed aperture photometry to reconstruct the dust continuum spectral energy distributions (SEDs; for a more detailed description of this method please see Sect.\,3 of \citealt{Konig2017}). We have used the fitting of the SEDs to provide information for 10 sources beyond the longitude range covered by \cite{Urquhart2018}. This method has also been used for a further 34 maser sources, which were either beyond the ATLASGAL survey coverage or were not detected within the survey itself. An example SED and corresponding fit is shown in Fig.\,\ref{fig:sed_fit}. 

\subsection{Clump Sizes}
\label{sect:clump_sizes}

In order to circumvent an observational bias when measuring the radius of clumps, which is discussed below, we have opted not to use the current sizes of clumps as found in \cite{Urquhart2018}. Instead the sizes of clumps is determined using the FWHM of the dust continuum emission, assuming that the flux from each clump is Gaussian in nature.

The clump FWHM angular radius is determined from the number of pixels, npix$_{\rm{fwhm}}$, with fluxes above the 50\,per\,cent peak flux contour:

\begin{equation}
R_{\rm{fwhm}} = \sqrt{\frac{A}{\pi}}
\end{equation}

\noindent where $A$ is npix$_{\rm{fwhm}}$ multiplied by the square of the pixel size in arcsec (i.e. 36\,sq.\,arcsec). The 50\,per\,cent flux level is below the detection threshold for sources with a SNR below 6$\sigma$, and consequently the npix$_{\rm{fwhm}}$ and $R_{\rm{fwhm}}$ for these sources are unreliable and are therefore excluded from this calculation. We also exclude clumps with fewer pixels than the beam integral ($\Theta_{\rm{fwhm}}^2\times 1.133 = 11.3$\,pixels, where $\Theta_{\rm{fwhm}}$ is 19.2\,arcsec) as the flux measurements will be overestimated. The $R_{\rm{fwhm}}$ that satisfy these two conditions are deconvolved from the ATLASGAL beam using:

\begin{equation}
R_{\rm{fwhm,decon}} = \sqrt{R_{\rm{fwhm}}^2-\left(\frac{\Theta_{\rm{fwhm}}}{2}\right)^2}
\end{equation}

The distances given in \cite{Urquhart2018} are used to convert the deconvolved radius into a physical radius, $R_{\rm{fwhm, pc}}$. 

In Fig.\,\ref{fig:radius_vs_lms} we present two distributions of clump radii versus evolutionary stages as determined by the clump \lmratio\ ratios. The upper panel shows previously calculated effective radii values (2.4$\times \sigma_{\rm{radius}}$; \citealt{Rosolowsky2010}) for dense clumps as taken from \cite{Urquhart2018}, with the lower panel showing the FWHM derived radii values from this study. The results from a Spearman's test are shown in the lower left corner of each panel. \citet{Urquhart2018} noted a correlation between the increasing size and decreasing surface density of clumps, and their evolutionary stage; the increase in radius is clearly seen in the upper panel of Fig.\,\ref{fig:radius_vs_lms}. This distribution has a Spearman's results of 0.36 and $p$-value of much less than our significance threshold with a best fit slope of 0.11$\pm$0.01. Similar trends for increasing radii and decreasing volume density as a function of evolution have been reported in a number of submm dust continuum studies (e.g., \citealt{Hill2005, Breen2010, Contreras2017}). These trends are generally explained as the expansion of the clumps towards the end of their evolution and although plausible for the final stages, due to mechanical and radiative feedback, this cannot explain the very smooth increase in radius throughout their evolution. Particularly given that at the earliest states we expect them to be collapsing and accreting material onto the evolving protostellar objects.

\citet{Urquhart2018} determined that this was due to an observational bias where the increasing average temperature of the clumps results in more of the extended envelope becoming bright enough to be detected by ATLASGAL, resulting in an apparent increase in the source sizes as the clumps evolve. In an effort to eliminate this bias we have recalculated the sizes and physical parameters using only the 870\,$\mu$m emission above the FWHM flux contour. In the lower panel of Fig.\,\ref{fig:radius_vs_lms} we show the relationship between the $R_{\rm{fwhm}}$ and our evolutionary tracer; this clearly demonstrates that the observational bias as been corrected for, as the corrected slope for this distribution is -0.02$\pm$0.004. Also the result from a Spearman's test gives a $r_{\rm s}$ statistic of -0.10, while this is significant (as shown by a corresponding $p$-value of less than our confidence threshold), it shows a poor correlation.

The distribution of these new radius values, for the entire ATLASGAL sample, can be found in the upper panels of Fig.\,\ref{fig:full_sample_radius_mass_volden}, along with the distribution of maser associated clumps. The upper right panel of Fig.\,\ref{fig:full_sample_radius_mass_volden} presents the cumulative distribution function of the radii values for a distance limited sample (2 to 4\,kpc), for both of these samples. It is clear from the CDF shown in the right panel that the methanol maser associated clumps are significantly more compact than the rest of the population ($p$-value $\ll 0.0013$). It appears that maser associated clumps have statistically significantly smaller FWHM radii values when compared to the full ATLASGAL sample. Therefore, dense clumps associated with a methanol maser are generally much more compact than those without this high-mass star formation signpost. 

\subsection{Clump masses and volume densities}
\label{sect:masses_vol_dens}

The mass within the 50\,per\,cent contour, $M_{\rm{fwhm}}$, is scaled from the total mass calculated in \cite{Urquhart2018} using the ratio of flux within the FWHM and the total flux of the clump. Figure\,\ref{fig:fwhm_mass_vs_lms} presents the distribution of the clump $M_{\rm{fwhm}}$ for the entire ATLASGAL sample and a distance limited sample (2 to 4\,kpc), it can be seen that the $M_{\rm{fwhm}}$ is independent of the evolution of a clump. The new mass distribution for the full sample and maser associated sample of clumps is shown in the middle panels of Fig.\,\ref{fig:full_sample_radius_mass_volden}. The middle right panel of Fig.\,\ref{fig:full_sample_radius_mass_volden} presents the distance limited sample CDF of these distributions. It can be seen that there is no overall difference in the masses of clumps associated with methanol maser emission and the general population of clumps, as confirmed by the results of a KS test ($p$-value = 0.387). This suggests that clump mass alone is not an important factor in determining whether a particular clump is likely to form a high-mass star.

We calculate the volume density, within the FWHM contour, by dividing the $M_{\rm{fwhm}}$ by the volume:

\begin{equation}
n\left({\rm H_2}\right) = \frac{3 }{4 \pi  }\frac{M_{\rm{fwhm}}}{\mu {\rm m_p} R_{\rm{fwhm}}^3} 
\end{equation}

\noindent where $n\left({\rm H_2}\right) $ is the hydrogen particle density per cm$^{-3}$, $\mu$ is the mean molecular weight per hydrogen atom (taken as 2.8; \citealt{Kauffmann2008}) and m$_{\rm p}$ is the mean proton mass, and $M_{\rm{fwhm}}$ and $R_{\rm{fwhm}}$ are as previously defined. We make the assumption that each clump is generally spherical and not extended along the line of sight. Volume densities are available for 81\,per\,cent of the clump sample presented in this paper, the distribution of this parameter can be seen in the lower panels of Fig.\,\ref{fig:full_sample_radius_mass_volden}. Inspection of these plots reveals that clumps hosting methanol masers have a significantly higher volume density. 


\subsection{Stability and free-fall collapse}

\begin{figure}
	\includegraphics[width=0.47\textwidth]{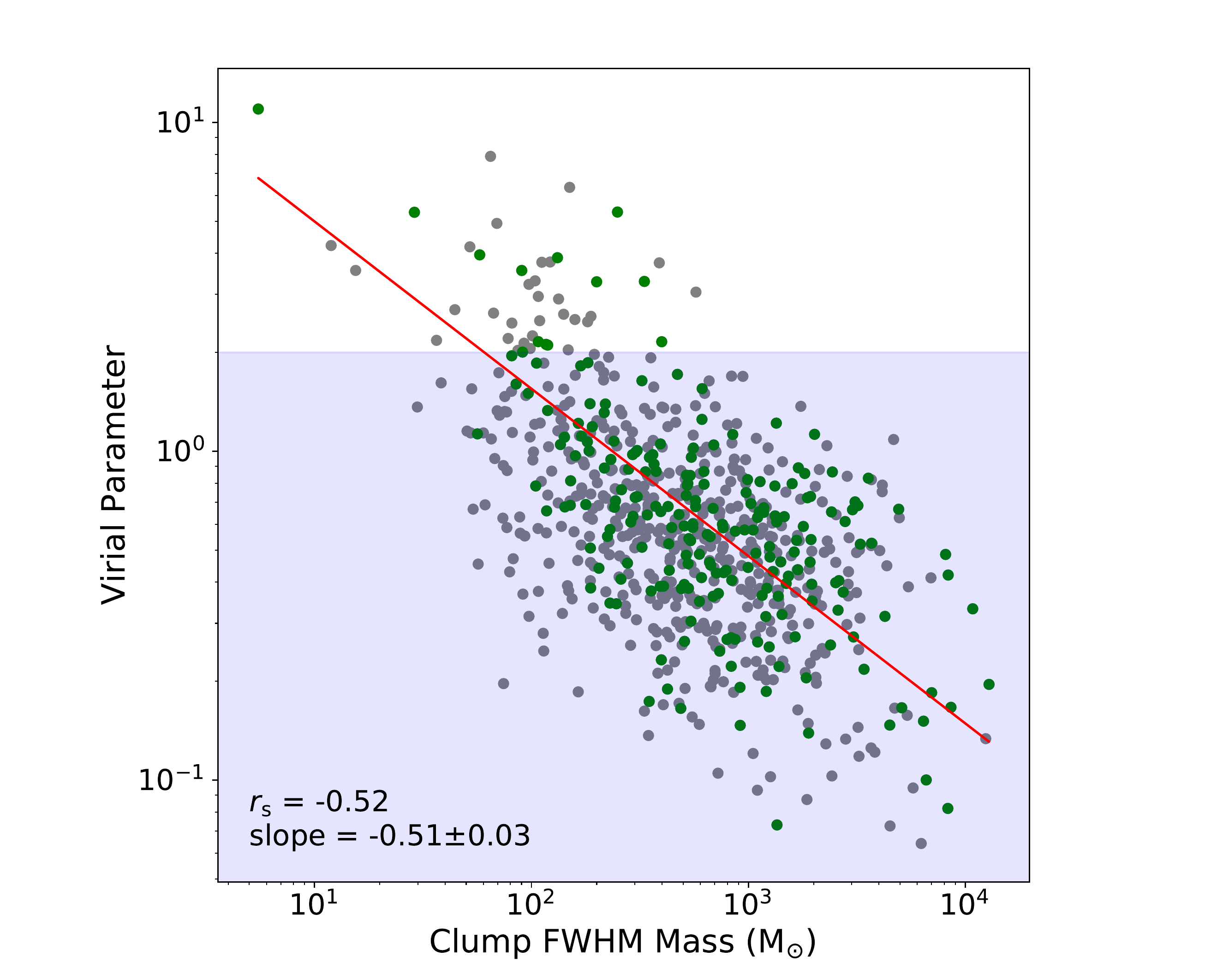}
	\caption{Virial parameter ($\alpha_{\rm{vir}}$) against clump FWHM mass. The maser associated clumps and full ATLASGAL sample are shown in green and grey respectively. The solid red line shows the results from an orthogonal distance regression of the maser associated sample, where $\alpha_{\rm{vir}} \propto M_{\rm{fwhm}}^{-0.5\pm0.03}$. The shaded region indicates unstable clumps where $\alpha_{\rm{vir}}$ < 2.}
	\label{fig:virial_parameter_vs_mass}
\end{figure}

In order to estimate the gravitational stability of our sample of clumps we have derived the virial parameter for each source. This parameter is a measure of the balance between internal energy and gravitational collapse and is defined as:

\begin{equation}
\alpha_{\mathrm{vir}}=\frac{5 \sigma_{v}^{2} R_{\mathrm{fwhm}}}{G M_{\mathrm{fwhm}}},
\end{equation}

\noindent where  $\sigma_{v}$ is the velocity dispersion, which is calculated from \nhthree (1,1) inversion transition observations (e.g., \citealt{dunham2011,urquhart2011_nh3,Wienen2012,wienen2018}), $G$ is the gravitational constant and all other parameters are as previously defined. We have calculated the virial parameters for 216 sources in our sample, with a mean and median of 0.84 and 0.61 respectively with a standard deviation of 1.02. A clump with a virial parameter of less than 2  indicates that it is unstable and will undergo global collapse in the absence of a supporting magnetic field (\citealt{kauffmann2013}); 203 clumps (94\,per\,cent) have a values of less than 2 and given that the majority of clumps are already associated with star formation (\citealt{Urquhart2018}) it is safe to assume that the majority of these clumps are collapsing. Figure\,\ref{fig:virial_parameter_vs_mass} presents the distribution of the virial parameter against the clump FWHM mass; this reveals a trend for the most massive clumps being the most gravitationally unstable. Fitting the data we find that $\alpha_{\rm{vir}} \propto M_{\rm{fwhm}}^{-0.5\pm0.03}$, which is consistent with the relations derived by \citet{larson1981}.

Free fall times have been derived from the calculated volume densities:

\begin{equation}
    t_{\rm ff} = \sqrt{\frac{3\pi}{32G\rho}} 
\end{equation}

\noindent where $\rho  = \frac{3M_{\rm fwhm}}{4\pi R_{\rm fwhm}^3}$ is the mean density of the clump. A free fall time has been derived for every clump with a corresponding density measurement, these values range from 2$\times$10$^{4}$ and 1$\times$10$^{6}$ years. We have not taken into account any thermodynamical changes the clump undergoes or any support mechanisms that might impede the global collapse. However, the change in temperature of the clumps during the embedded evolution stages is quite modest ($\sim15$\,K) and given the low virial parameters for the majority of the sample and the fact that all of the high density clumps are already associated with star formation it is unlikely any support mechanism is sufficient to counteract the global collapse on clump scales.

\subsection{Bolometric and maser luminosities}
\label{sect:luminosities}

\begin{figure}
	\includegraphics[width=0.47\textwidth]{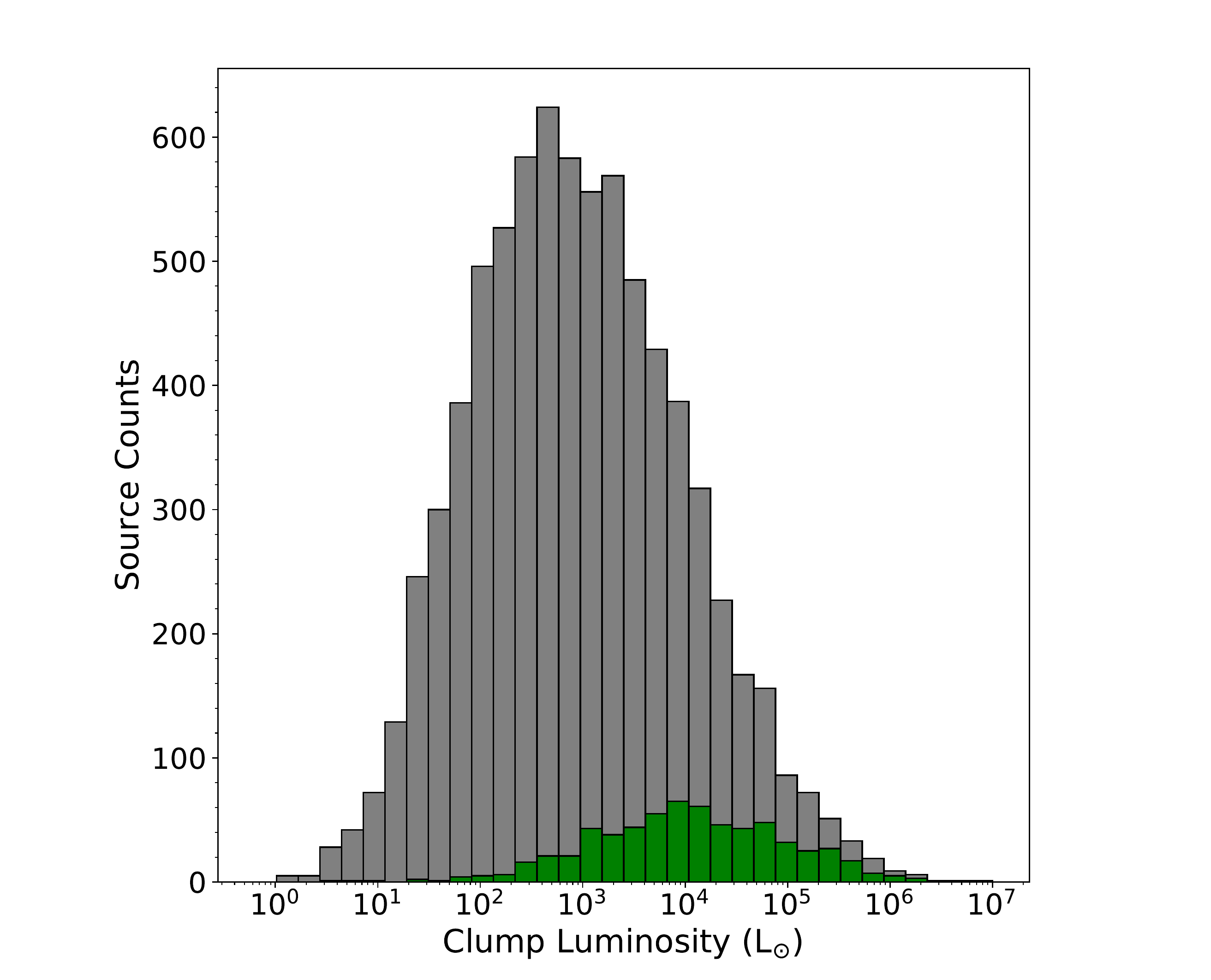}
	\includegraphics[width=0.47\textwidth]{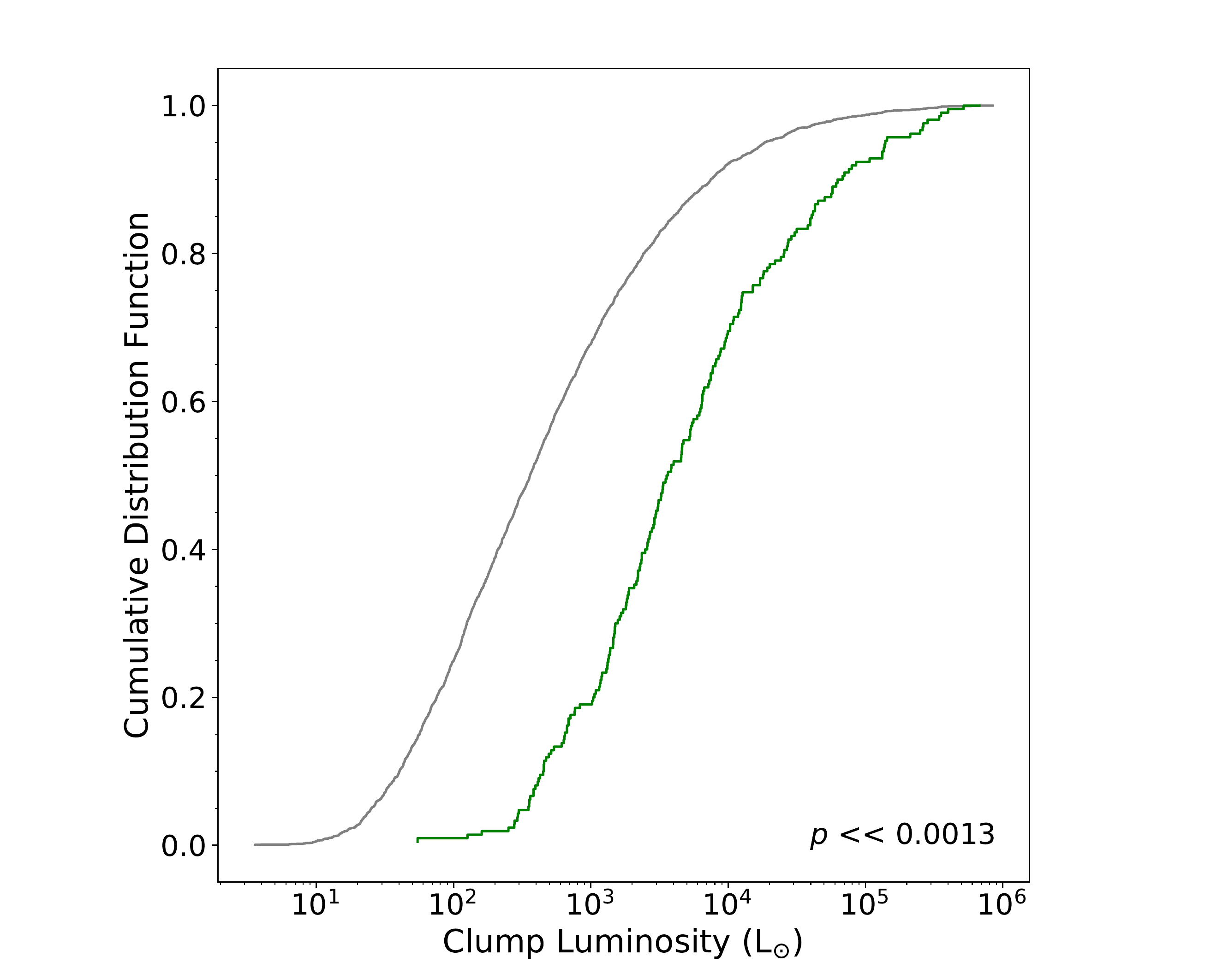} \\
	\caption{Clump luminosity parameter distributions. The histogram in the upper panel presents the entire distribution of the maser associated clumps and ATLASGAL sample in green and grey respectively, whereas the cumulative distribution function in the lower panels presents a distance limited sample. The $p$-value result from a KS test are shown in the lower right of the lower panel.}
	\label{fig:full_sample_lums}
\end{figure}

\begin{figure}
	\includegraphics[width=0.47\textwidth]{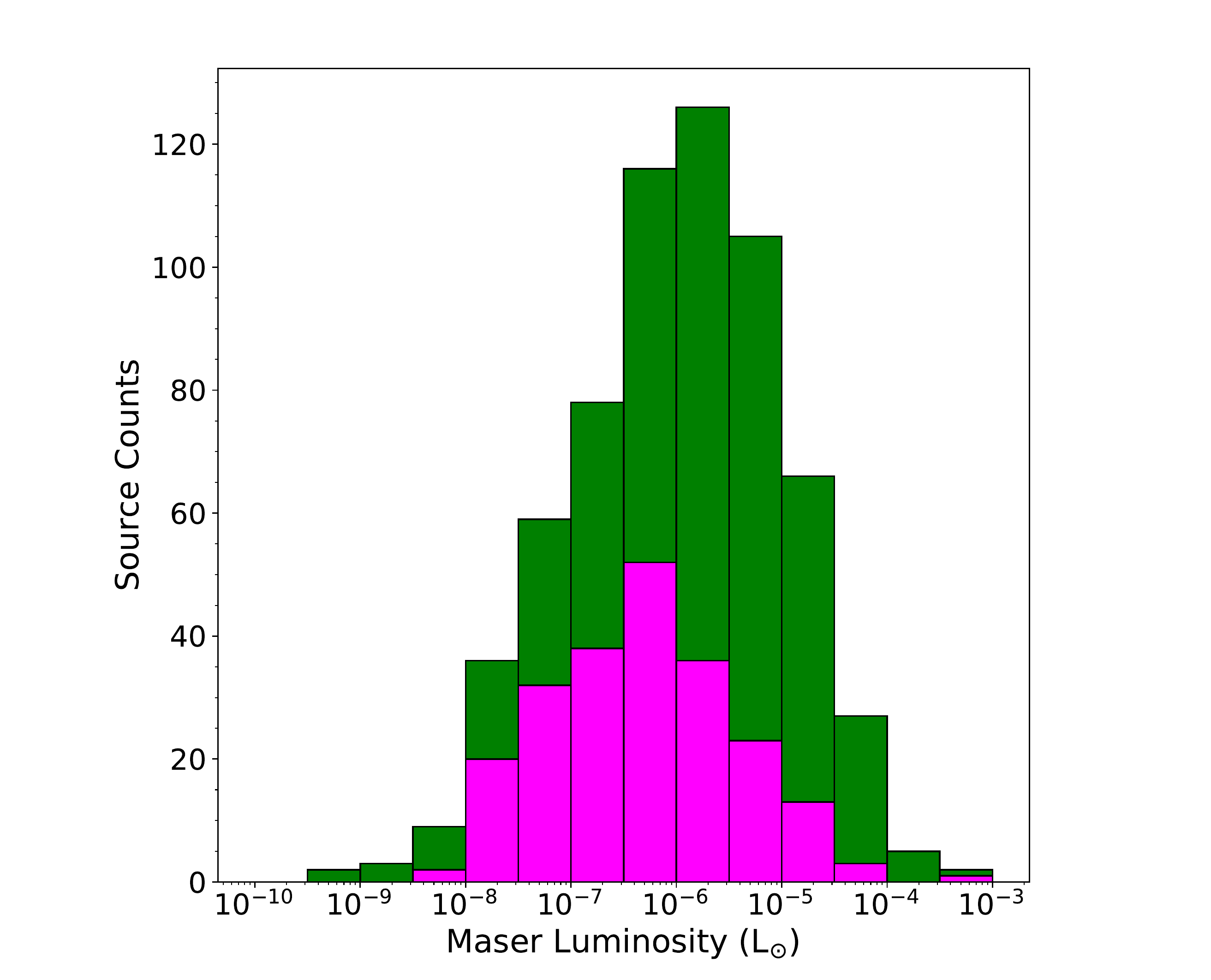} \\
	\includegraphics[width=0.47\textwidth]{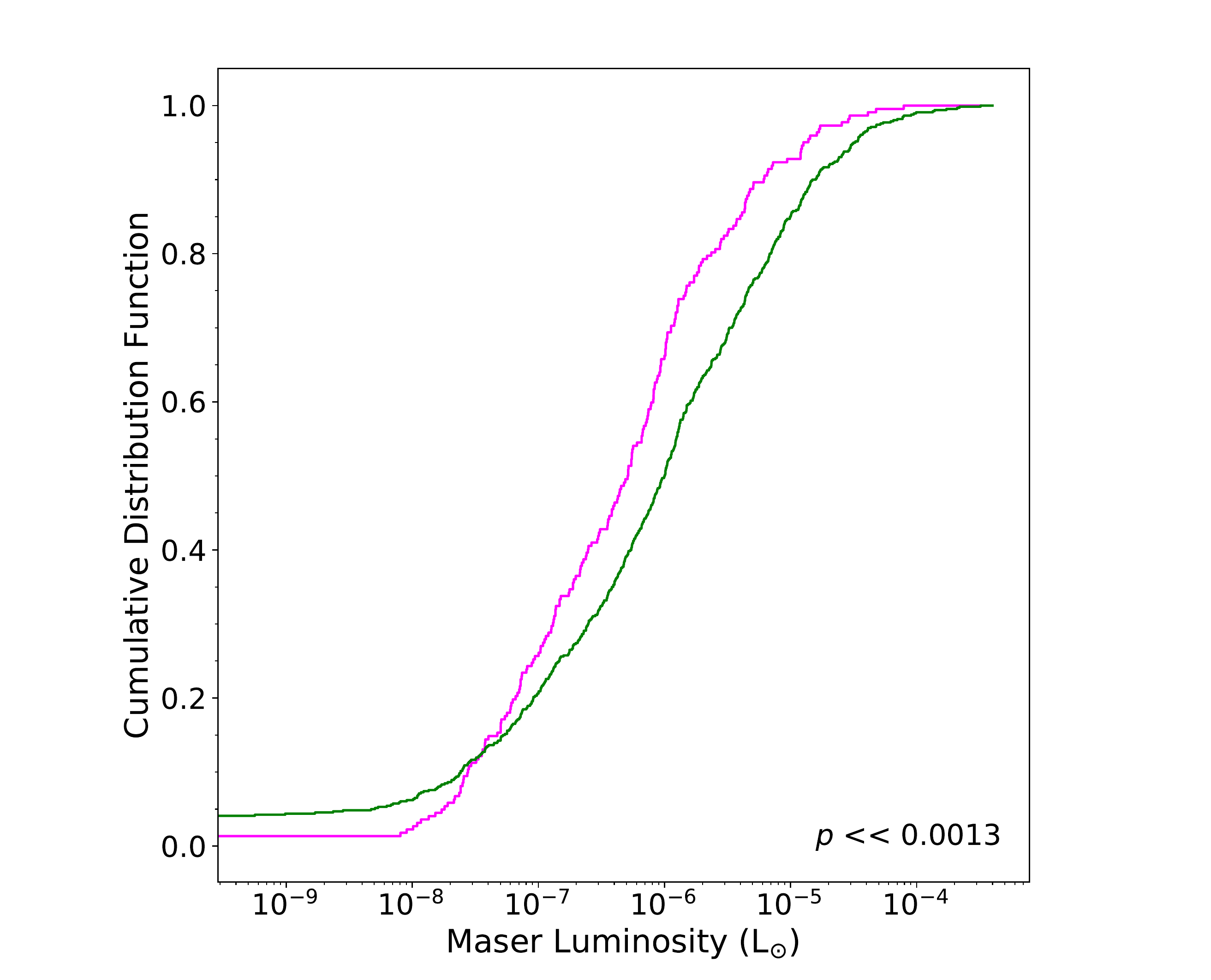}
	\caption{The upper presents a histogram of the methanol maser luminosities. A distance limited sample is shown overplotted in magenta, the bin size is 0.5 dex. The lower panel presents the CDF of the same distribution.}
	\label{fig:mmb_lums}
\end{figure}

\begin{figure}
	\includegraphics[width=0.47\textwidth]{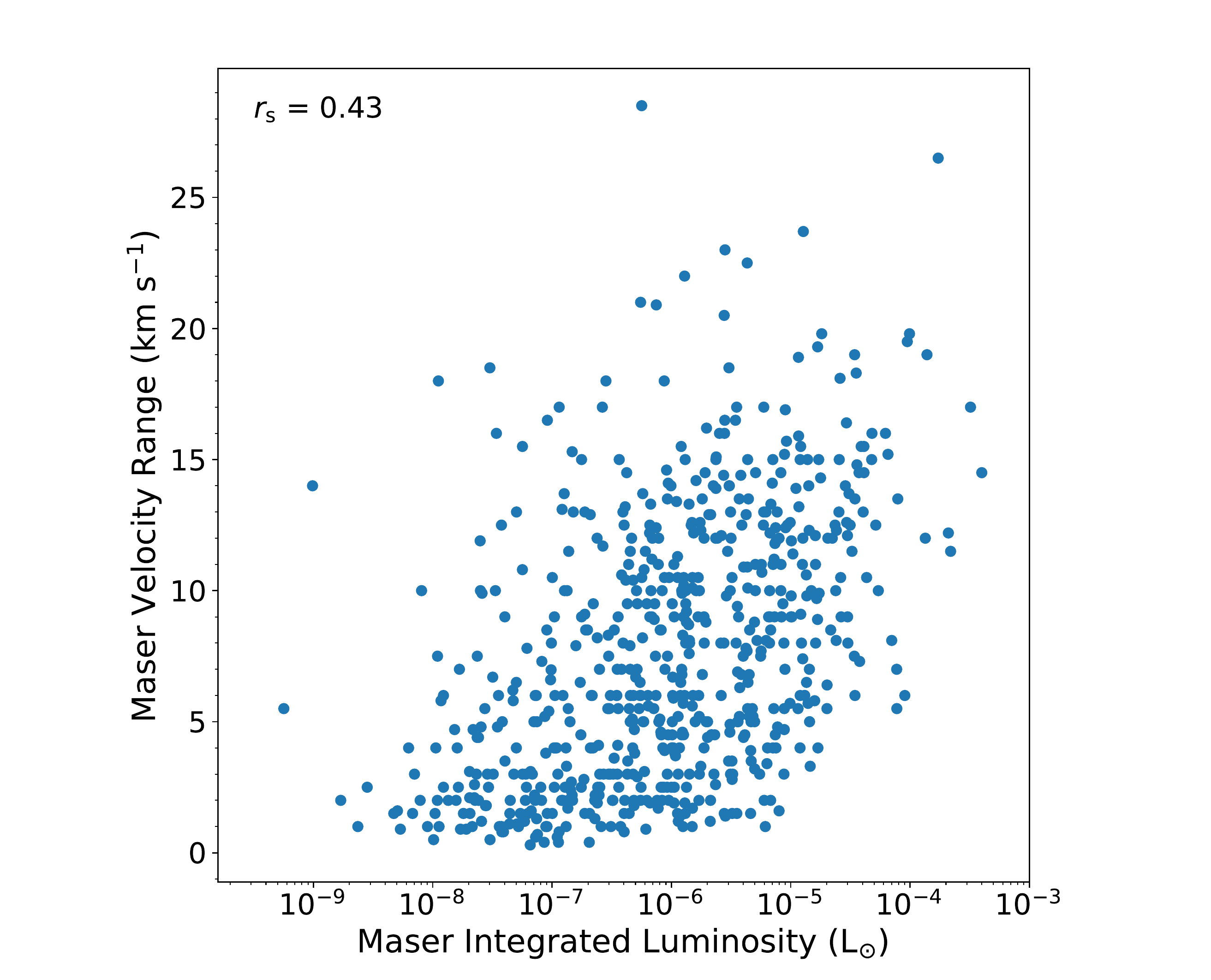}
	\caption{Scatter plot presenting the distribution of maser integrated luminosity against the maser velocity ranges. The result from a Spearman's rank correlation test is shown in the upper left, with a $p$-value of $\ll$ 0.0013.}
	\label{fig:mmb_int_mmb_velo_range}
\end{figure}


The clump bolometric luminosities are taken unchanged from \cite{Urquhart2018}. The distribution of clump luminosity can be found in Fig.\,\ref{fig:full_sample_lums}. It can be seen from the upper panel of this figure that the luminosity values for maser associated clumps are, on average, increased when compared to the full ATLASGAL sample. This is expected as the 6.7\,GHz methanol maser is thought to be produced in the dense envelopes surrounding high-mass protostellar objects and YSOs. Naturally, the majority of a clump's luminosity will be due to any associated high-mass embedded object. Lower luminosity clumps, especially those that are quiescent, with no known embedded objects, are unlikely to produced maser emission, which is supported by the analysis presented here.

The difference between clumps with and without maser emission can be more clearly seen in the lower panel of Fig.\,\ref{fig:full_sample_lums}. This panel presents a cumulative distribution function of luminosity and shows both distance limited samples of the full ATLASGAL sample and maser associated clumps. There is a clear difference between these two samples, as confirmed by a KS test ($p$ $\ll$ 0.0013).

The maser luminosities have been calculated using:

\begin{equation}
L = 4\,\pi\,D^{2}\,S_{\rm int,6.7\,GHz},
\end{equation}

\noindent where $S_{\rm int,6.7\,GHz}$ is the integrated maser flux and $D$ is the heliocentric distance to the host clump. We assume that maser emission from any particular source is isotropic and the flux of maser sources decreases following an inverse square law. The units of this parameter are Jy\,\kms\,kpc$^2$ and are, therefore, somewhat arbitrary. A conversion factor has been derived to convert the maser luminosities from Jy\,\kms\,kpc$^2$ into \lsun:

\begin{equation}
\Delta f = \frac{f \Delta v}{c} = \frac{6.7\,{\rm GHz}\times 1\,{\rm km\,s^{-1}} }{3\times 10^5\,{\rm km\,s^{-1}}} = 22.3\,{\rm kHz},
\end{equation}

\noindent where $f$ is the base frequency of the maser emission (6.7\,GHz) and $c$ is the speed of light. This shows that a change in velocity of 1\,\kms\ is proportional to a frequency change of 22.3\,kHz. Figure\,\ref{fig:mmb_lums} presents a histogram of the methanol maser integrated luminosities along with the corresponding CDF, both the full and distance limited sample are shown. The difference in maser luminosity between the full and distance limited sample (2 to 4\,kpc) are found to be significant ($p$-value $\ll 0.0013$) as shown by the result from a KS test. 

Figure\,\ref{fig:mmb_int_mmb_velo_range} shows how maser velocity ranges change depending on the maser luminosity. It can be seen that as maser luminosity increases, the velocity range also increases, the results from a Spearman's rank test indicate there is a reasonable positive correlation between these two variables ($r_{\rm s}$ = 0.43). It is likely that the increase in velocity range can be attributed to an increase in individual spots within the maser spectra, this has also been stated in previous studies (e.g. \citealt{Green2017}), and with other maser species (e.g. \citealt{anglada1996}). There appears to be a large amount of scatter in Fig.\,\ref{fig:mmb_int_mmb_velo_range}, and so it is not necessarily true that the most luminous masers also have the largest velocity ranges.

\subsection{Uncertainties in the physical parameters}
\label{sect:uncertainties}

Dust temperature values are obtained through the fitting of the spectral energy distributions with a mean error of $\sim$10\,per\,cent. The uncertainties in the distances are estimated from the Bayesian distance algorithm presented in \cite{Reid2016} and are of order $\pm0.5$\,kpc and the fractional uncertainty in the radius will be the same as for the distance as these are linearly related ($\sim$30\,per\,cent at 1\,kpc, but only a few per\,cent at distances greater than 10\,kpc). The fractional uncertainty in the maser luminosity is $\sqrt{2}$ times the fractional uncertainty in the distance, however, in calculating this quantity we are assuming the maser emission is isotropic and so the uncertainty on these measurements is hard to estimate. The error on the bolometric fluxes are approximately 50\,per\,cent but when combined with the uncertainty in the distance we estimate the total uncertainty is approximately a factor of 2. The uncertainty in the mass, surface and volume density calculations are likely to be dominated by the uncertainty in the dust to mass ratio and the value of $\kappa_v$ (dust absorption coefficient; taken as 1.85\,cm$^{2}$\,g$^{-1}$ interpolated from \citealt{Schuller2009}) both of which are poorly constrained and therefore the uncertainties for these is likely to be a factor of 2-3. However, while the uncertainties associated with the physical parameters may be quite large, these affect the entire sample uniformly and will increase the scatter in distributions but will still allow statistical trends to be identified and robustly analysed, especially when considering a distance limited sample.

\section{Discussion}
\label{sect:discussion}

The sample presented in this study is the largest and most representative of maser associated dense clumps to date. Our statistical analysis will focus on the central 80\,per\,cent of sources in each parameter space, this will remove any potential outliers arising from incorrect distance measurements or due to extreme sources that might skew the results.

\subsection{Luminosity, mass and radius correlations}

\begin{figure}
	\includegraphics[width=0.47\textwidth]{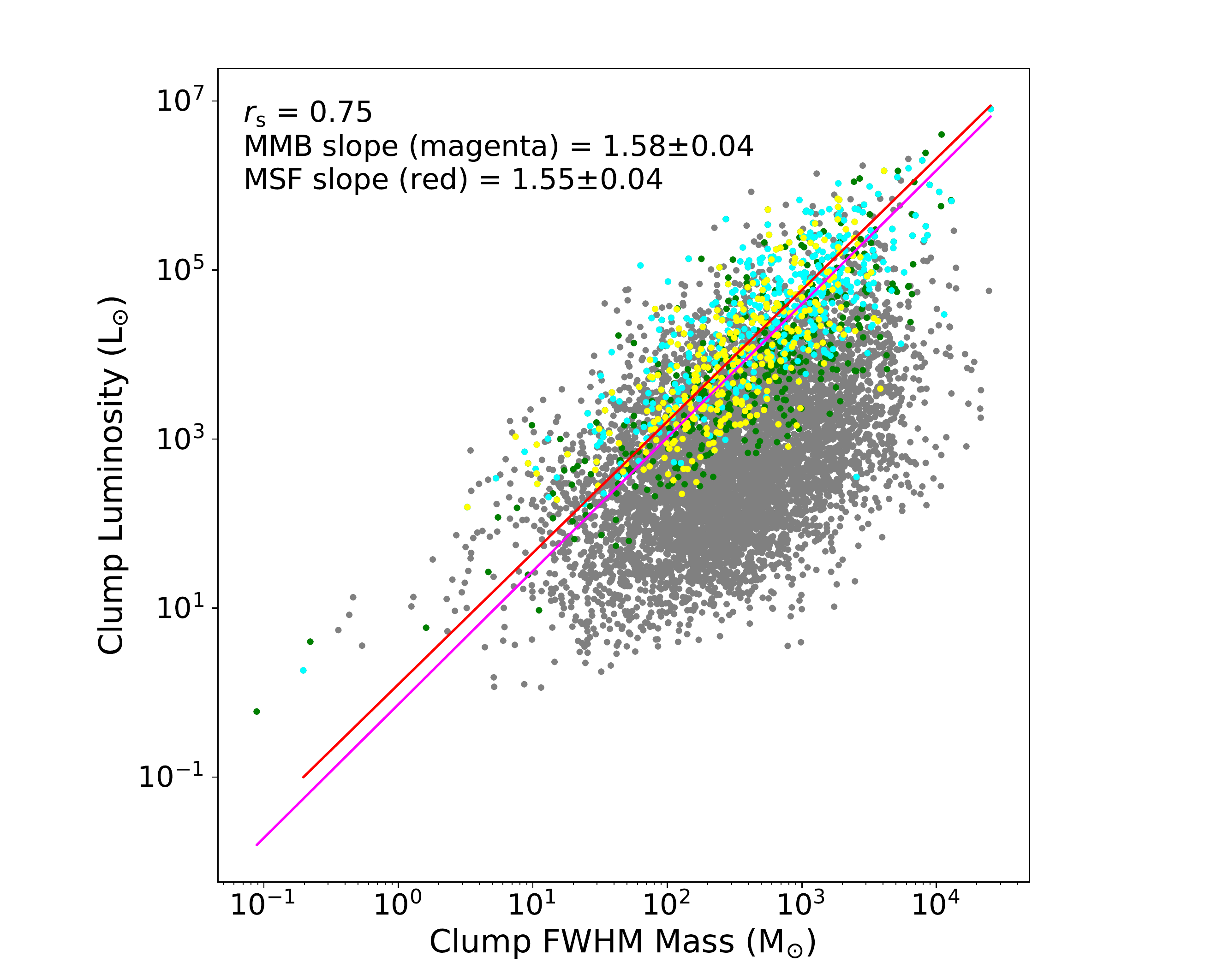}
	\includegraphics[width=0.47\textwidth]{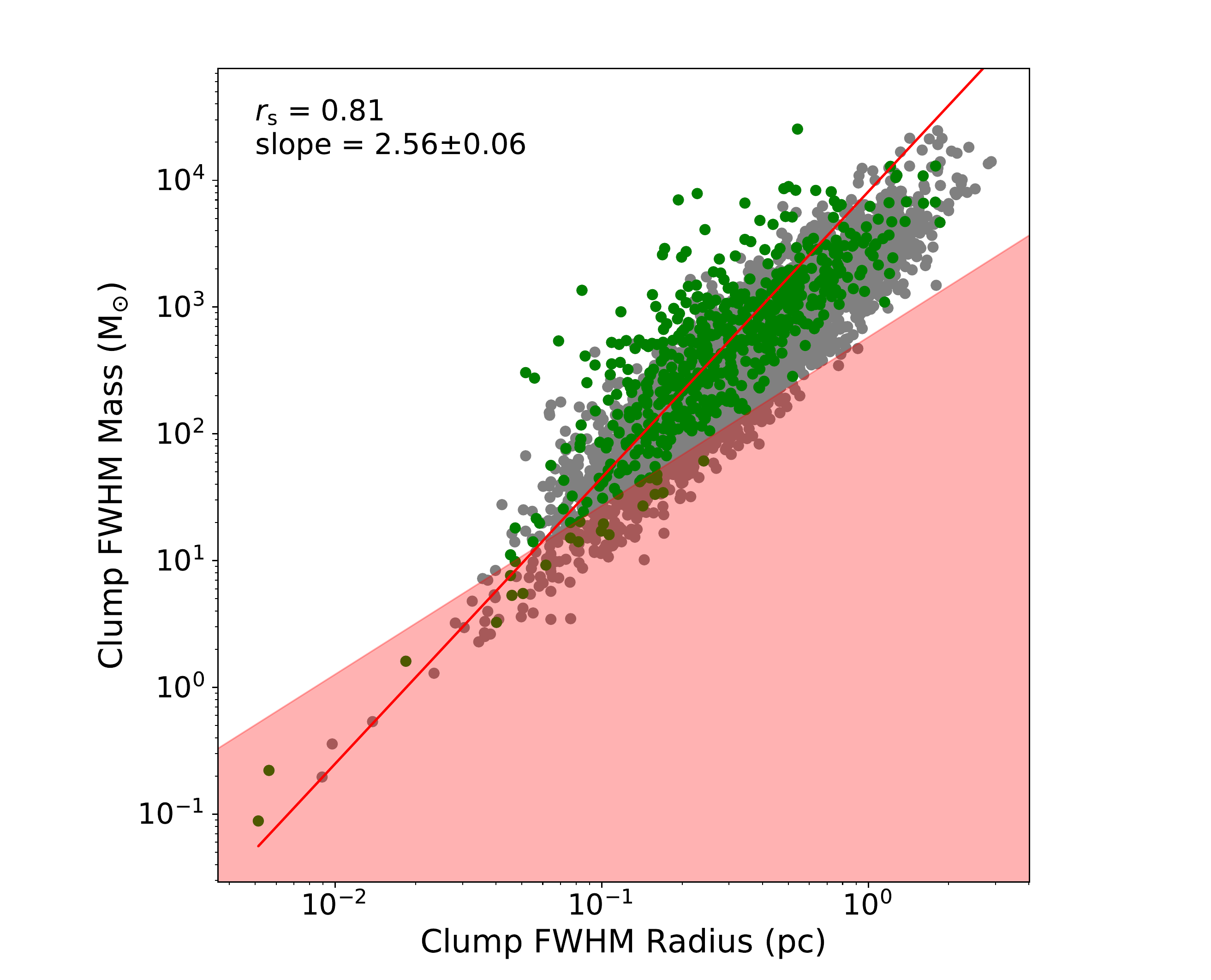}
	\caption{The upper panel presents the distribution of clump bolometric luminosity versus clump FWHM mass, maser associated clumps, YSO associated clumps, \hii\ region associated clumps and the full ATLASGAL sample are shown in green, yellow, cyan and green respectively. The lower panel presents clump FWMH mass versus clump FWHM radius with maser associated clumps and the full ATLASGAL sample in green and grey respectively. Orthogonal distance regression has been used with the clump samples to produce best fit lines, shown in red and magenta. The slope of the best fit lines and the corresponding Spearman's coefficient are shown in the upper left of each panel, the $p$-values for both distributions is less than our significance threshold. The shaded area in the lower panel shows the area devoid of massive star formation as derived by \protect\cite{Kauffmann2010a}.}
	\label{fig:lumsmass_massradius}
\end{figure}

In Fig.\,\ref{fig:lumsmass_massradius} we present plots showing the relationship between luminosity, mass and radius for the ATLASGAL clumps. These are similar to plots presented in \citet{Urquhart2018} but have been updated using the FWHM clump masses and radii in order to remove any bias due to evolution. The overall correlations between these parameters are actually very similar, but the slopes are significantly steeper. The previously slope for $L_{\rm bol}$ vs. $M_{\rm clump}$ was $1.31\pm0.02$ compared to  $L_{\rm bol}$ vs. $M_{\rm fwhm}$ which is $1.55\pm0.04$ for the full sample of massive star forming clumps (i.e. those hosting massive (M)YSOs and \hii\ regions; \citealt{urquhart2014_csc}) and so the smaller FWHM masses have resulted in a modestly steeper relationship. The slope of the clumps associated with methanol masers is $1.58\pm0.04$ and so is not  significantly different to that of the massive star forming clumps. The slopes, however, for the distance limited sample of clumps and clumps associated with a methanol maser are significantly steeper at $2.58\pm 0.17$ and $2.35\pm 0.19$, respectively. This is likely due to the fact that we are mapping clouds at larger distances rather than clumps and so capturing more mass that is not actually contributing to the star formation process.

In the lower panel of Fig.\,\ref{fig:lumsmass_massradius} we show the mass-radius relationship of the clumps. Comparing the results again with those given in \citet{Urquhart2018} we find that the  slope of $2.12\pm0.01$ for the $M_{\rm fwhm}$ vs. $R_{\rm fwhm}$ for the whole sample is significantly larger than the slope of $1.65\pm0.01$ derived from the $M_{\rm clump}$ vs. $R_{\rm eff}$ (where $R_{\rm eff}$ is the effective radius of the clump). The change in the slope  is because a modest increases in temperature  has a much more significant influence on the radius (a change in temperature between 12 to 45\,K can result in change in radius by a factor of 4.2; as discussed in Sect.\,8 of \citealt{Urquhart2018}). For comparison the slope of clumps determined by a set of targeted recent high-resolution 350\,\mum\ observations towards ATLASGAL clumps selected from the GaussClump catalogue (\citealt{csengeri2014}) reported a mass-radius slope between 2 and 2.3 ({\color{red}Lin et al. 2019, subm.}). Analysis of the CHIMPS $^{13}$CO (3-2) survey (\citealt{rigby2016}) finds the mass-radius slope to be $2.47\pm0.02$ ({\color{red}Rigby et al. 2019, subm.}), while \citet{roman-duval2010} reported a slope of $2.36\pm0.04$ from anlysis of $^{13}$CO (1-0) emission drawn from the Galactic Ring Survey (GRS; \citealt{jackson2006}). There is, therefore, a significant amount of variation in the measured slope of the relationship between the mass and radius in the literature, which is probably due to differences in the way the radius is determined ($R_{\rm eff} = \eta R_\sigma$, $\eta$ is a multiplication factor (between 1.9-2.4) that scales the $R_\sigma$, which is the standard deviation of the intensity-weighted emission profile), the tracer (CO or dust), and the assumptions used to determined the clump masses (dust emissivity, dust-to-gas ratio and H$_2$ to CO ratio) and the method used to fit the data (least-square fitting, orthogonal distance regression etc.). 

The slope for the mass-radius relation for the maser associated clumps is significantly steeper than found for the full sample of clumps ($2.56\pm 0.06$). This is to be expected given that the methanol maser associated clumps are, while having a similar mass to the general population of clumps, significantly smaller and have, on average, higher volume densities. The slope for the distance limited samples of clumps and maser associated clumps are both significantly larger but also have much larger uncertainties; they are $2.56\pm0.06$ and $4.37\pm0.48$, respectively.

\subsection{\lmratio\,Ratio Correlations}
\label{sect:l/m_ratios}

\begin{figure}
	\includegraphics[width=0.47\textwidth]{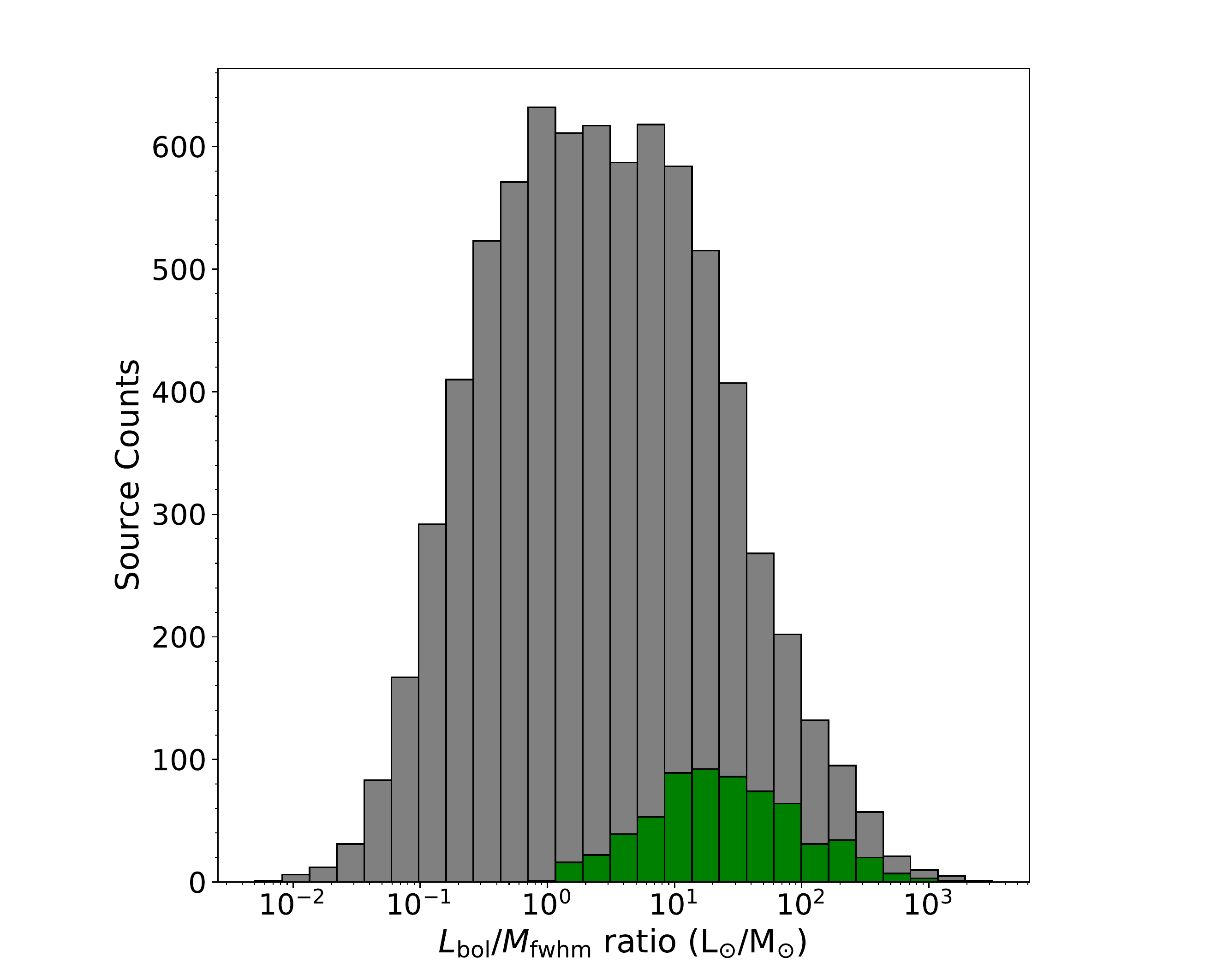}
	\includegraphics[width=0.47\textwidth]{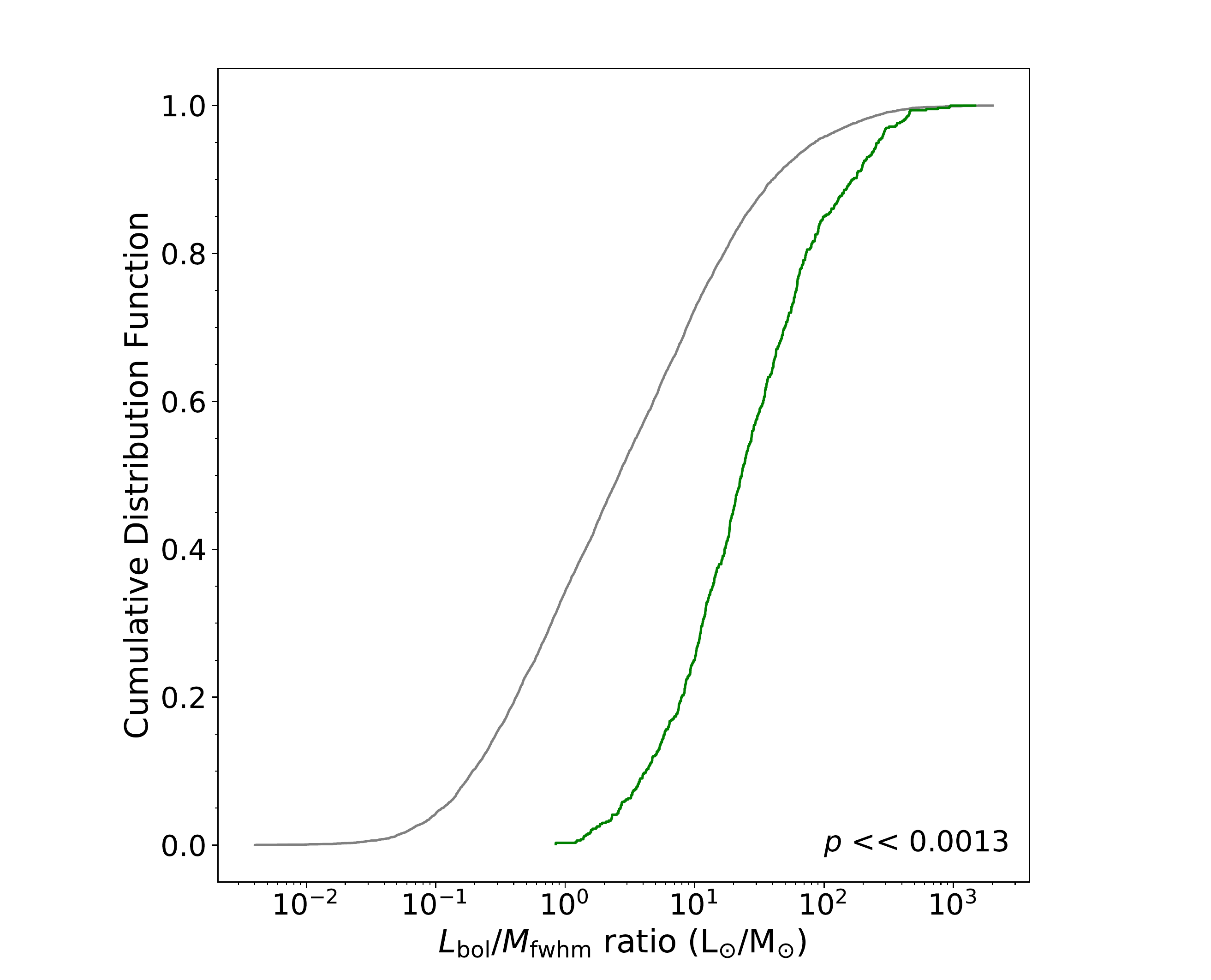} \\
	\caption{\lmratio\ ratio distributions. The histograms in the upper panel presents the entire distribution of the maser associated clumps and ATLASGAL sample in green and grey respectively, with he corresponding cumulative distribution function shown in the lower panel. The $p$-value results from a KS test is shown in the lower right of the CDF panel.}
	\label{fig:full_sample_lms}
\end{figure}

Luminosity-mass ratios have proven to be an effective indicator of the evolutionary stages of dense Galactic clumps, as mentioned in Sect.\,\ref{sect:physical_parameters}. As with the previous physical parameters, Fig.\,\ref{fig:full_sample_lms} presents the histogram and distance limited CDF of the \lmratio\ ratios of the maser associated clumps and the entire ATLASGAL sample. The clumps associated with a maser appear to occupy a distinct region of the parameter space, from $\sim$10$^{0.6}$ to 10$^{2.2}$ (central 80\,per\,cent). Therefore, masers are associated with a specific part towards the end, of the evolutionary process of star formation within dense clumps. 
 
\begin{figure}
	\includegraphics[width=0.47\textwidth]{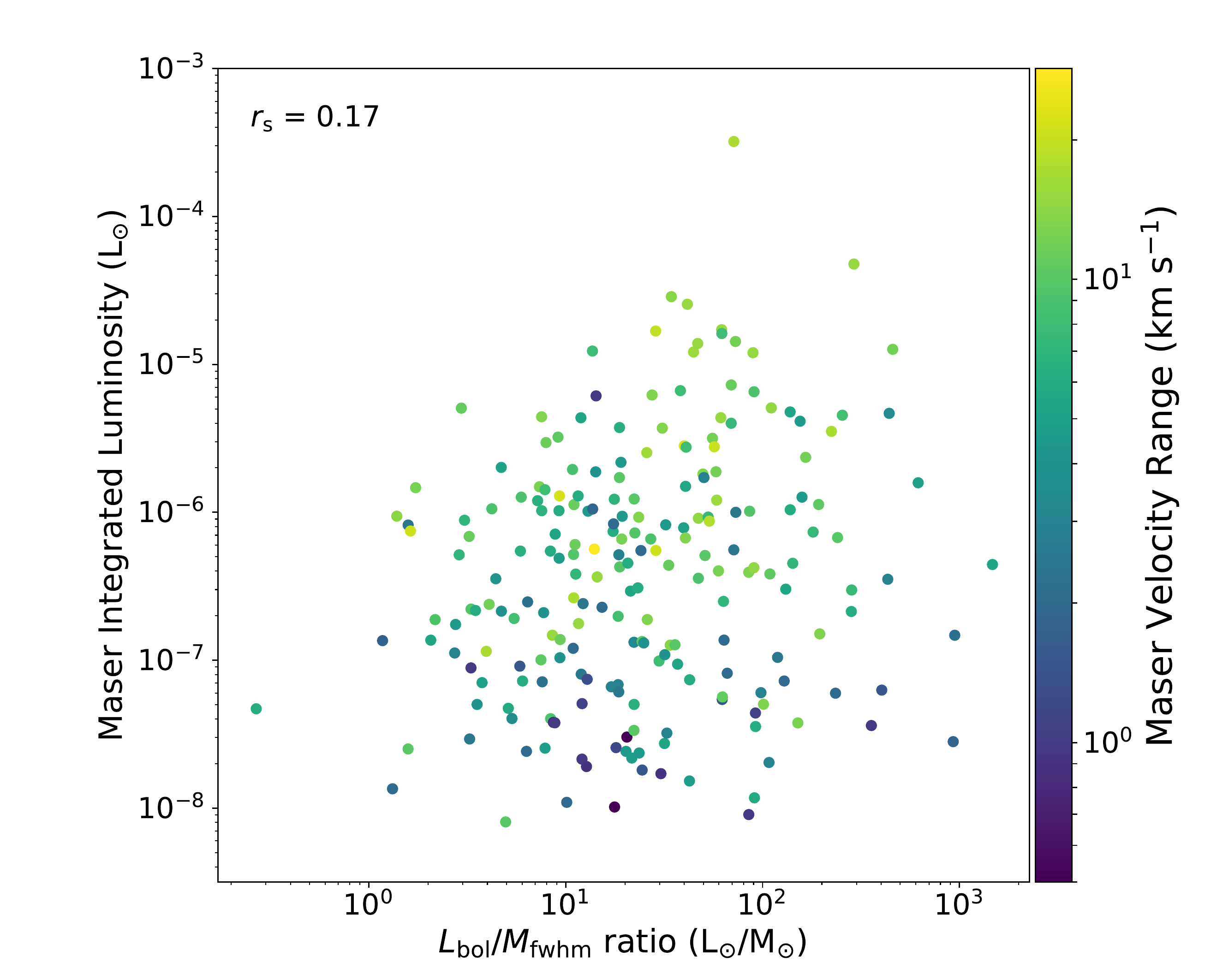}
	\includegraphics[width=0.47\textwidth]{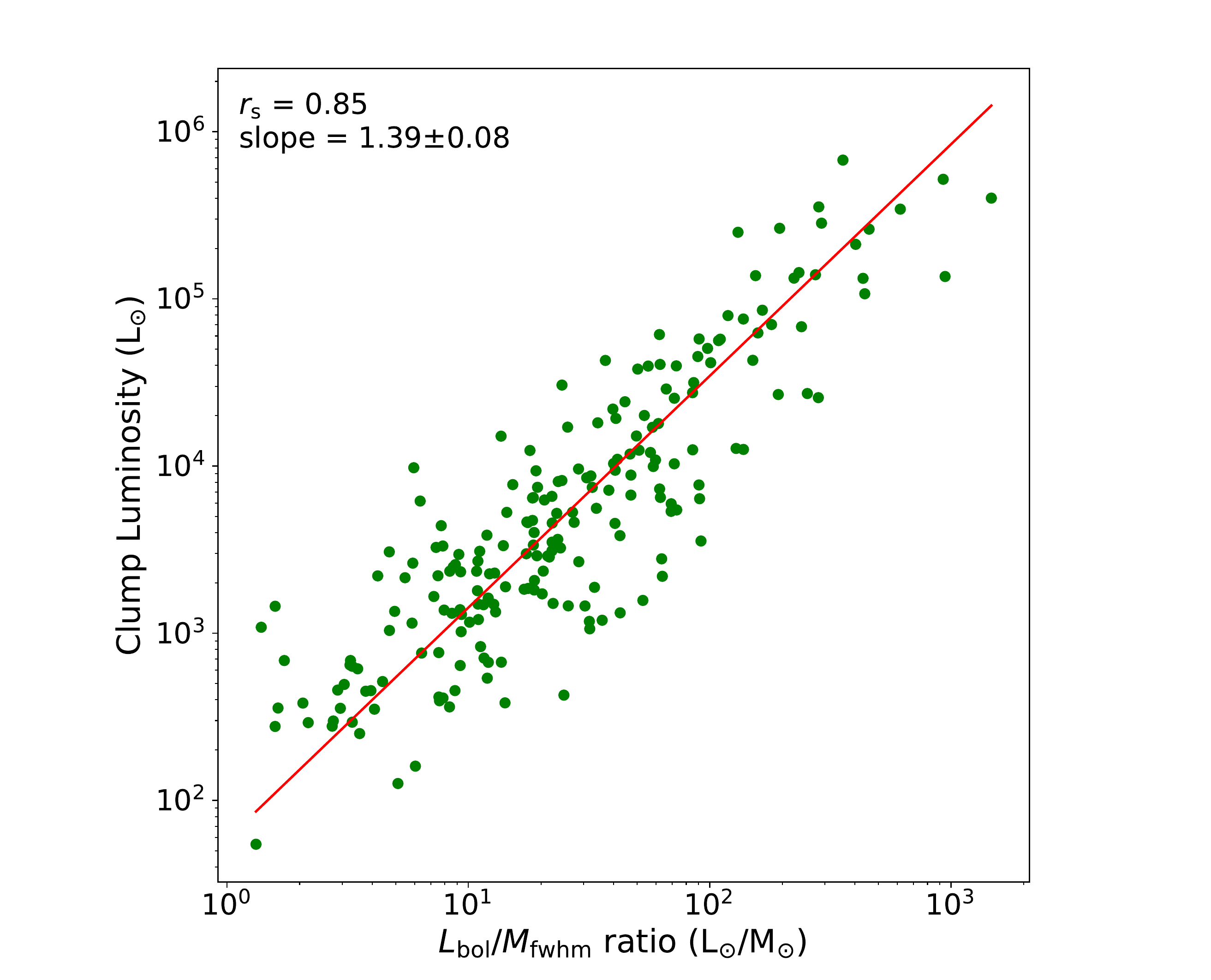}
	\includegraphics[width=0.47\textwidth]{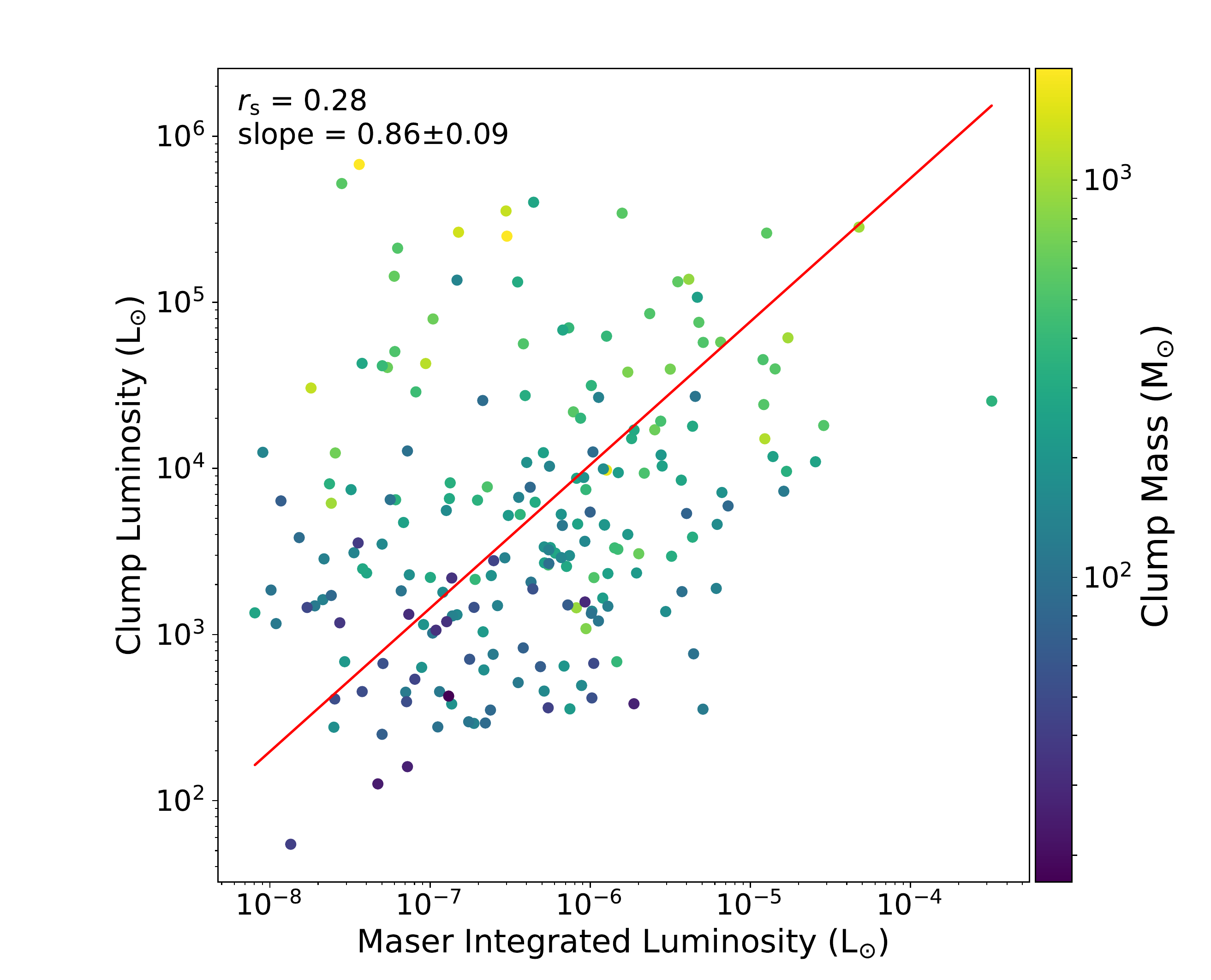}
	\caption{The upper panel presents the maser integrated luminosity distribution against clump \lmratio\ ratio with maser velocity range as a third parameter. The middle panel presents the clump bolometric luminosity distributed against \lmratio\ ratio. The lower panel presents the clump bolometric luminosity versus the maser integrated luminosity, with clump FWHM mass as a third parameter. All three plots have been fitted using orthogonal distance regression, shown as solid red lines. The corresponding Spearman's rank coefficients are shown in the upper left of each panel, and all associated $p$-values are less than our confidence threshold ($p$ < 0.0013). }
	\label{fig:mmb_lums_clump_lms}
\end{figure}

As \lmratio\ ratios can be used to study the evolutionary stages of clumps, we can use the derived parameters to test current theories of maser evolutionary stages. Previous studies have related the luminosity of different maser species with specific stages in the ongoing evolution. The upper panel of Fig.\,\ref{fig:mmb_lums_clump_lms} presents the distribution of methanol maser luminosity against the parent clump \lmratio\ ratio with the maser velocity range being shown as a third parameter. There is no significant correlation between the integrated luminosity of methanol masers and evolutionary stage of their corresponding clumps, and there is a large amount of scatter in this distribution. This scatter may be due to a number of reasons, such as the viewing angle of individual maser sources or their intrinsic periodic variability ($\sim$ an order of magnitude; e.g. \citealt{Goedhart2003, VanDerWalt2009, durjasz2019}), flux variations of pumping sources, variations in density, abundance and temperature, or the fact that the methanol masers are only present over a very limited range of the protostar's evolution. 


The middle panel of Fig.\,\ref{fig:mmb_lums_clump_lms} shows how the clump luminosity changes with increasing \lmratio\ ratio. The expected strong correlation is seen ($L_{\rm{bol}}$ $\propto$ \lmratio$^{1.39\pm0.08}$, $r_{\rm{s}}$ = 0.85) as the luminosity of evolving clumps generally determines the \lmratio\ ratio, due to the mass being relatively constant over the lifetime of a clump. In the lower panel Fig.\,\ref{fig:mmb_lums_clump_lms} we present the distribution of maser and clump luminosities for the distance limited sample (2 to 4\,kpc). It can be seen that as the luminosity of a clump evolves, the strength of corresponding maser emission increases. 

We find that $L_{\rm{bol}} \propto L_{\rm{maser}}^{0.86\pm0.09}$ with a Spearman's rank coefficient of $0.28$, this is consistent with the slope reported in our previous paper ($L_{\rm{bol}} \propto L_{\rm{maser}}^{0.93\pm0.08}$; \citealt{Urquhart2015}) that was determined from the peak flux. So there is no significant difference between estimating the maser luminosity from the peak or integrated flux. We also find this relationship to be true for the non-distance limited sample ($L_{\rm{bol}} \propto L_{\rm{maser}}^{0.90\pm0.04}$). The maser and bolometric luminosity are almost linearly correlated, this is consistent with the hypothesis that methanol masers are radiatively pumped \citep{Sobolev1997}. If, as we expect, the methanol maser is exclusively associated with massive stars, the linear correlation would suggest that the majority of the bolometric luminosity is the result of a single star within a clump \citep{Walsh2001}. 

The distribution found in the lower panel of Fig.\,\ref{fig:mmb_lums_clump_lms} has also been colour-coded based on clump FWHM mass. As it has been shown in Fig.\,\ref{fig:fwhm_mass_vs_lms} that the mass of clumps does not change during evolution, the lower panel of Fig.\,\ref{fig:mmb_lums_clump_lms} shows that the most intense masers are associated with the most massive clumps. We find that $L_{\rm{maser}} \propto M_{\rm{fwhm}}^{0.50\pm0.02}$. It is clear that maser strength is dependent on clump mass, however, this correlation is less significant than the maser dependence on bolometric luminosity. Therefore, it follows that the brightness of masers is affected more by the luminosities of embedded central objects rather than the mass of the surrounding material.

\subsection{Lower limits on physical conditions}
\label{sect:lower_limits}

\subsubsection{Stellar and Clump Mass}
\label{sect:stellar_mass}

It has been shown in previous studies that the class II 6.7\,GHz maser emission is only associated with intermediate to high mass stars \citep{Minier2003}. These authors performed a search for the class II 6.7\,GHz methanol maser towards 123 low-mass young stellar objects and protostellar condensations. The study failed to find any strong methanol masers associated with low-mass star formation, and found a lower protostellar mass limit of $\sim$3\,\msun, leading them to conclude that methanol masers are exclusively associated with intermediate and high-mass protostellar objects. 


We can estimate the minimum mass of embedded objects in our sample using the clump luminosity measurements that are mainly attributed to these central objects. Luminosity can be approximated from the mass of a star (L $\sim$ M$^{3.5}$; \citealt{Kuiper1938}). We find this minimum luminosity to be 10$^{2.77}$\,\lsun\ (590\,\lsun). Using this value we estimate the corresponding minimum mass of an embedded object to be $\sim$6\,\msun, assuming that the luminosity is primarily coming from a single source \citep{Walsh2001}. The uncertainty in this value is dominated by the error in the derived bolometric luminosities, a value of $\sim$42\,per\,cent (see Sect.\,\ref{sect:uncertainties}). Therefore, we find the uncertainty on our lower mass limit to be $\sim$2.5\msun. Our results strongly support the finding of \cite{Minier2003}.

\subsubsection{Volume Density}

\begin{figure}
	\includegraphics[width=0.47\textwidth]{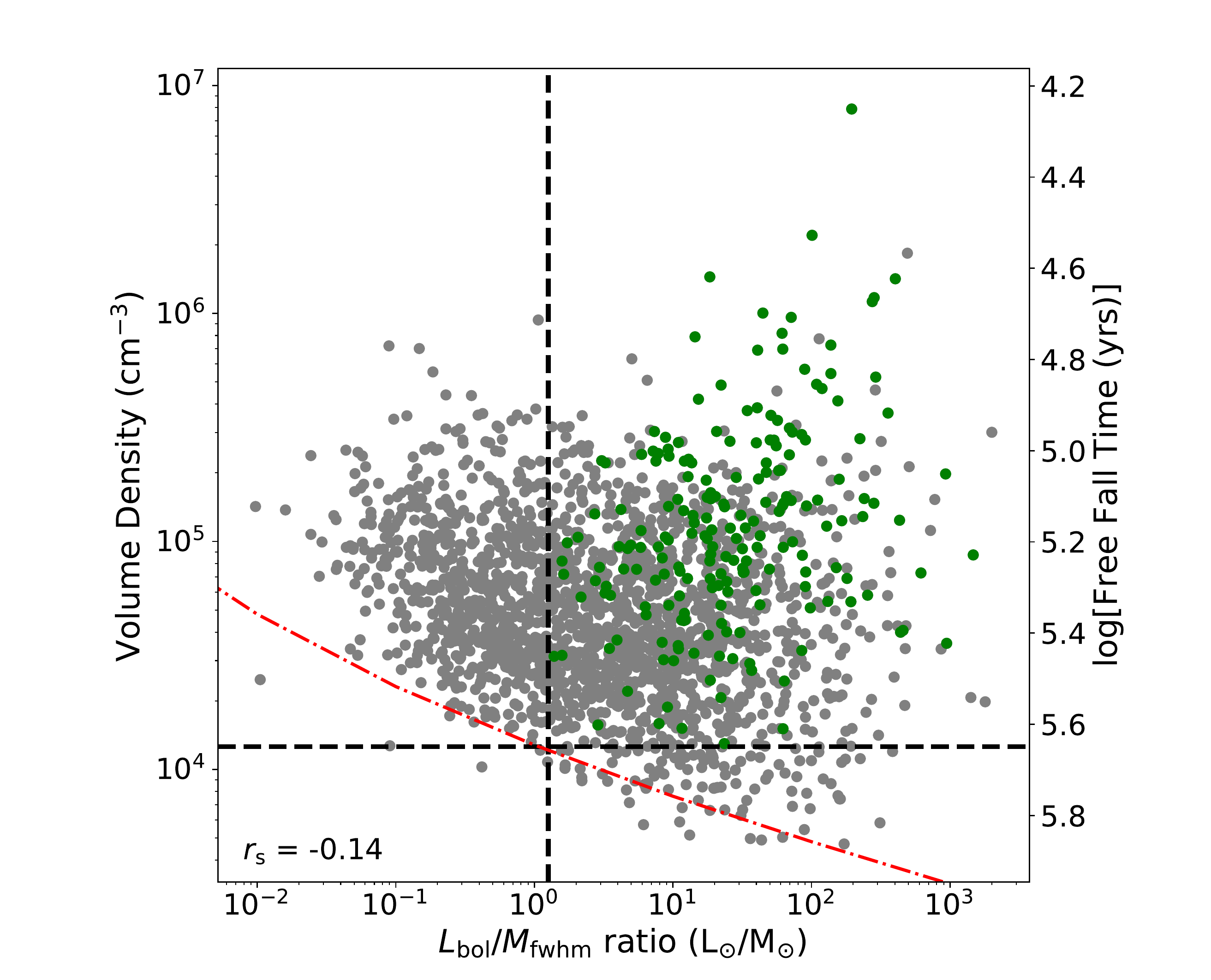}
	\caption{Distance limited sample of volume density versus \lmratio\ ratio. The entire ATLASGAL sample is shown in grey with the maser associated sample shown in green, the lower limits of each parameter are shown as dotted black lines, with the observational sensitivity limit of ATLASGAL shown as a red dash-dotted line. The results from a Spearman's correlation test for the full ATLASGAL sample is shown in the lower left of the panel, and the corresponding $p$-value is less than our significance threshold.}
	\label{fig:vol_den_lms}
\end{figure}

\begin{figure}
	\includegraphics[width=0.47\textwidth]{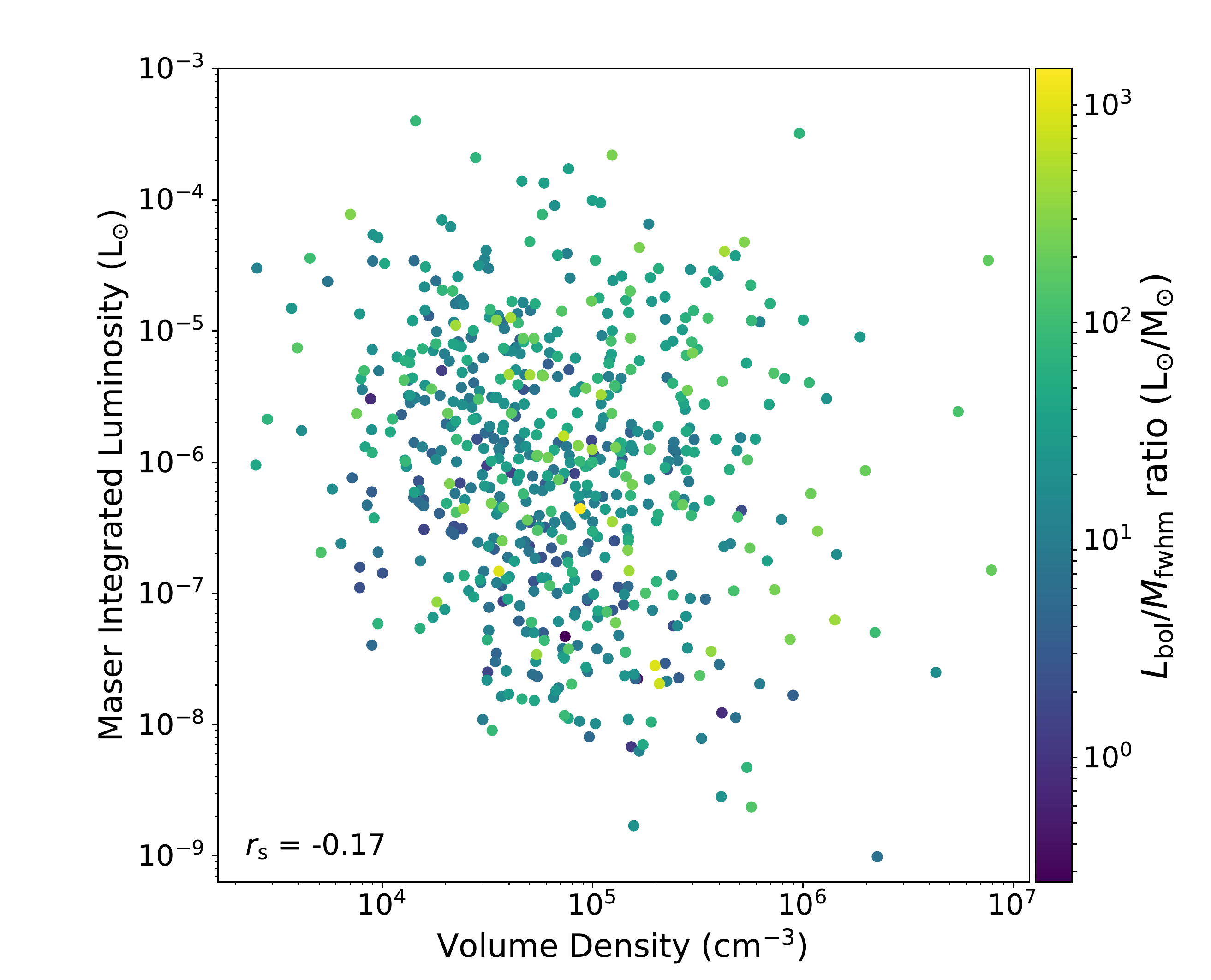}
	\caption{Distance limited sample of methanol maser luminosity versus volume density with the data coloured based on the \lmratio\ ratios. The results from a Spearman's correlation test is shown in the lower left of the panel.}
	\label{fig:mmb_int_vol_den}
\end{figure}

Section\,\ref{sect:physical_parameters} describes how the radii and masses of clumps have been recalculated in order to remove any temperature bias that may be present within the observations. Furthermore, new values for volume density have also been calculated to align with these new radius and mass values. In general, volume densities for maser associated clumps lie between 10$^{4}$ and 10$^{6}$ particles per cm$^{3}$ as seen in the lower left panel of Fig.\,\ref{fig:full_sample_radius_mass_volden}. Figure\,\ref{fig:vol_den_lms} presents the distribution of clump volume density against the corresponding \lmratio\ ratios for the distance limited sample, with the entire ATLASGAL and the maser associated samples shown in grey and green respectively. Given that the both the mass and radius of the clumps have been shown to be constant during evolution, it should follow that volume density is also constant. There is a small negative correlation between these two parameters, as confirmed by a Spearman's rank coefficient test, with an $r_{\rm s}$ value of $-0.14$. However, this is likely due to an observational bias as we are less sensitive to low volume densities at low temperatures (the sensitivity limit is shown but the dash-dotted curve in Fig.\,\ref{fig:vol_den_lms}). As a consequence, ATLASGAL is more sensitive to more evolved sources.

It appears that the maser associated clumps occupy a well defined region in this parameter space, \lmratio\ ratio values of $\geq$1, as discussed above (see Sect.\,\ref{sect:l/m_ratios}), and volume densities above $\sim$10$^{4.1}$\,\cmthree. This suggests that there is a lower density limit for the production of maser emission and that there is a clear maser turn on stage during protostellar evolution. It may be that the measured mass of a clump is not as important in the production of maser emission as the volume density, as shown by the middle right panel of Fig.\,\ref{fig:full_sample_radius_mass_volden}. We find this lower limit on volume density to be 10$^{4.1}$\,\cmthree\ and any clump with a value lower than this is unlikely to produce maser emission. Masers associated with compact, dense clumps have a specific turn-on point and must be driven by an intermediate to high-mass protostar.

We have also tested the maser luminosities against the corresponding volume densities. Figure\,\ref{fig:mmb_int_vol_den} shows the relationship between volume density and the integrated maser luminosity. It is clear from the distribution of these parameters that there is a weak correlation between them ($r_{\rm s} = -0.17$; $p$-value $\ll 0.0013$); this is consistent with the findings of \citet{Urquhart2013} and \citet{Urquhart2015}. \citet{Breen2010} and \citet{Breen2011a} also investigated the correlation between the peak and integrated luminosities of the 6.7\,GHz methanol maser transition and reported trends of decreasing volume density with increasing maser luminosity ($r_{\rm s} = -0.46$ and -0.39 for the peak and integrated luminosities for samples sizes of 113 and 46 respectively). While these previous studies and the work presented here both find negative correlations within this distribution, the result determined here is likely to be more robust due to the significantly larger sample (613 sources).

\subsection{Maser Lifetimes}

The lifetime of the methanol maser phase has been investigated in a number of previous studies (e.g. \citealt{Codella2004,VanDerWalt2005}). In this subsection we use our much larger sample in an effort to provide a more concrete estimate on this maser lifetime. It is likely that the maser lifetime is not a constant across every single star forming region of the Galaxy and is dependent on a number of factors, including the mass of the central object, clump volume density, methanol abundance and the amount of ionising radiation over time, although our large sample should provide strong statistics to calculate a firm lifetime for maser emission.

Free fall times have been derived for each clump in our sample with a corresponding volume density measurement (see Sect.\,\ref{sect:physical_parameters}). These free fall times range from $\sim2\times 10^4$ to 10$^6$\,yrs, and masers will exist for a portion of these time scales. It is likely that maser emission is seen once a relatively efficient pumping mechanism is formed and will continue to exist until an UC \hii\ region forms and begins to disrupt the local environment. Therefore, to calculate how long this maser phase lasts, we have taken the ratio of the number of maser clumps against the total number of dense clumps within the ATLASGAL sample, to derive a clump ratio at free fall time intervals. These ratios can then be multiplied by the clump free fall times to give an absolute value for statistical lifetime for the maser phase at specific clump volume densities (i.e. $t_{\rm{stat}} = \frac{N_{\rm{maser}}}{N_{\rm{all}}} \times t_{\rm{ff}} $). 
\begin{figure}
	\includegraphics[width=0.47\textwidth]{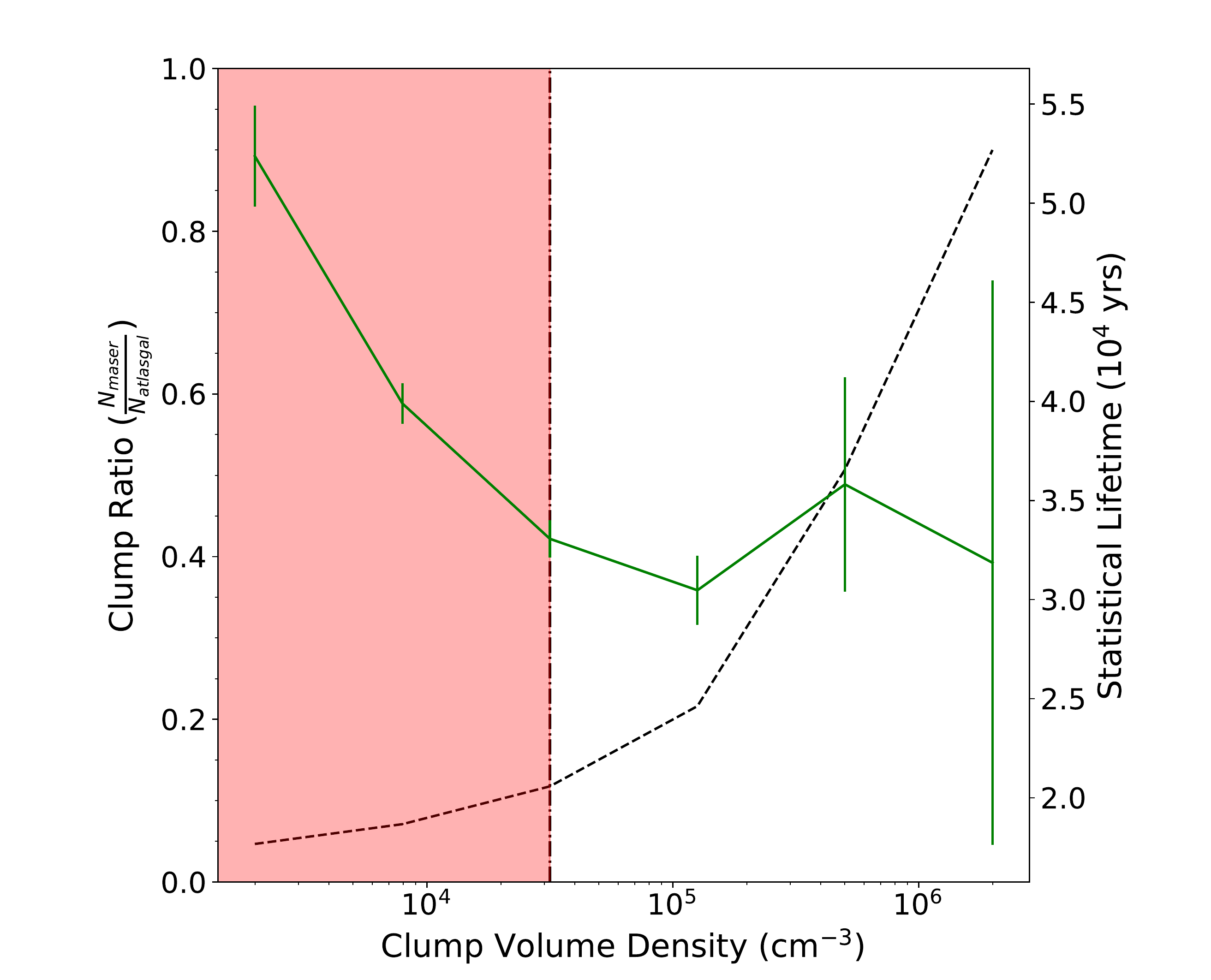}
 	\caption{Plot presenting the clump ratio ($\frac{N_{\rm{maser}}}{N_{\rm{all}}}$), and statistical lifetime as a function of clump volume density. The black dotted line shows the increasing number of maser associated clumps with respect to volume density, whereas the solid green line presents the drop in statistical lifetime as volume density increases. The shaded region represents the parameter space where our sample is incomplete. Errors shown are derived from Poisson statistics.}
	\label{fig:lifetimes}
\end{figure}

Figure\,\ref{fig:lifetimes} presents how these lifetimes change through the volume density parameter space, the errors on this plot are derived using Poisson statistics. Volume densities appear to be complete above 10$^{4.5}$\,\cmthree\ (see Fig.\,\ref{fig:mmb_int_vol_den}), and therefore, we only consider the lifetimes of clumps with a density value above this threshold.

The lifetimes range between 3$\times$10$^{4}$ and 3.6$\times$10$^{4}$\,years, with a mean error of $\sim$10\,per\,cent. We find the average maser statistical lifetime to be $\sim$3.3$\times$10$^{4}$\,years, with a standard deviation of 0.23$\times$10$^{4}$ years. While volume density is dependent on distance, we find that as the statistical lifetimes depend on the ratio between two volume density samples, the overall calculation for these lifetimes is distance independent. 

As the radii of clumps are constant across their lifetime, volume density is dependent on the clump mass, our results show that the most dense clumps, and therefore, the most massive will have associated maser emission with the shortest lifetimes. This falls within the currently accepted theory whereby denser clumps will evolve on much shorter timescales and will quickly produce \hii\ regions, disrupting masing material much faster than lower density clumps. The least dense and massive clumps take longer to produce \hii\ regions and so, potentially, the conditions for maser emission lasts for a greater period of time.  The proportion of clumps with maser emission increases significantly with density (see dashed line in Fig.\,\ref{fig:lifetimes}), and the majority of masers are associated with YSOs (63\,per\,cent). Therefore, it is likely that pre-stellar and protostellar stages are extremely short-lived in high-density clumps and so are less observed, resulting in a low association rate between protostellar objects and methanol maser sources.

\begin{figure*}
	\includegraphics[width=0.47\textwidth]{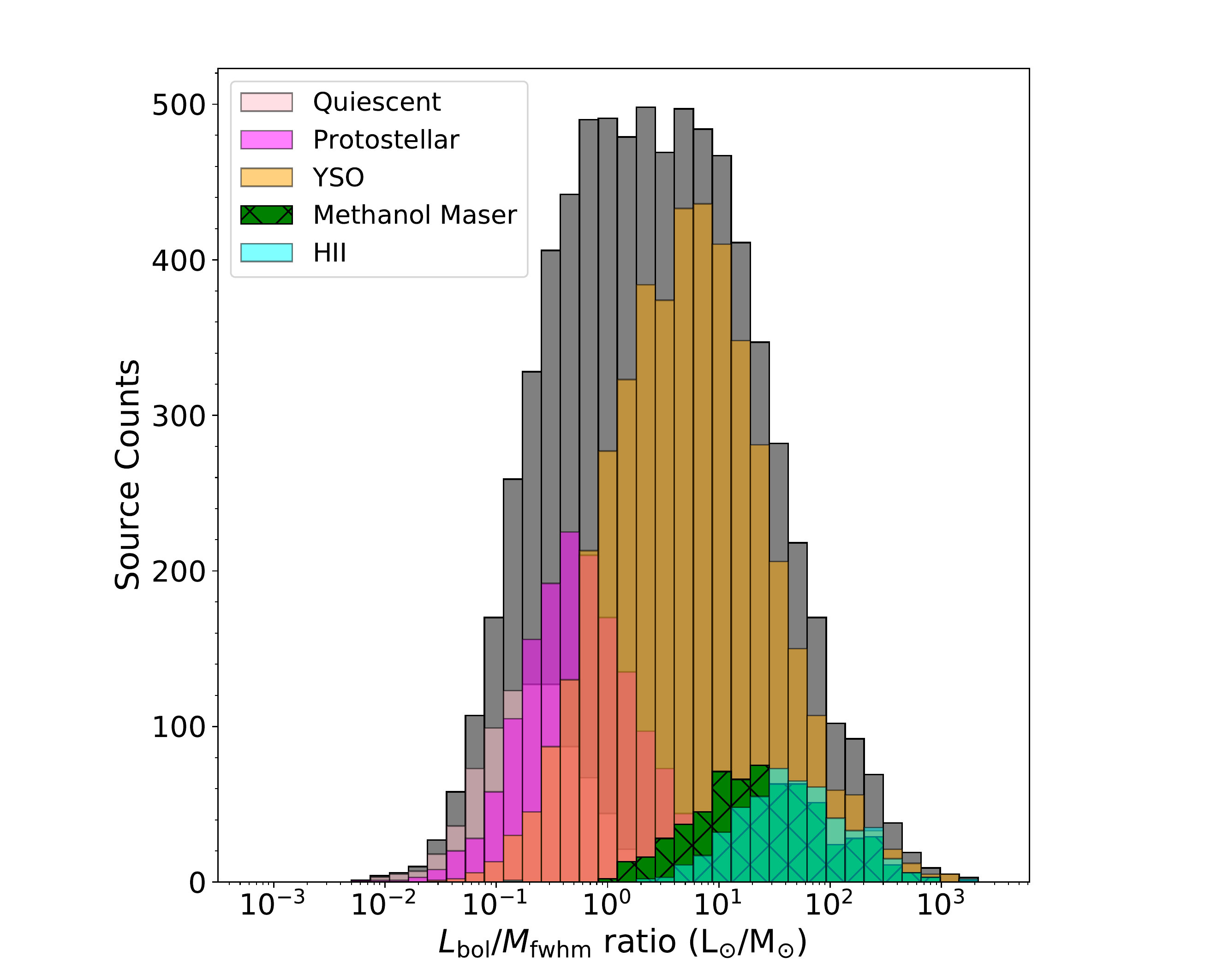}
	\includegraphics[width=0.47\textwidth]{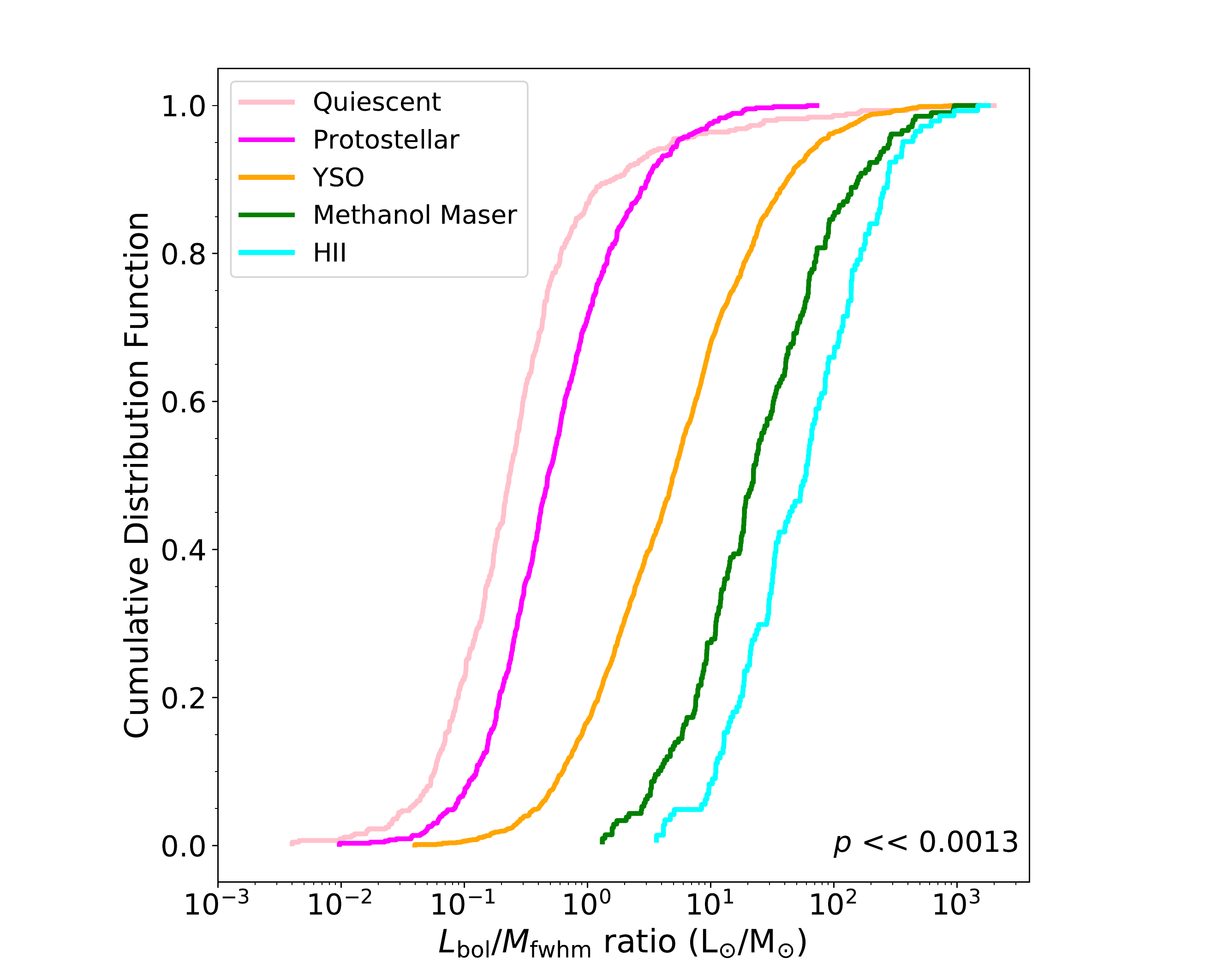}
	\caption{The left panel presents a histogram of \lmratio\ ratios for each evolutionary stage defined by the ATLASGAL survey, with the entire ATLASGAL sample being shown in grey. Maser associated clumps are shown with black hatching to clearly show where these objects lie within this parameter space. The right panel presents the cumulative distribution function for this histogram, for a distance limited sample (2 to 4\,kpc). Legends with evolutionary classification are shown in the upper left of each panel.}
	\label{fig:full_sample_lms_all}
\end{figure*}

\cite{VanDerWalt2005} undertook a statistical study to estimate the lifetimes of the class II 6.7\,GHz methanol masers. This estimate was based on the current known number of masers at the time. This work applied a completeness correction to this number to estimate the potential number of maser detections that should exist within the Galaxy. At the time only 519 methanol masers sources were known, and he estimated that the minimum number of maser detections present across the Galaxy should be 850, this value is slightly lower than the number of maser sites detected within the MMB survey. \cite{VanDerWalt2005} found that the maser lifetimes should lie between 2.5$\times$10$^{4}$ and 4.5$\times$10$^{4}$ years, the variation within this result was mainly due to the different initial mass functions used in the analysis. 

The lifetimes we have determined from the ratio of maser associated and unassociated clumps, and the free-fall times lie within the range calculated by \cite{VanDerWalt2005}. Our results, therefore,  provide strong observational support for theoretical predictions of the class II 6.7\,GHz methanol maser lifetimes.

\subsection{Embedded Evolutionary Stages}
\label{sect:evo_stages}

We have already found that the methanol maser phase covers a relatively narrow range in the evolution of clumps (see Fig.\,\ref{fig:full_sample_lms}). It is likely that the masers are associated with nearby young massive embedded objects \citep{Norris1993,Norris1998,Lee2001} that themselves cover a range of evolutionary stages. The ATLASGAL catalogue of dense clumps is likely to include the full range of evolutionary stages and comparing the properties of the full ATLASGAL sample with those associated with methanol masers can provide a way to observe how the physical parameters of dense clumps change during the ``maser phase''. 

The ATLASGAL survey has classified each detected clump based on the most likely associated embedded object. Clumps have been classified as one of the following: quiescent, protostellar, young stellar object and \hii\ region (see Sect.\,4.1 \citealt{Urquhart2018} for a complete description of the evolutionary sequence). We aim to evaluate where sources with maser emission fit within this classification scheme. Figure\,\ref{fig:full_sample_lms_all} presents a histogram and cumulative distribution function of \lmratio\ ratios based on the clumps' evolutionary classification. The expected distribution is seen for the different classifications; quiescent clumps have the lowest \lmratio\ ratios, which increase when the protostellar object forms and continue to increase as more material is accreted onto the protostar and its luminosity increases. As the bolometric luminosities increase the peak of the SED shifts to shorter wavelengths and when it becomes detectable at mid-infrared wavelengths they are classified as young stellar objects (YSO). The highest \lmratio\ ratios are attributed to \hii\ regions, which are signposts of the final embedded stages in the formation of the highest-mass stars ($> 10$\,\msun). This distribution can be more clearly seen in the CDF (see right panel of Fig.\,\ref{fig:full_sample_lms_all}). 

The ``maser phase'' can be defined using the luminosity-mass ratios of the dense dust clumps. The methanol masers are associated with \lmratio\ ratios of between 10$^{0.6}$ and 10$^{2.2}$ and therefore occupy a distinct part of this parameter space as discussed in Sect.\,\ref{sect:l/m_ratios}. The maser \lmratio\ distribution also indicated (green histogram and curves in the plots shown in Fig.\,\ref{fig:full_sample_lms_all}). Generally, the maser associated clump \lmratio\ ratios overlap with those of protostellar and YSO stages but are most similar to that of the \hii\ region associated clumps. This is expected, as masers are thought to turn on once a protostellar source of sufficient mass has been formed ($\gtrsim 6$\,\msun; \citealt{Minier2003} and Sect.\,\ref{sect:stellar_mass} of this paper), and continue until the physical conditions required are disrupted when the feedback from the central source (i.e., expanding \hii\ regions and dispersion of the host clump). This hypothesis is supported by the results reported by a number of studies that have investigated the association between methanol masers and \uchii\ regions (e.g. \citealt{Walsh1997, Walsh1998, VanDerWalt2003}). The CDF of the \lmratio\ methanol maser associated clumps (shown in the right panel of Fig.\,\ref{fig:full_sample_lms_all}) further provides strong evidence for this hypothesis.

\cite{Walsh1998} found that there was a $\sim$20-25\% coincidence of 6.7\,GHz masers and radio continuum emission, which would suggest that the conditions required for the masers emission persist for a short period after the formation of the \uchii\ region. A similar association rate has also been reported in other studies (i.e. \citealt{Hu2016}). Where there is an association, it is likely that the \hii\ region is not yet developed enough to disrupt the masing material. Given the mean lifetime of the methanol maser is $\sim3\times 10^4$\,yrs this period is likely to be $\sim7\,500$\,yrs.

\begin{figure}
	\centering
	\begin{tikzpicture}
	\draw[black, thick] (-1,0) circle (2cm);
	\node at (-1,2.3) {Methanol maser};
	\node at (-1.7,0.2) {603};
	\node at (-1.7,-0.2) {(187)};
	\draw[black, thick] (1,0) circle (2cm);
	\node at (1,2.3) {\hii\ region};
	\node at (1.7,0.2) {623};
    \node at (1.7,-0.2) {(131)};
	\node at (0,0.2) {252};
	\node at (0,-0.2) {(46)};
	\end{tikzpicture}
	\caption{Venn diagram presenting the overlap between methanol masers and \hii\ regions. Values determined by \protect\cite{Walsh1998} are shown in brackets and they find 20\,per\,cent of methanol masers are coincident with \hii s while a larger proportion (25\,per\,cent) of \hii s are associated with methanol masers. Results from this study show these values to be 30\,per\,cent and 29\,per\,cent respectively.}
	\label{fig:venn_hii_matches}
\end{figure}
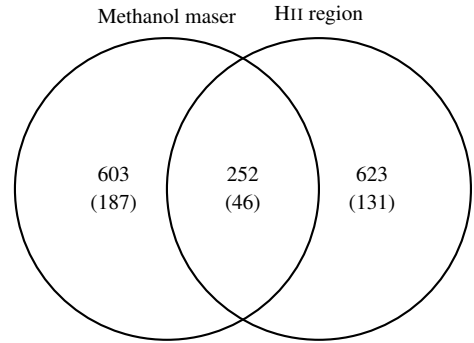

Excluding the Galactic centre region ($5\degr < |\ell| < 60\degr$), we have used the full sample of matched masers presented in this study, 855 maser emission sources, to evaluate the association rate between methanol masers and \hii\ regions. We find that for the 855 maser sources, 252 are associated with a clump that harbours a \hii\ region, either ultra compact or extended, as identified by the ATLASGAL survey follow up observations. Within the ATLASGAL survey, 875 dust continuum sources are found to be associated with a \hii\ region and therefore 29\,per\,cent of \hii\ regions are also associated with a methanol maser. Figure\,\ref{fig:venn_hii_matches} presents a Venn diagram of the maser - \hii\ region matches from \cite{Walsh1998} and this work. The association rate between masers and \hii\ regions found in this study (30\,per\,cent) is slightly higher than that found by \cite{Walsh1998} (20-25\,per\,cent), and more similar to the association rate presented in \cite{Hu2016}.


\cite{Breen2010} defined an evolutionary sequence based on 1.2\,mm emission, 6.7\,GHz methanol maser emission and 8\,GHz radio continuum emission. The results presented in this work support the \cite{Breen2010} model and, using the \lmratio\ ratio, we are able to put a quantitative result to their sequence: 

\begin{enumerate}
	
	\item Clumps initially begin with no detectable maser or radio continuum emission. 
	
	\item Once a clump has an \lmratio\ ratio of $\sim$10$^{0.6}$, 6.7\,GHz maser emission turns on, and we show, in Sect.\,\ref{sect:lower_limits}, that these clumps must also have a protostellar core of $>6$\,\msun\ and volume density above 10$^{4.1}$\,\cmthree.
	
	\item As the \lmratio\ ratio approaches a value of $\sim$10$^{0.65}$, radio continuum emission produced from \hii\ regions is seen in conjunction with maser emission. 
	
	\item As \lmratio\ ratios increase towards 10$^{2.2}$, maser emission begins to decline as the effects of \hii\ regions begin to disrupt the local environment, and the \hii\ regions themselves become extended. 
	
\end{enumerate}

\cite{Breen2011a} demonstrated that, during this evolutionary sequence, the integrated luminosity and velocity ranges of the 6.7\,GHz methanol maser increase, which is supported by our work. However, due to the amount of scatter in these distributions, the maser luminosity and velocity range is not a particularly useful indicator of evolution.

\section{Conclusions}
\label{sect:conclusions}

We have compiled the largest and most representative sample of class II 6.7\,GHz methanol maser associated dust clumps to date. In this latest study, we have matched 257 out of 265 methanol maser sources in the region $20\degr < l < 60\degr$, to submm emission in either JPS or ATLASGAL. In total, 958 out of 972 methanol maser sources identified by the MMB survey across the Galactic plane have been matched to submm dust emission, an association rate of 98.6\,per\,cent. Two of the unmatched masers lie beyond the coverage of both JPS and ATLASGAL, with the remaining 12 appearing to have no corresponding submillimetre emission.

Physical parameters (sizes, masses, densities, luminosities, virial parameters, statistical lifetimes) have been derived for each maser associated clump for which a distance measurement is available and correlations have been investigated. A summary of the main findings of this study are listed below:

\begin{enumerate}
    
    \item We have recalculated the radii, masses and volume densities of clumps within the ATLASGAL survey, and we find that previous effects seen during the evolution of star-forming clumps (increasing radius, mass and density) can be attributed to an observational bias. By using the FWHM values for these parameters, we have removed this bias and find that these physical properties remain constant across the lifetime of individual clumps. Therefore, the mass and radius of clumps, and by extension density, are independent of evolutionary stage and these properties are determined at the earliest stages in the formation of these dense clumps. \\
    
    \item There is an almost ubiquitous association between the 6.7\,GHz methanol maser and dust continuum sources, with 958 (99\,per\,cent) of sources within the MMB survey catalogue being matched to either 850 or 870\,\micron\ emission across the Galactic plane ($186 \degr \leq \ell \leq 60\degr$). This is strong evidence for masers being tightly associated with star formation, and in nearly all cases the clumps show evidence of being in an advanced stage of star formation, due to their associations with either a YSO or \hii\ region ($\sim$95\,per\,cent). \\

    \item The bolometric luminosities of clumps has been found to be proportional to the luminosities of associated maser sources with a power-law exponent of order unity ($L_{\rm{bol}} \propto L_{\rm{maser}}^{0.86\pm0.09}$). This provides strong support for the currently accepted theory that the 6.7\,GHz methanol maser transition is radiatively pumped. Furthermore, the linear correlation would also suggest that most of the bolometric luminosity can be attributed to a single star that is driving the methanol maser. \\
    
    \item Using the bolometric luminosities of embedded objects associated with maser emission, we have derived a minimum mass for stellar objects associated with methanol maser emission. We find this limit to be $\geq6$\,\msun, which is consistent with previous studies that have concluded that these masers are exclusively associated with intermediate and high-mass star formation (\citealt{Minier2003}). \\
    
    \item The distribution of the masses of clumps associated with methanol masers is indistinguishable from the general population of dense clumps and so mass alone is not an important parameter in determining the current or future occurrence of a methanol maser. However, the radius and density of maser associated clumps are significantly different from the values determined for the full sample of clumps, but as radius is shown not to alter during evolution, it is likely that the density of clumps is the most significant indicator of current or future maser emission. Our results suggest that there is a lower density threshold below which no masers are found ($n$(H$_2$)$ \geq10^{4.1}$\,cm$^{-3}$), and could indicate that intermediate and high-mass star formation, and maser emission, is inefficient below this threshold.\\
    
    \item The evolutionary sequence for clumps spans over five orders of magnitude of the \lmratio\ ratio parameter space (0.01 to 10$^3$\,\lsun\,\msun$^{-1}$). We find that the maser associated clumps occupy a relatively narrow region ($\sim1.5$ orders of magnitude) within this parameter space. Therefore, there is a well defined maser turn-on and turn-off stage in the evolutionary process of high-mass star formation. The low association rates with protostellar objects is likely to be linked to the minimum stellar mass required to drive the maser and the time needed to accrete sufficient material to attain this mass. We quantify this evolutionary process in Sect\,\ref{sect:evo_stages} and present the necessary physical parameters for the production of maser emission.\\
    
    \item We have used the free-fall times and the fraction of clumps associated with a methanol maser to estimate a statistical lifetime of $\sim$3.3$\times$10$^{4}$\,years for the maser phase. There is a 30\,per\,cent coincidence between masers and \hii\ regions, and maser emission persists for a short time after the creation of an \hii\ region, we find this time period to be $\sim$7\,500 years. The \lmratio\ ratios of maser associated clumps, and \hii\ region associated clumps supports previous work in the development of a evolutionary sequence based on maser and radio continuum emission in the Galactic plane \citep{Breen2010}.

\end{enumerate}

Overall, we provide a firm calculation of the lifetime for the 6.7\,GHz methanol maser ($\sim$3.3$\times$10$^{4}$\,yrs) and the required parameters for its production (core mass $\geq$6\,\msun\ and volume densities above 10$^{4.1}$\,\cmthree). This work provides a concrete quantification for the ``straw man'' model present by \cite{Ellingsen2007}. In future work, we will develop this model further by deriving statistical lifetimes of multiple maser species using the physical parameters provided by our continuing analysis of the ATLASGAL compact source catalogue.

\section*{Acknowledgements}

S.\,J.\,Billington wishes to acknowledge an STFC (Science and Technology Facilities Council) PhD studentship for this work. We have used the collaborative tool Overleaf available at: https://www.overleaf.com/.




\bibliographystyle{mnras}
\bibliography{library,urquhart2016}







\bsp	
\label{lastpage}

\end{document}